\documentclass[prx,aps,twocolumn,a4paper,floatfix,showpacs,nofootinbib]{revtex4-1}

\usepackage[utf8]{inputenc}  
\usepackage[T1]{fontenc}
\usepackage{graphicx,psfrag}
\usepackage{mathrsfs}
\usepackage{amsmath,amsfonts,amssymb,pifont,gensymb}
\usepackage{multirow,enumerate}
\usepackage{comment,hyperref}
\usepackage{color}
\usepackage{acronym}
\usepackage{xspace}
\usepackage[normalem]{ulem}
\usepackage{mathtools}
\usepackage{subfigure}
\usepackage{makecell}
\usepackage{enumitem}

\newacro{BH}{black hole}
\newacro{NS}{neutron star}
\newacro{PN}{Post-Newtonian}
\newacro{BBH}{binary black hole}
\newacro{BNS}{binary neutron star}
\newacro{EOB}{effective-one-body}
\newacro{NR}{numerical relativity}
\newacro{GW}{gravitational wave}
\newacro{EOS}{equation-of-state}

\newcommand{\be}{\begin{equation}}
\newcommand{\ee}{\end{equation}}
\newcommand{\bea}{\begin{eqnarray}}
\newcommand{\eea}{\end{eqnarray}}
\newcommand{\bel}{\begin{align}}
\newcommand{\eel}{\end{align}}

\newcommand{\linf}{\texttt{LALInference}}

\newcommand{\ptr}{p_{\small\textrm{tr}}}

\def\GMc2{{\rm G M_{\odot} c^{-2}}}

\def\SEOBNRv4T{\texttt{SEOBNRv4T}\xspace}

\usepackage{color}
\definecolor{cyan}{rgb}{0,0.9,0.9}
\definecolor{orange}{rgb}{0.9,0.5,0}
\definecolor{magenta}{rgb}{1,0,1}
\definecolor{purple}{rgb}{0.8,0.4,0.8}
\definecolor{gray}{rgb}{0.5,0.5,0.5}
\definecolor{mygreen}{rgb}{0.1,0.8,0.1}
\definecolor{darkblue}{rgb}{0.0,0.0,0.6}

\begin{document}

\title{Parameter estimation for strong phase transitions in supranuclear matter using gravitational-wave astronomy}

\author{Peter T.\ H.\ \surname{Pang}$^{1,2}$}\thanks{thopang@nikhef.nl}
\author{Tim \surname{Dietrich}$^3$}
\author{Ingo \surname{Tews}$^4$}
\author{Chris \surname{Van Den Broeck}$^{1,2}$}

\affiliation{${}^1$ Nikhef, Science Park, 1098 XG Amsterdam, The Netherlands} 
\affiliation{${}^2$ Department of Physics, Utrecht University, Princetonplein 1, 3584 CC Utrecht, The Netherlands}
\affiliation{${}^3$ Institute for Physics and Astronomy, University of Potsdam, Karl-Liebknecht-Str. 24/25, 14476, Potsdam, Germany}
\affiliation{${}^4$ Theoretical Division, Los Alamos National Laboratory, Los Alamos, New Mexico 87545, USA}

\date{\today}

\begin{abstract}
At supranuclear densities, explored in the core of neutron stars, a strong phase transition from hadronic matter to more exotic forms of matter might be present. 
To test this hypothesis, binary neutron-star mergers offer a unique possibility to probe matter at densities that we can not create in any existing terrestrial experiment. 
In this work, we show that, if present, strong phase transitions can have a measurable imprint on the binary neutron-star coalescence and the emitted gravitational-wave signal. 
We construct a new parameterization of the supranuclear equation of state that allows us to test for the existence of a strong phase transition and extract its characteristic properties purely from the gravitational-wave signal of the inspiraling neutron stars. 
We test our approach using a Bayesian inference study simulating 600 signals with three different equations of state and find that for current gravitational-wave detector networks already twelve events might be sufficient to verify the presence of a strong phase transition. 
Finally, we use our methodology to analyze GW170817 and GW190425,  
but do not find any indication that a strong phase transition is present at densities 
probed during the inspiral.  
\end{abstract}

\maketitle

\section{Introduction}
\label{sec:introduction}

Neutron Stars (NSs) are remnants of core-collapse supernovae and contain matter at the highest densities that we can observe in the Universe, up to several times nuclear saturation density, $n_{\rm sat}=0.16$ fm$^{-3}$, which corresponds to a mass density of $2.7 \times 10^{14}$ g cm$^{-3}$. 
Hence, NSs are perfect laboratories to determine the unknown equation of state (EOS) of dense matter. 
The EOS relates the pressure with the energy density in the NS interior and is determined by the fundamental degrees of freedom inside the NS and their interactions among each other. 
Each possible EOS determines the global structure of NSs, i.e., their masses and radii, in a unique way.
Therefore, detailed astronomical observations of NSs, in particular of binary NS (BNS) coalescences, are of extreme importance to nuclear physicists and allow us to constrain the dense-matter EOS.
To date, most EOS constraints stem from NS mass measurement~\cite{Demorest:2010bx, Antoniadis:2013pzd, Cromartie:2019kug} or radius extractions from X-ray observations~\cite{Steiner:2012xt, Miller:2019cac, Riley:2019yda}. 
The latter however, suffer from relatively large statistical~\cite{Miller:2019cac} or systematic~\cite{ozel:2016oaf} uncertainties. 
In addition to electromagnetic (EM) observations, the remarkable observation of gravitational waves (GWs) from a BNS merger in 2017, GW170817, by Advanced LIGO~\cite{TheLIGOScientific:2014jea} and Advanced Virgo~\cite{TheVirgo:2014hva} provided another avenue to determine NS properties and the EOS~\cite{TheLIGOScientific:2017qsa,GBM:2017lvd,Monitor:2017mdv}. 
Numerous efforts have been made to extract information on the EOS from the GW signal of BNS mergers, see e.g., Refs.~\cite{Read:2009yp, Markakis:2010mp, Markakis:2011vd, Read:2013zra, DelPozzo:2013ala, Lackey:2014fwa, Agathos:2015uaa, Annala:2017llu, Fattoyev:2017jql, Most:2018hfd, Lim:2018bkq, Tews:2018chv, Greif:2018njt, Raaijmakers:2019dks, PhysRevD.101.063007, Landry:2020vaw, Chatziioannou:2019yko, Essick:2020flb, wysocki2020inferring, Dietrich:2020lps, Landry:2020vaw}; cf. Ref.~\cite{Chatziioannou:2020pqz} for a recent review and further references. 

Constraints from GW170817 arise either purely from the analysis 
of the GW signal, e.g., Refs.~\cite{TheLIGOScientific:2017qsa, Dai:2018dca, De:2018uhw, Abbott:2018wiz, Abbott:2018exr, LIGOScientific:2018mvr}, from a combination of GW and EM information, e.g, Refs.~\cite{Radice:2017lry,Bauswein:2017vtn,Margalit:2017dij,Coughlin:2018miv, Radice:2018ozg, Coughlin:2018fis, Dietrich:2020lps}, or from analyses of large sets of possible EOSs constrained by 
nuclear-physics theory at low densities, 
e.g.,~Refs.~\cite{Annala:2017llu, Most:2018hfd, Tews:2018chv, Capano:2019eae, Raaijmakers:2019dks, Dietrich:2020lps, Essick:2020flb}.
Furthermore, the NS mass distribution, and therefore also the maximum density in the NS core, is bounded from above by the NS maximum-mass configuration. 
This is the highest NS mass that can be supported against gravitational collapse by the dense matter in the NS interior and depends on the EOS.
While this maximum mass can in principle be as high as $3-4\, M_{\odot}$ (see e.g.~Ref.~\cite{Tews:2018chv}), the EM observation of the kilonova associated with GW170817 has constrained the maximum mass to be much smaller, of the order of $2.2-2.3\, M_{\odot}$~\cite{Margalit:2017dij, Rezzolla:2017aly, Shibata:2017xdx}.
Most NS observations so far have explored NSs in a mass range of $1.4-2.1\, M_{\odot}$, and hence below the maximal possible density. 
In contrast, the coalescence of two typical NSs of approximately $1.4\, M_{\odot}$ creates an object that is likely above the maximum mass, and truly explores the EOS at the highest densities in the Universe.

In the context of the dense-matter EOS, an important problem is to determine the nature of matter inside of NSs. 
For example, at very high energy densities the fundamental theory of strong interactions, Quantum Chromodynamics, predicts that matter undergoes a phase transition to quark matter but it is unknown at what densities such a transition occurs. 
A long-standing question is whether NSs explore such phase transitions to new and exotic forms of matter in their cores or whether they solely consist of nucleonic matter~\cite{Glendenning:1992vb, Alford:2001dt, Alford:2006vz}, and whether these transitions are observable~\cite{Alford:2004pf, Alford:2013aca}. 
Among NSs that explore a phase transition, ``twin stars'' are a particular family. 
For these stars, the phase transition in the EOS is sufficiently strong so that the mass-radius curve around the phase transition is disconnected and contains two or more branches~\cite{Benic:2014jia, Alford:2017qgh}. 

For EOSs where a phase transition is present, two scenarios can be distinguished based on its onset density.
First, it is possible that the transition happens at very high densities, i.e., only in heavy stars. 
In this case, the phase transition is probed only after the collision of the two individual stars; see, e.g., Refs.~\cite{Bauswein:2018bma, Most:2018eaw, Bauswein:2019skm, Most:2019onn, Weih:2019xvw,Blacker:2020nlq}.
A second possibility is that the onset density of the phase transition is at lower densities explored in typical NS around $1.4\,M_{\odot}$, that are probed already during the inspiral phase of the NS merger, e.g., Refs.~\cite{Montana:2018bkb, Gieg:2019yzq}.
In the latter case, one could imagine scenarios in which a mass asymmetry between the two stars in the BNS leads to one (lighter) star containing only nuclear matter, while in the other (more massive) star a quark core is already present. 
In such a case the two individual stars could have very different radii and tidal deformabilities while their masses are comparable\footnote{The tidal deformability $\Lambda_2 = (2/3)k_2 (c^2R/GM)^5$ with the Love number $k_2$, radius $R$, and mass $M$ determines the deformation of the star in an external 
gravitational field~\cite{Flanagan:2007ix,Hinderer:2007mb,Hinderer:2009ca}.}.
This is of particular importance since a number of existing GW analyses, e.g., Refs.~\cite{De:2018uhw, Abbott:2018exr}, and multi-messenger constraints on the EOS, e.g., Refs.~\cite{Radice:2017lry, Bauswein:2017vtn, Coughlin:2018miv, Radice:2018ozg, Coughlin:2018fis}, rely on the assumption that both inspiraling objects are NSs following some given quasi-universal relations~\cite{De:2018uhw, Yagi:2016bkt, Chatziioannou:2018vzf}. 
In the presence of a phase transitions those quasi-universal relations might be violated, in which case their employment in GW analyses are likely to lead to biases of the determined binary properties and the EOS. 

While the current analysis of GW170817 seems to disfavor NSs 
with too large radii and tidal deformabilities, consistent with the appearance of phase transitions, the data from this single event is insufficient to conclusively answer this question; see, e.g., Ref.~\cite{PhysRevD.101.063007}.
Many recent works have addressed the question whether GWs allow to constrain the existence of hybrid stars, i.e., NSs that explores strong phase transitions to exotic forms of matter in their cores, and in particular twin stars~\cite{Bauswein:2018bma, Montana:2018bkb, Most:2018eaw, Annala:2019puf, Chatziioannou:2019yko, Weih:2019xvw, Essick:2020flb}.
For example, Ref.~\cite{Chatziioannou:2019yko} searched for the presence of a phase transition by applying quasi-universal relations. In particular, the presence of a strong phase transition was probed via observing the breakdown of quasi-universal relations. 
Ref.~\cite{Chatziioannou:2019yko} found that the mass at which the phase transition occurs, $M_t$, can be measured with $50-100$ detections and the corresponding microscopic parameters can be estimated via quasi-universal relations.
Furthermore, Refs.~\cite{PhysRevD.101.063007, Landry:2020vaw} looked for indications of phase transitions in GW data using a non-parametric inference approach to the EOS. 
By combining heavy pulsar observations, GW170817, and the recent NICER observation~\cite{Miller:2019cac}, a Bayes factor in favor of the presence of multiple stable branches of $1.8\pm0.2$~\cite{Landry:2020vaw} was found.

When looking for the imprint of a phase transition in the GW170817 data, these previous works have mainly searched for the presence of multiple stable branches in the mass-radius relation.
However, this is only one among various scenarios. 
In this work, we aim at quantifying whether we are capable of determining the presence of a strong phase transition from GW data even when only one stable branch is present.
In particular, we ask the question how many GW observations are necessary to observe a phase transition and recover the parameters of an injected EOS from GW data.
We focus on three different EOS that experience a phase transition in the typical mass range explored in BNS systems and which show 3 different behaviors in the mass-radius relation.

For this purpose, we introduce a novel method, based on a new parameterization for EOSs at supranuclear densities, of testing GW data from the inspiral phase of a BNS merger for the appearance of a strong phase transition.
This new approach is based on Bayesian inference methods, and can be used with current GW detectors. 
Simulating 600 signals for three different EOSs, we find that already 12 events might be sufficient to confidently find the presence of a phase transition. 
However, when analyzing the signals GW170817 and GW190425 with our method, we do not find a hint of a strong phase transition.

The major differences between previous studies and our work are that we simultaneously
(i) analyze EOSs with different phase-transition signatures, i.e., one EOS with a twin-star solution which is commonly searched for, but also two EOSs with phase transitions leading to single-branch solutions,
(ii) analyze both simulated data and actual events with state-of-the-art Bayesian GW data analysis techniques, which allows for hypothesis testing and parameter estimation at once, and
(iii) explicitly demonstrate that our method is able to measure the microscopic characteristics of strong phase transitions, by comparing injected with recovered parameters.
Hence, our method allows us to make statistically robust statements on the presence of strong phase transitions.

The paper is structured as follows. 
We describe our methods and our mock data setup in Sec.~\ref{sec:eos} and \ref{sec:methods}, respectively. 
Main results are shown in Sec.~\ref{sec:search}, and in Sec.~\ref{sec:real_GW} 
we apply our method to GW170817 and GW190425. We conclude in Sec.~\ref{sec:conclusion}. 

\section{Phase transitions and their imprint on the GW signal}
\label{sec:eos}

\subsection{The equation of state of NS matter}

\begin{figure*}[t]
\begin{center}
\includegraphics[width=0.49\textwidth]{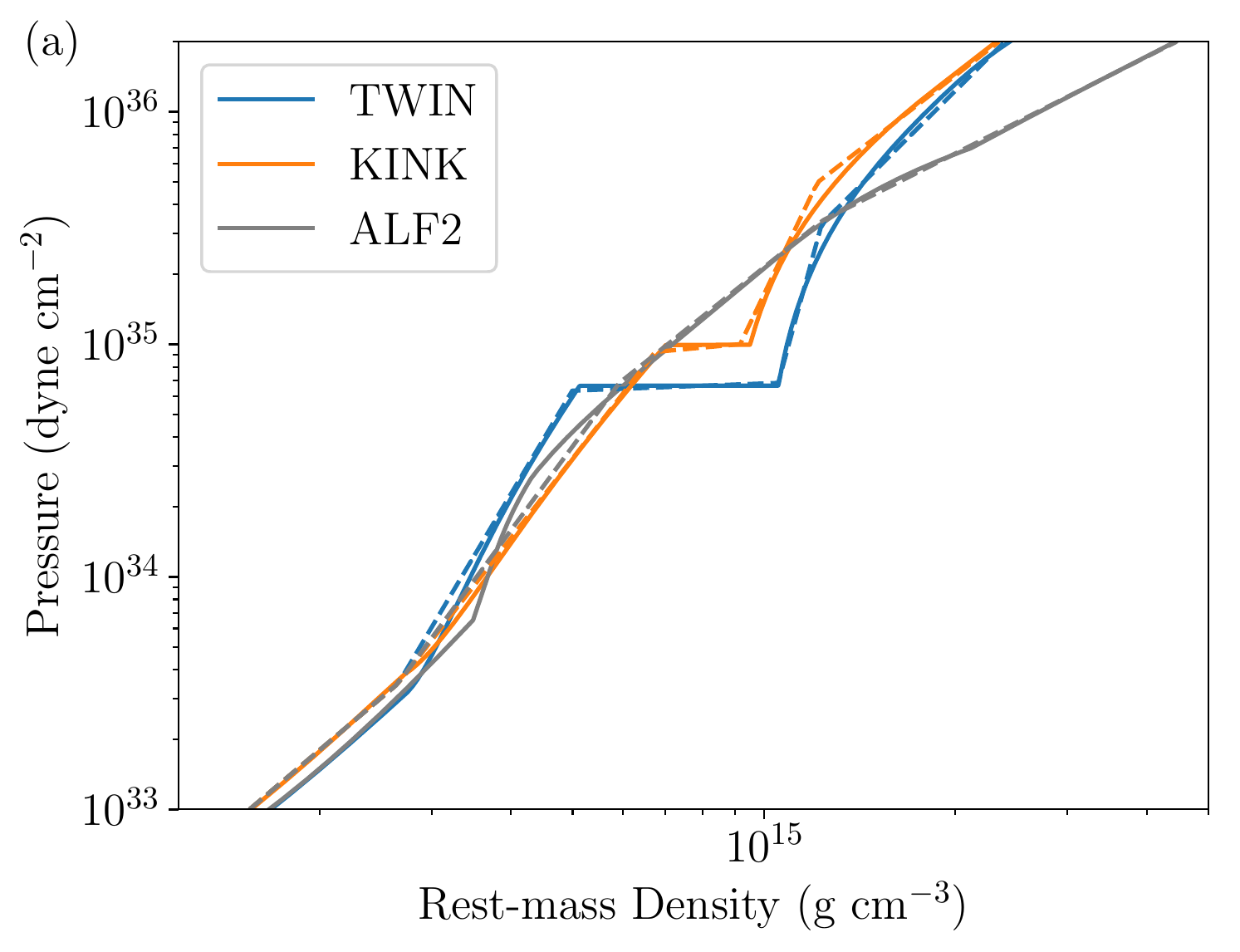}
\includegraphics[width=0.49\textwidth]{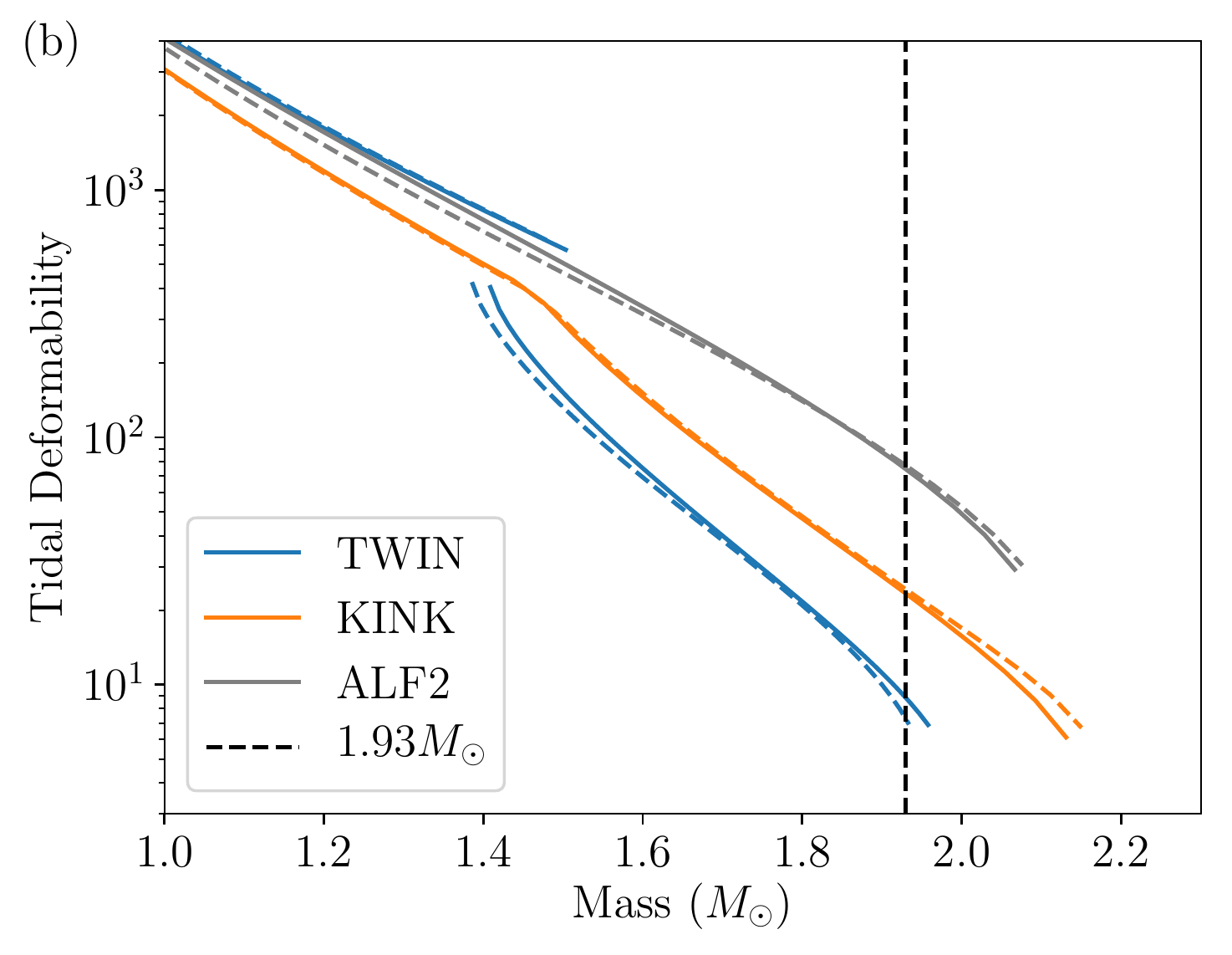}
	\caption{Density-pressure relations (a) and mass--tidal-deformability relations (b) of the three EOSs (solid) used in this work, and the least-squares Maxwell fits (dashed) to the the EOSs. 
The Maxwell parameterization successfully captures all features, including the phase transition, for both the EOS and the mass-radius curve. 
Moreover, both the EOSs and the EOSs' best-fits support heavy NSs.
}
\label{fig:EOS}
\end{center}
\end{figure*}

The structure of NSs is completely determined by solving the Tolman-Oppenheimer-Volkoff (TOV) equations. 
The only necessary input is the EOS, a relation between the pressure, energy density, temperature, and composition inside the NS.
The EOS is determined by the microscopic degrees of freedom in the NS interior and their interactions. 
At lower densities, these are mostly neutrons with a few percent protons interacting via nuclear forces, but at higher densities new degrees of freedom might appear.
With typical radii of the order of $12$ km and masses of $1-2\, M_{\odot}$, the densities inside NSs are so large that thermal energies are much smaller than typical Fermi energies, except in the most violent astrophysical scenarios. 
Hence, for isolated NSs and NSs during the inspiral phase of a BNS merger, finite temperature effects can be neglected and the EOS is simply a relation of pressure and energy density for a given composition.

From the theoretical side, the EOS of cold dense matter at the densities explored in the NS core is very uncertain. 
There exists a multitude of models for the EOS which explore a wide range of pressures at densities beyond $n_{\rm sat}$, leading to large uncertainties for the 
radii of typical NSs.
The models can differ both in the degrees of freedom that they assume and in the effective interactions among them.
At low densities explored in the NS crust and outermost core, where experimental input on, e.g., saturation properties or extractions of the symmetry energy are available to constrain models, the EOS can instead be constrained rather reliably.
In this density regime, approaches ranging from density functionals~\cite{Douchin:2001sv} or relativistic mean-field models~\cite{Typel:2009sy, steiner:2012rk} to ab initio calculations using a variety of models for the nuclear interactions, e.g., Refs.~\cite{Gandolfi:2011xu, Hebeler:2013nza}, lead to consistent results.

In recent years, important constraints on the EOS of NS matter at low densities have been obtained from microscopic calculations of neutron matter using systematic interactions from chiral effective field theory (EFT)~\cite{Hebeler:2013nza, Carbone:2013eqa, Hagen:2013yba, Holt:2016pjb, Drischler:2017wtt, Tews:2018kmu}. 
Chiral EFT~\cite{Epelbaum:2008ga, Machleidt:2011zz} represents a systematic low-momentum expansion of nuclear forces, which is connected to the symmetries of the fundamental theory of strong interactions, Quantum Chromodynamics. 
Instead of using quark and gluon degrees of freedom, it uses the fact that quarks are confined to hadrons at low densities and models interactions in terms of effective degrees of freedom, nucleons and pions. 
Chiral EFT naturally includes two-body and many-body interactions among nucleons, and provides an order-by-order scheme for the interactions among nucleons in terms of contact interactions, whose couplings are fit to experimental data, and long-range pion-exchange interactions. 
By going to higher orders in this expansion, calculations increase in difficulty, but results become more precise and accurate. 
In addition, the systematic order-by-order scheme can be used to obtain theoretical uncertainty estimates~\cite{Epelbaum:2014efa, Drischler:2020yad}. 
Hence, calculations of neutron matter with chiral EFT interactions provide constraints on the EOS with reliable uncertainties. 

However, chiral EFT calculations are valid only within the radius of convergence of the theory. 
Typically, the momentum expansion breaks down at momenta of the order of $500-600$ MeV; see, e.g., Ref.~\cite{Drischler:2020yad} for a recent analysis. 
Hence, in neutron matter the chiral EFT approach might be 
reliable only up to $(1-2) n_{\rm sat}$, but most likely fails beyond that~\cite{Tews:2018kmu, Essick:2020flb}.
Therefore, at densities beyond two times saturation density, currently no reliable statement about the EOS can be made from microscopic nuclear theory.
At these densities, models suffer from the absence of available experimental data and our ignorance of strong interactions in this regime. 
In particular, while we know the relevant degrees of freedom at lower densities to be nucleons, it is not clear which degrees of freedom appear at larger densities. 
While many astrophysical EOSs assume nucleonic degrees of freedom to be valid in the whole NS, a phase transition to new degrees of freedom, e.g., quark matter or exotic condensates, might occur~\cite{Annala:2019puf}.
This phase transition might be strong and of first order, in which case it can lead to interesting features in the mass-radius relation, like kinks or disconnected branches~\cite{Alford:2013aca}.

To extend microscopic low-density results for the EOS to higher densities, the uncertainty in the degrees of freedom must be taken into account. 
This is typically done by applying general extension schemes, e.g., by using sets of polytropes for the energy density and pressure~\cite{Read:2008iy, Hebeler:2013nza, Annala:2017llu} or an expansion in the speed of sound~\cite{Tews:2018chv, Greif:2018njt, Chatziioannou:2019yko}. 
An extension of this approach are nonparametric inference schemes which have prior support for all possible EOS curves~\cite{Landry:2018prl} and have recently been combined with chiral EFT calculations~\cite{Essick:2020flb}.
Such extension schemes abandon explicit assumption about the degrees of freedom at higher densities but instead model all EOS curves permitted by the chosen functional form in the case of parametric extensions\footnote{Because no particular choice for the functional form of the EOS is made in nonparametric inference schemes, they are less limited in this sense. However, in this work we choose a parameterized approach for the EOS because it is straightforward to implement a $c_s^2=0$ segment that appears in case of a strong first-order phase transition.}  
and general physics considerations such as causality.
By sampling all allowed functions, uncertainties at low densities can be systematically extended to high densities. 

These general EOS sets contain both smooth EOSs, i.e., EOSs for which the change of pressure with energy density is continuous at all densities, as well as EOSs with drastic changes in the pressure.
While EOSs of the first type might be obtained by using a purely nucleonic description of NSs, the latter type contains EOSs with strong first-order phase transitions.
Such transitions can be modeled within a Maxwell or Gibbs construction, depending on the properties of the considered phases. 

In a Maxwell construction, no mixed phases appear and the phase transitions can be modeled by an EOS segment where the speed of sound, $c_S^2=\partial p/\partial \epsilon=0$, vanishes.
This EOS segment, and hence the phase transition, can be described by its onset density, where the speed of sound becomes 0, and its width, i.e., the density jump between the two phases with nonvanishing speed of sound.
Depending on these properties, different features might be observed in the $M$-$R$ relation (or the $M$-$\Lambda$ relation). 
In Fig.~\ref{fig:EOS}, we show two examples of such phase transitions, which we have selected from the EOS set of Ref.~\cite{Tews:2018chv}. 
This EOS set was constrained by microscopic chiral EFT calculations below nuclear saturation density, including a consistent NS crust~\cite{Tews:2016ofv}, and extended to larger densities by using a speed-of-sound extension scheme. 
Hence, it ensures NS stability ($c_S>0$) and causality ($c_S<c$, with $c$ the speed of light) by design.
For one of the two chosen EOS (labeled KINK, orange line), an intermediate width is chosen for the phase transition which leads to a visible kink in the mass-radius curve. 
For the other EOS (labeled TWIN, blue line), a larger width leads to a stronger phase transition which results in the appearance of two disconnected branches in the mass-radius relation, a so-called twin-star solution. 
For both EOS, the onset density is chosen such that the interesting feature appears already in typical NSs.

In the Gibbs construction, on the other hand, a mixed phase appears and smoothens the resulting EOS around the phase transition.  
In that case an EOS with a phase transition might be indistinguishable from a purely nucleonic EOS, which is known as the masquerade problem~\cite{Alford:2004pf}.
We compare the two EOSs TWIN and KINK with the model ALF2~\cite{Alford:2004pf}, which is a hybrid EOS with a phase transition to quark matter that leads to the formation of a mixed phase.
For all three EOS models, the maximum mass is greater than $1.93 \,M_{\odot}$ and, hence, consistent with observed masses of heavy NSs~\cite{Antoniadis:2013pzd, Cromartie:2019kug}.

\subsection{Imprint of phase transitions on the GW signal}

A possible phase transition can imprint itself upon the GW signal in different ways during the inspiral and during the postmerger phase. 

\textit{Inspiral:} 
During the inspiral, the GW signal depends on the properties of the two binary stars (masses, spins, and tidal deformabilities), as well as on the source location and orientation. 
Of particular importance for the description of tidal effects are the parameters
\begin{align}
 \tilde{\Lambda} = & \frac{16}{13} \Lambda_1 \frac{m_1^4}{M^4} \left(12 - 11 \frac{m_1}{M}) \right) + \\ \nonumber
                   & \frac{16}{13} \Lambda_2 \frac{m_2^4}{M^4} \left(12 - 11 \frac{m_2}{M}) \right)\,,
\end{align}
that captures the leading-order contribution at fifth Post-Newtonian (PN) order, and 
\begin{align}
 \delta \tilde{\Lambda} = & \left( \frac{1690}{1319} \nu -\frac{4843}{1319} \right) \left( \frac{m_1^4}{M^4} \Lambda_1 - 
 \frac{m_2^4}{M^4} \Lambda_2 \right) \\ 
  & + \frac{6162}{1319} \sqrt{1 -  4 \nu }  \left( \frac{m_1^4}{M^4} \Lambda_1 +  \frac{m_2^4}{M^4} \Lambda_2 \right)\,, \nonumber
\end{align}
with the symmetric mass ratio $\nu = m_1 m_2 (m_1+m_2)^{-2} = m_1 m_2 / M^2 $, which captures additional contributions at sixth PN order. 

\begin{figure}[t]

\begin{center}
\includegraphics[width=1.0\columnwidth]{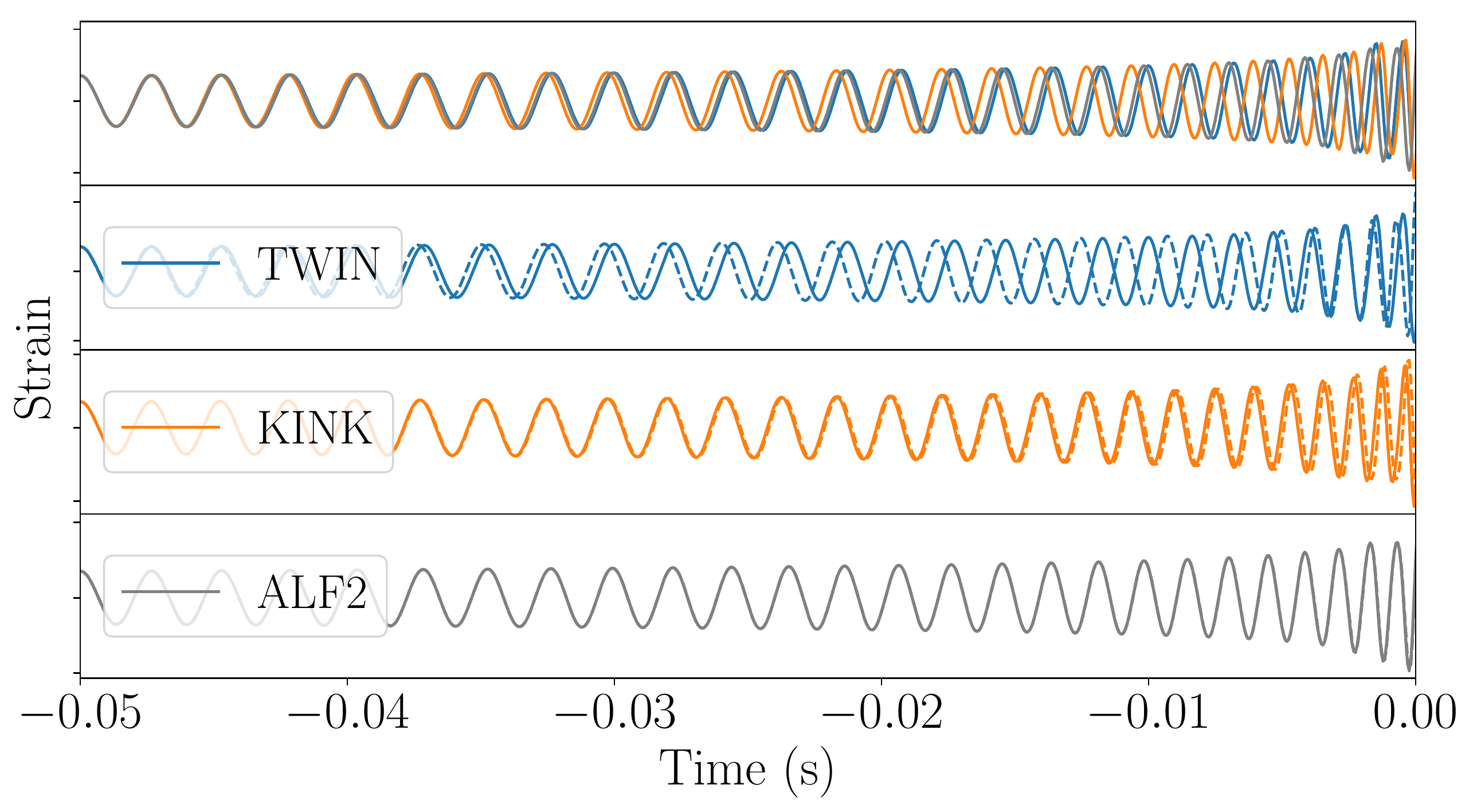}
	\caption{The GW waveforms for a non-spinning BNS system with component masses of $1.50M_\odot$ and $1.45M_\odot$ for the three EOS used in this work. 
The dashed lines are the waveforms assuming that the Binary-Love relation~\cite{Yagi_2016, Chatziioannou:2018vzf} holds.}
\label{fig:GW_example}
\end{center}
\end{figure}

The presence of a phase transition in the EOS might lead to a significant change in the radii and tidal deformabilities for almost equal mass systems, if in any of the two stars the onset density for the transition is already reached in the core, cf.~Fig.~\ref{fig:EOS}.
In Fig.~\ref{fig:GW_example}, we show the expected GW signal for a non-spinning BNS system with component masses of $1.50\,M_\odot$ and $1.45\,M_\odot$ for the three EOSs that we have chosen in this work.
For the two EOSs with strong first-order phase transitions, due to the different tidal deformabilities of both stars, 
we find a dephasing of the waves compared to the ALF2 EOS.
Since the leading-order tidal contribution enters at the fifth PN order~\cite{Flanagan:2007ix, Hinderer:2007mb, Hinderer:2009ca, Damour:2012yf}, the dephasing is most prominent in the late-inspiral phase.   

In addition to the GWs for the EOSs described above, we also present as dashed lines the waveforms when we assume that the quasi-universal relation of Refs.~\cite{Yagi_2016,Chatziioannou:2018vzf} holds for TWIN, KINK, and ALF2. 
To apply this relation, we fix the tidal deformability of the lower-mass star and compute the tidal deformability of the primary component by using the quasi-universal relation. 
We find that the resulting waveform significantly deviates from the full waveform for EOSs with a very strong phase transitions, i.e., TWIN, but approximates the waveform well in the other cases. 
These differences suggest the failure of the quasi-universal relation with respect to EOSs with strong phase transitions like TWIN, while the relation holds approximately for the KINK EOS.
For smooth EOSs, like ALF2, there is no observable difference and the quasi-universal relation seems to be valid. This suggests that detecting a phase transition that does not result in a twin-star solution will be challenging if a methodology purely based on quasi-universal relations is employed.

\textit{Postmerger:} 
In cases in which the phase transition happens at densities beyond the ones probed during the inspiral, the postmerger signature can change if a phase transition is present, as outlined in, e.g., Refs.~\cite{Bauswein:2018bma, Montana:2018bkb, Most:2018eaw, Annala:2019puf, Chatziioannou:2019yko, Weih:2019xvw,Blacker:2020nlq}. 
In most cases, the presence of a phase transition will lead to a different (often shorter) lifetime of the remnant and a shift of the main postmerger GW emission mode. 
However, due to the missing sensitivity of existing GW detectors in the high-frequency range~\cite{advLIGOcurves, TheVirgo:2014hva} and the absence of high-quality GW models describing the postmerger evolution of BNS mergers -- see Refs.~\cite{Chatziioannou:2017ixj, Bose:2017jvk, Torres-Rivas:2018svp,Easter:2018pqy,Tsang:2019esi, Breschi:2019srl} for some first attempts -- it seems natural to investigate, at the current stage, possible phase transition effects that can be extracted from the GW signal during the inspiral. 

\subsection{EOS parameterization for phase transitions}
\label{sec:maxwell}

When analyzing NS observations, one needs to assume an EOS describing the relation between pressure and energy density. 
For the ``true'' NS EOS realized in nature, however, the functional form is unknown.  
Hence, an EOS parameterization needs to be flexible enough to capture the various effects one might encounter in nature, in particular phase transitions.
In this work, we consider the three EOSs of Fig.~\ref{fig:EOS} as three possible ``true'' EOSs.
In order to capture all features of these EOSs, in particular the first-order phase transitions, here we propose to use a 5-parameter piecewise-polytrope EOS parameterization scheme, which we refer to as Maxwell parameterization.
This scheme is similar to the parameterization proposed in Ref.~\cite{Read:2008iy, Alvarez_Castillo_2017}.

In our Maxwell parametrization, at low densities up to nuclear saturation density, we use an EOS constrained by the chiral EFT calculation of Ref.~\cite{Tews:2018kmu} ($V_{E, 1}$ parametrization).
This EOS contains a consistent inner crust and uses the BPS model for the outer crust.
For the high-density part beyond $n_{\rm{sat}}$, we use a modified 5-parameter 4-piece polytrope.
Each polytrope is characterized by the starting pressure $p_i$ and the adiabatic index $\Gamma_i$.
Therefore, our extension starts with 8 free parameters: $\{p_1, p_2, p_3, p_4\}$ and $\{\Gamma_1, \Gamma_2, \Gamma_3, \Gamma_4\}$. 

\begin{figure}[t]
\begin{center}
\includegraphics[width=1.0\columnwidth]{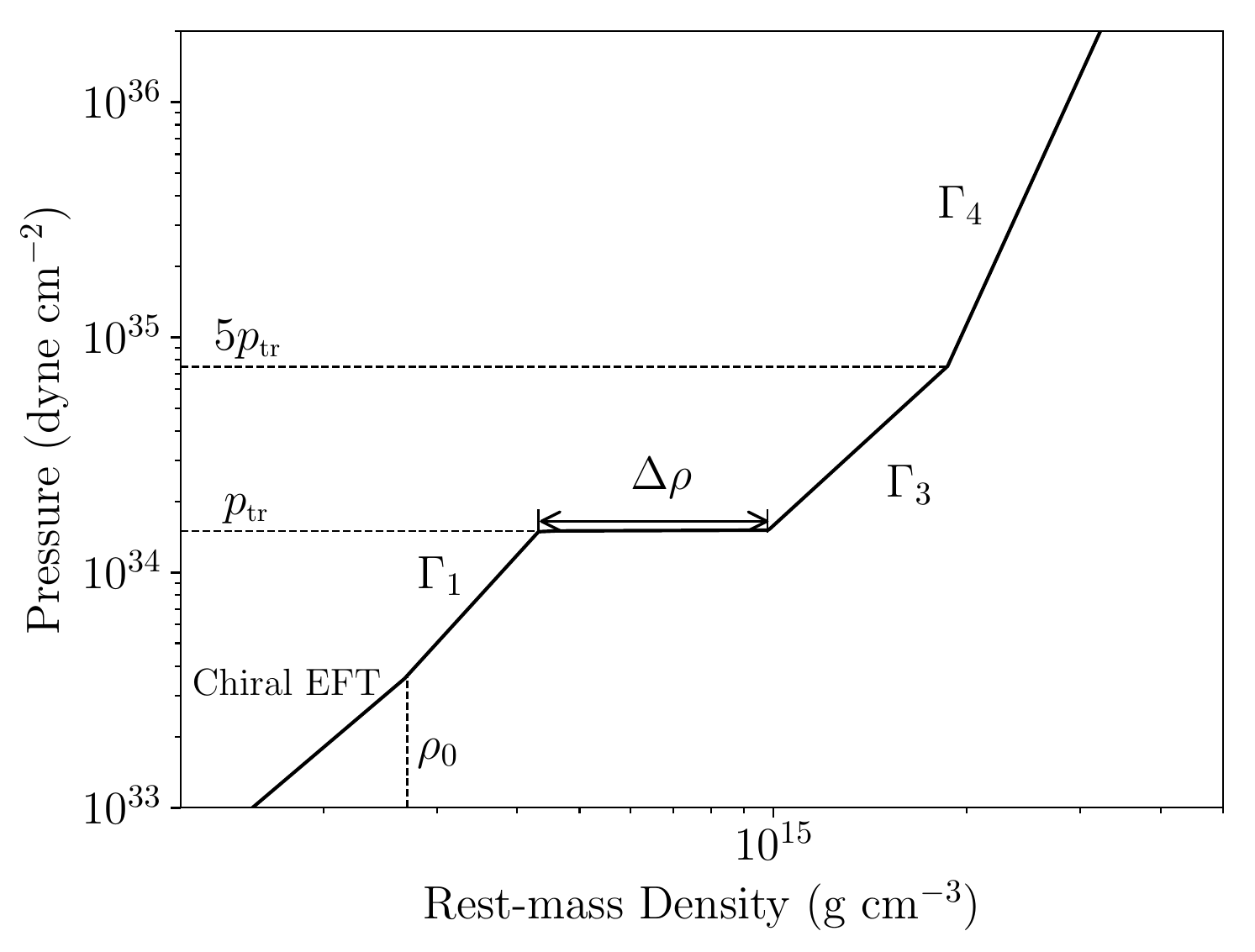}
	\caption{Sketch of the Maxwell parameterization, which is characterized by the adiabatic indices $\{\Gamma_1, \Gamma_3, \Gamma_4\}$, the phase transition onset pressure $\ptr$, and the transition density jump $\Delta\rho$.}
\label{fig:Maxwell_sketch}
\end{center}
\end{figure}

To ensure continuity for the first polytrope, the starting pressure $p_1$ is chosen to be the pressure of the chiral EFT EOS at the nuclear saturation density, $p_{\small\textrm{CEFT}}(\rho_0)$. 
The adiabatic index of the second polytrope, $\Gamma_2$, is set to be zero to represent a Maxwell construction for a phase transition extended across a density gap of $\Delta\rho$.
Therefore, $p_2 = p_3 = p_{\textrm{tr}}$, where $\ptr$ is the phase transition pressure. 
Furthermore, we choose the transition pressure between the third and fourth polytropes to be 5 times the phase transition pressure, $p_4 = 5\ptr$. 
Fixing $p_4$ reduces the numbers of free parameters and, therefore, helps during the recovery. 
The particular value of $5\ptr$ is chosen ad-hoc by comparison against various different EOSs.
We have also explored leaving this parameter free, but this only improved the fitting performance marginally. Therefore, to reduce dimensionality, we have fixed this parameter. 
A sketch of the parameterization is shown in Fig.~\ref{fig:Maxwell_sketch} and 
its capability of representing the three EOSs is shown in Fig.~\ref{fig:EOS}. 
We find that the Maxwell parametrization works well for EOSs where phase transitions appear between $1-4n_{\textrm{sat}}$, corresponding to a NS in a typical mass range.
As suggested in~\cite{PhysRevD.98.063004,Gamba_2019}, we have found that our parametrization introduces systematic uncertainties due to imperfect fits to the ``true'' EOSs. 
Yet, as shown in the right panel of Fig.~\ref{fig:EOS}, the error is overall small and will be below the statistical uncertainty of an EOS measurement~\cite{Wade:2014vqa}. 
However, it might be necessary to verify the generality of this assumption assuming different parametrizations. 
Because of the large computational cost, this is not part of this work.
Furthermore, we do not expect systematics induced by our parametrization to significantly affect the present study and its main goal, namely to identify strong phase transitions and their parameters.

In total, our parametrization is described by five parameters $\Gamma_1, \Gamma_3, \Gamma_4$, the phase transition onset pressure $\ptr$, and the transition density jump $\Delta\rho$.
In practice, in case of an EOS with a twin-star solution, it is possible to have NSs with the same mass but different radii, i.e., NSs that live on different branches of the $M$-$R$ relation. 
These NSs have central densities around the onset density of the phase transition.
Because in this case the NS mass cannot be used to distinguish the individual stars -- see Fig.~\ref{fig:EOS} (right panel) -- we need to add two extra parameters $B_1$ and $B_2$ for each of the two NSs to indicate on which branch the star lives.
Therefore, the EOS parameters $\vec{E}$ are given by
\begin{equation}
\label{eq:EOSParameters}
\begin{aligned}
	\vec{E} &= \{\log_{10} p_{\small\textrm{tr}}, \log_{10}\Delta\rho, \Gamma_1, \Gamma_3, \Gamma_4, B_1, B_2\}\\
	&= \{\vec{E}_c, \{B_i\}\},
\end{aligned}
\end{equation}
where $\vec{E}_c$ denotes the common parameters of all stars, assuming they follow the same EOS. 

\section{Mock Data Simulation}
\label{sec:methods}

\subsection{Bayesian Analysis}
\label{ssec:bayesian_analysis}

According to Bayes' theorem, the posterior $p(\vec{\theta} | d,\mathcal{H}) \equiv \mathcal{P}(\vec{\theta})$ on the parameters $\vec{\theta}$ under hypothesis $\mathcal{H}$ and with data $d$ is given by
\begin{equation}
	\mathcal{P}(\vec{\theta}) = \frac{p(d|\vec{\theta},\mathcal{H})p(\vec{\theta}|\mathcal{H})}{p(d|\mathcal{H})} \equiv \frac{\mathcal{L}(\vec{\theta})\pi(\vec{\theta})}{\mathcal{Z}(d)}\,,
\end{equation}
where $\mathcal{L}(\vec{\theta})$, $\pi(\vec{\theta})$, and $\mathcal{Z}(d)$ are the likelihood, prior, and evidence, respectively. 
The prior describes our knowledge of the source or model parameters prior to the experiment or observation. 
The likelihood and evidence quantify how well the hypothesis describes the data with the given set of parameters and over the whole parameter space, respectively.
By assuming Gaussian noise, the likelihood $\mathcal{L}(\vec{\theta})$ that the data $d$ is a sum of noise and a GW signal $h$ with parameters $\vec{\theta}$ is given by~\cite{Veitch:2014wba}
\begin{equation}
	\mathcal{L}(\vec{\theta}) \propto \exp\left(-\frac{1}{2}\langle d-h|d-h\rangle\right),
\end{equation}
where the inner product $\langle a|b\rangle$ is defined by
\begin{equation}
	\langle a | b \rangle = 4\Re\int_{f_\textrm{low}}^{f_\textrm{high}}\frac{\tilde{a}(f)\tilde{b}^*(f)}{S_n(f)}df.
\end{equation}
Here, $\tilde{a}(f)$ is the Fourier transform of $a(t)$, ${}^*$ denotes complex conjugation, and $S_n(f)$ is the one-sided power spectral density of the noise. In our study we will set $f_\textrm{low}$ and $f_\textrm{high}$ to $20$ Hz and $2048$ Hz, respectively.
The evidence $\mathcal{Z}$ is given by
\begin{equation}
	\mathcal{Z} = \int\mathcal{L}(\vec{\theta})\pi(\vec{\theta})d\vec{\theta},
\end{equation}
which is the normalization constant for the posterior distribution.

Moreover, we can compare the plausibilities of two hypotheses, $\mathcal{H}_1$ and $\mathcal{H}_2$, by using the odd ratio, which is given by
\begin{equation}
\mathcal{O}^1_2 \equiv \frac{p(d|\mathcal{H}_1)}{p(d|\mathcal{H}_2)}\frac{p(\mathcal{H}_1)}{p(\mathcal{H}_2)} \equiv \mathcal{B}^1_2\Pi^1_2,
\end{equation}
where $\mathcal{B}^1_2$ and $\Pi^1_2$ are the Bayes factor and prior odds, respectively. If $\mathcal{O}^1_2 > 1$, $\mathcal{H}_1$ is more plausible than $\mathcal{H}_2$, and vice versa. Throughout this study, we have the prior odds set to $1$, in which case the Bayes factor is the same as the odd ratio.

Within the Bayesian framework we can combine the information from multiple detections. 
For parameters that are expected to be the same across several detections (e.g. EOS parameters), the combined posterior for common parameters $\mathcal{P}_c(\vec{\theta}_c)$ can be obtained as
\begin{equation}
	\mathcal{P}_c(\vec{\theta}_c) = \pi(\vec{\theta}_c)^{1-N}\prod_{i=1}^N\mathcal{P}_i(\vec{\theta}_c)\,,
\end{equation}
where $\vec{\theta}_c$ are the common parameters and $\mathcal{P}_i(\vec{\theta}_c)$ is the posterior including the $i$-th detection. 
We can also combine the odds ratios into a catalog odds ratio $\mathcal{O}^{1\textrm{ (cat)}}_{2}$, which is given by\footnote{ 
By cataloging odd ratios with simple multiplication, the information that some parameters 
are shared across detections is not included.
This conservative choice is dictated by computational limitations; see the discussion in Sec.~IV.B.}
\begin{equation}
	\mathcal{O}^{1\textrm{ (cat)}}_{2} = \Pi^{1}_{2}\prod_{i=1}^N\mathcal{B}^{1}_{2,i} = \Pi^{1}_{2}\mathcal{B}^{1\textrm{ (cat)}}_{2},
\end{equation}
where $\mathcal{B}^{1}_{2,i}$ is the Bayes factor for the $i$-th detection and $\mathcal{B}^{1\textrm{ (cat)}}_{2}$ is the catalog Bayes factor. 
As we have the prior odds set to 1, the catalog odd ratio $\mathcal{O}^{1\textrm{ (cat)}}_{2}$ is the same as the catalog Bayes factor $\mathcal{B}^{1\textrm{ (cat)}}_{2}$.

\subsection{Waveform approximants}

In this paper, we restrict our studies to the PN model 
\texttt{TaylorF2}. 
This model was also employed in the analysis of GW170817 and GW190425 by the LIGO and Virgo Collaborations; see e.g.~Refs.~\cite{Abbott:2018wiz,LIGOScientific:2018mvr,Abbott:2020uma}.

The version of \texttt{TaylorF2} we use is based on on a 3.5 PN order point-particle description~\cite{Sathyaprakash:1991mt,Blanchet:1995ez,Damour:2001bu,Goldberger:2004jt,Blanchet:2005tk} that includes spin-orbit effects~\cite{Bohe:2013cla} and spin-spin effects~\cite{Arun:2008kb,Mikoczi:2005dn,Bohe:2015ana,Mishra:2016whh}. 
Tidal effects are added up to 7.5 PN following Refs.~\cite{Vines:2011ud,Damour:2012yf}. 
We note that we also incorporate the EOS-dependency of the 2PN and 3PN spin-spin contributions. 
For this purpose, we use quasi-universal relations outlined in~\cite{Yagi:2016bkt} to connect the spin-induced quadrupole moments to the tidal deformability of the NS stars. 
This approach is commonly used for GW data analysis and was by default used here, but a phase transition could also affect the employed quasi-universal relation. This might introduce additional biases. 
Sampling over the individual quadrupole moments of the NSs (see, e.g., Ref.~\cite{Samajdar:2020xrd}), would cause an increase of the computational costs and it seemed more appropriate to employ the quasi-universal relations
of~\cite{Yagi:2016bkt} than simply neglecting the EOS imprint on the quadrupole moment.
However, since all simulated signals are non-spinning, we do not expect that any significant biases appear during our analysis and refer to the study of Ref.~\cite{Samajdar:2019ulq}, where it was shown that potential biases only arise for high spins. 

\subsection{Injection setup}
\label{ssec:inj}

We choose to use an astrophysically motivated distribution for the parameters of the simulated sources. 
We distribute the sources uniformly in a co-moving volume with the optimal network SNR range $\rho \in [30, 100]$. 
Thus, a relatively high lower bound on SNR is assumed. Indeed, to probe the phase transition, an accurate measurement of $\Lambda_i$ is needed, which can not be achieved with BNS signals that have low or medium SNR~\cite{Wade:2014vqa}.
 
The orientation $(\iota, \psi)$ and the sky location $(\alpha,\delta)$ of the sources are placed uniformly on a sphere. 
Since NS spins are expected to be small~\cite{O_Shaughnessy_2008}, we set them to zero for all simulated sources. 
The component masses of the binaries are sampled from the uniform distribution $[1\,M_\odot,M_{\textrm{TOV}}]$, where $M_{\textrm{TOV}}$ is the maximum allowed mass of a NS with the given EOS.

For the EOSs, we want to investigate under what circumstances one can distinguish between the existence and the absence of a strong phase transition. 
Therefore, we choose two EOSs that have phase transitions with different onset pressure and density jumps but that lead to an observable and distinct feature within the mass range of $[1,2]M_\odot$ in the mass-radius curve, labeled as TWIN and KINK; cf.~Sec.~\ref{sec:eos}. 
In addition, we choose another EOS with phase transition but a smooth density dependence of the pressure. 
As the phase transition masquerades, this model is indistinguishable from a purely nucleonic EOS model, see Fig.~\ref{fig:EOS}. 
Here, we choose the ALF2 EOS described in Ref~\cite{Alford:2006vz} because of its plausibility based on the multi-messenger analysis of GW170817~\cite{Coughlin:2018fis, Radice:2017lry}.
The three EOSs are all able to support the observed heavy NSs~\cite{Demorest:2010bx, Antoniadis:2013pzd, Cromartie:2019kug}. 

The simulated GW signals are injected coherently into the data of the Advanced LIGO and Advanced Virgo detectors. 
The detector noise is simulated as stationary Gaussian noise with the power spectral density to be that of the design sensitivities of each detectors~\cite{advLIGOcurves,TheVirgo:2014hva}.

\subsection{Implementation}
\label{sec:implementation}

Our analysis follows a similar approach as previous works~\cite{Lackey:2014fwa,PhysRevD.101.063007,wysocki2020inferring}. The analysis consist of a two-stage process:
\begin{enumerate}[label=\Roman*.]
	\item Estimation of the posterior of the macroscopic parameters $\mathcal{P}(\vec{\theta}_{\small\textrm{macro}})$ based upon a GW analysis.
	\item Estimation of the posterior of the microscopic parameters (\textit{i.e.} the EOS parameters) $\mathcal{P}(\vec{E})$ with $\mathcal{P}(\vec{\theta}_{\small\textrm{macro}})$ given. 
\end{enumerate}

For stage I, the posterior $\mathcal{P}(\vec{\theta}_{\small\textrm{macro}})$ is estimated with the Nested Sampling algorithm~\cite{skilling2006} implemented in \linf~\cite{lalsuite} with a prior of $m_i \in [0.5,3.0]\,M_{\odot}$ and $\Lambda_i\in[0,5000]$.
For stage II, the posterior $\mathcal{P}(\vec{E})$ is given by
\begin{equation}
\begin{aligned}
	\mathcal{P}(\vec{E}) &\propto \pi(\vec{E}) \mathcal{L}(\vec{E})\\
	&= \pi(\vec{E})\int d\vec{\theta}_{\small\textrm{macro}} \frac{\pi(\vec{\theta}_{\small\textrm{macro}}|\vec{E})}{\pi(\vec{\theta}_{\small\textrm{macro}}|I)}\mathcal{P}(\vec{\theta}_{\small\textrm{macro}})\,,
\end{aligned}
\end{equation}
where $\pi(\vec{\theta}_{\small\textrm{macro}}|\vec{E})$ and $\pi(\vec{\theta}_{\small\textrm{macro}}|I)$ are the priors on $\vec{\theta}_{\small\textrm{macro}}$ with and without the EOS given, respectively. For our study, the macroscopic parameters of interest are the component masses $m_{1,2}$ and the corresponding tidal deformabilities $\Lambda_{1,2}$. Therefore the likelihood $\mathcal{L}(\vec{E})$ is given by
\begin{equation}
\begin{aligned}
	\mathcal{L}(\vec{E}) &= \int dm_id\Lambda_i \frac{\pi(m_i,\Lambda_i|\vec{E})}{\pi(m_i,\Lambda_i|I)} \mathcal{P}(m_i,\Lambda_i)\\
	&= \int d\Lambda_idm_i \frac{\prod_i\delta(\Lambda_i-\Lambda(m_i; \vec{E}))}{\pi(\Lambda_i| m_i, I)}\frac{\pi(m_i|\vec{E})}{\pi(m_i|I)} \mathcal{P}(m_i,\Lambda_i)\\
	&= \int dm_i \left.\frac{\mathcal{P}(m_i,\Lambda_i)}{\pi(\Lambda_i|, m_i, I)}\right\vert_{\Lambda_i=\Lambda(m_i;\vec{E})}\,,
\end{aligned}
\end{equation}
where $\Lambda(m, \vec{E})$ is the tidal deformability as a function of mass with an EOS given. We have also chosen the priors $\pi(m_i|\vec{E})$ and $\pi(m_i|I)$ to be the same.

Therefore, the joint posterior $\mathcal{P}(\vec{E},m_i)$ is given by
\begin{equation}
	\mathcal{P}(\vec{E},m_i) \propto \pi(\vec{E}) \left.\frac{\mathcal{P}(m_i,\Lambda_i)}{\pi(\Lambda_i| m_i, I)}\right\vert_{\Lambda_i=\Lambda(m_i,\vec{E})}.
\end{equation}
The joint posterior is estimated with the Nested Sampling algorithm Multinest~\cite{Feroz_2009} implemented in \texttt{PyMultinest}~\cite{pymultinest1}. The posterior $\mathcal{P}(\vec{E})$ is then obtained via marginalizing $\mathcal{P}(\vec{E},m_i)$. 

For the stage II process, we choose the prior for the parameters to be $m_i\in[0.5,3.0]\,M_{\odot}$, $\Gamma_i\in(1,10]$, $\log_{10}\ptr[\textrm{dyne cm}^{-2}]\in[33.7,38.0]$ and $\log_{10}\Delta\rho[\textrm{g cm}^{-3}]\in[13.85, 16]$. We also impose the constraints of $M_{\textrm{TOV}}\geq1.93\,M_{\odot}$ as part of the prior. 
To increase efficiency, sampling over masses is done in terms of the chirp mass $\mathcal{M}$ and $\ln\Delta\nu$ rather than individual masses. The chirp mass $\mathcal{M}$ and $\ln\Delta\nu$ are given by
\begin{align}
	\mathcal{M} &= \frac{(m_1m_2)^{3/5}}{(m_1+m_2)^{1/5}}\,,\\
	\ln\Delta\nu &= \ln\left(\frac{1}{4} - \nu\right)\,.
\end{align}

\section{Locating phase transitions from GW signals}
\label{sec:search}

\subsection{Method description}
\label{sec:methodology}

Because the pressure $p$ within a compact star is monotonically decreasing from $p_c$ in the center to $p=0$ at the surface, only the part of the EOS with pressures below the central pressure $p_c$ is observable. 
With this in mind, we define the hypotheses to be tested as follows:
\begin{itemize}
	\item $\mathcal{H}_{\small\textrm{PT}}:$ The phase transition pressure $\ptr$ is below $p_{c}$ for one or both of the stars and the phase transition experiences a density jump $\Delta\rho > 0$;
	\item $\mathcal{H}_{\small\textrm{NPT}}:$ The phase transition density jump $\Delta\rho$ is zero or $\ptr$ is larger than $p_{c}$, and therefore the transition is not observable.
\end{itemize}
For $\mathcal{H}_{\small\textrm{NPT}}$, we found that it is sufficient to test for the condition $\Delta\rho=0$. 
Within our parameterization, the condition $\ptr>p_{c}$ is equivalent to fitting the whole observable EOS with a single polytrope (instead of 3-4 polytropes in the case of $\ptr<p_{c}$), which is penalized by the fit.
Moreover, a significant decrease of the number of fit degrees of freedom complicates our interpretation of the evidence and, correspondingly, the Bayes factors.

The evidences for the two hypotheses are given by
\begin{equation}
\begin{aligned}
	\mathcal{Z}_{\small\textrm{PT}} = &\int d\vec{E} \, dm_i\, \mathcal{L}(\vec{E},m_i)\\
	&\times\pi(\vec{E},m_i\mid (\ptr<p_{c,1}\textrm{ or } \ptr<p_{c,2})\, \& \, \Delta\rho>0)\,,\\
\end{aligned}
\end{equation}
and
\begin{equation}
	\mathcal{Z}_{\small\textrm{NPT}} = \int\mathcal{L}(\vec{E}_c,m_i) \, \pi(\vec{E}_c\vert\Delta\rho=0) \, d\vec{E}_c \, dm_i,
\end{equation}
where the central pressure $p_{c,i}$ is estimated via interpolation of $m_i$-$p_{c,i}$ for given EOS parameters $\vec{E}$.

The Bayes factor $\mathcal{B}^{\textrm{PT}}_{\textrm{NPT}}$ between $\mathcal{H}_{\small\textrm{PT}}$ and $\mathcal{H}_{\small\textrm{NPT}}$ is given by
\begin{equation}
\begin{aligned}
	\mathcal{B}^{\textrm{PT}}_{\textrm{NPT}} =\frac{\mathcal{Z}_{\small\textrm{PT}}}{\mathcal{Z}_{\small\textrm{NPT}}}.
\end{aligned}
\end{equation}
By examining the Bayes factor $\mathcal{B}^{\textrm{PT}}_{\textrm{NPT}}$, one can deduce if a phase transition is observed. 

\subsection{Method Validation}

\begin{figure}[t]
\begin{center}
\includegraphics[width=0.95\columnwidth]{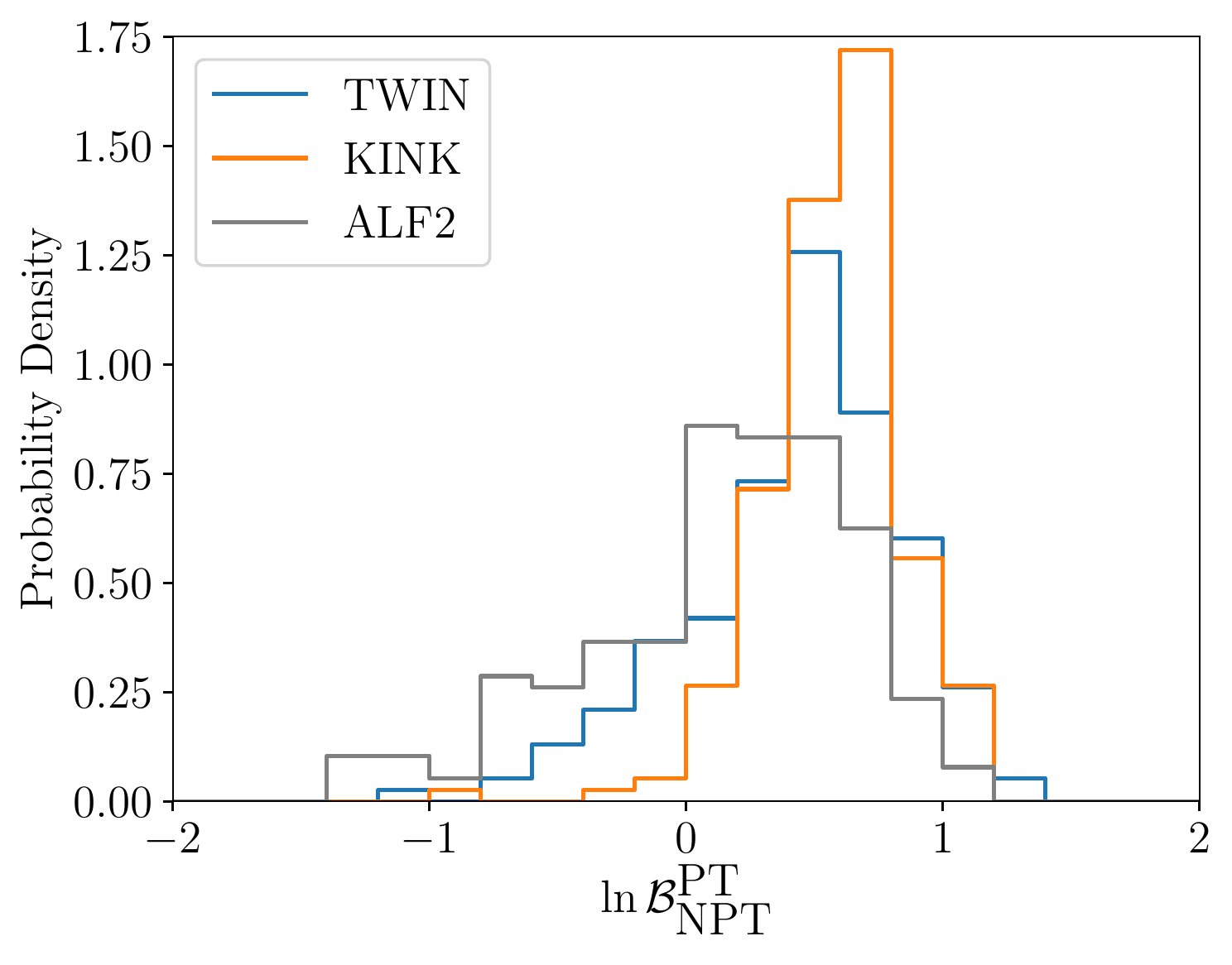}
\caption{The distribution of $\ln\mathcal{B}^{\textrm{PT}}_{\textrm{NPT}}$ for injections with the TWIN, KINK, and ALF2 EOSs. The presence of a strong phase transition does shift the distribution of $\ln\mathcal{B}^{\textrm{PT}}_{\textrm{NPT}}$ towards larger values.}
\label{fig:PT_NPT_bayes}
\end{center}
\end{figure}

\begin{figure*}[t]
\begin{center}
\includegraphics[width=0.99\textwidth]{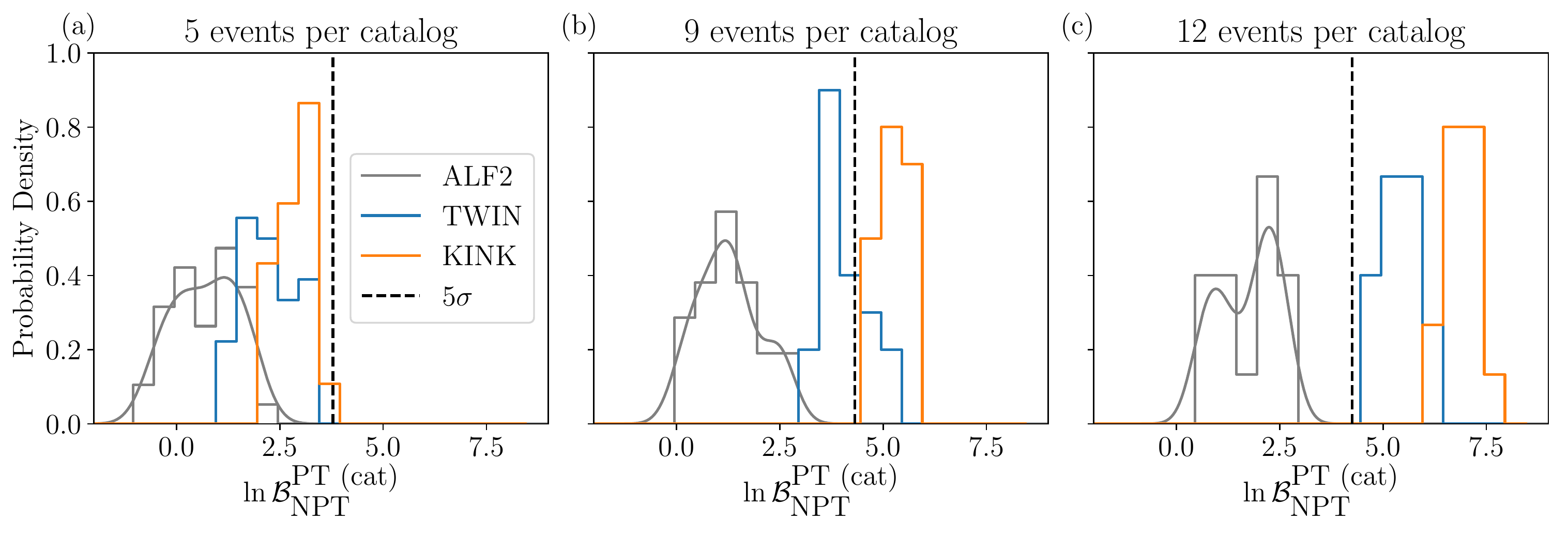}
	\caption{The distributions of $\ln\mathcal{B}^{\textrm{PT (cat)}}_{\textrm{NPT}}$ for injections with the TWIN, KINK, and ALF2 EOSs. Each catalog consists of 5 events (a), 9 events (b), or 12 events (c). The presence of a strong phase transition can be recognized from the distributions for 9 or more events. 
	The $5\sigma$ threshold, indicated by the black dashed line, is estimated with respect to the log Bayes factor distribution for the ALF2 EOS using a Gaussian kernel density estimation (the smooth grey curve).}
\label{fig:PT_NPT_bayes_cat}
\end{center}
\end{figure*}

With 200 BNS merger signals for each EOS, the parameter estimation is performed and the evidences are estimated as described in Sec.~\ref{sec:methodology}.

The probability distribution functions of the log Bayes factors $\ln\mathcal{B}^{\textrm{PT}}_{\textrm{NPT}}$ obtained from all 200 injections for each of the three EOSs are shown in Fig.~\ref{fig:PT_NPT_bayes}.
In our simulations, 85\%, 98\% and 70\% of the injections lead to a positive $\ln\mathcal{B}^{\textrm{PT}}_{\textrm{NPT}}$ for the TWIN, KINK, and ALF2 EOSs, respectively. 
Even though the EOSs with a strong phase transition, KINK and TWIN, do shift the distribution of $\ln\mathcal{B}^{\textrm{PT}}_{\textrm{NPT}}$ towards larger values, the shift is not pronounced enough to draw a statistically robust conclusion. 
Furthermore, while there is no strong first-order phase transition in the ALF2 EOS, $\mathcal{H}_{\textrm{PT}}$ is favored over $\mathcal{H}_{\textrm{NPT}}$ for the ALF2 injections. 
However, in $\mathcal{H}_{\textrm{PT}}$, the EOS below $p_c$ is fitted with $3-4$ polytropes while in $\mathcal{H}_{\textrm{NPT}}$ it is fitted with only $1-3$ polytropes. The additional degrees of freedom for $\mathcal{H}_{\textrm{PT}}$ improve the fit to the $M$-$\Lambda$ curve and lead to a higher evidence.

To improve on the situation, we follow the catalog technique described in Sec.~\ref{ssec:bayesian_analysis}, and estimate the catalog log Bayes factors $\ln\mathcal{B}^{\textrm{PT (cat)}}_{\textrm{NPT}}$.
In Fig.~\ref{fig:PT_NPT_bayes_cat}, we show the distributions of $\ln\mathcal{B}^{\textrm{PT (cat)}}_{\textrm{NPT}}$ with 5, 9, and 12 events per catalog for the three EOSs. 
We find that the presence of a strong phase transition in the EOS can be clearly recognized from the distributions for 9 or more events per catalog. 
For both the TWIN and KINK EOSs, with 12 events per catalog, all the catalogs result in a higher than $5\sigma$ statistical significance with respect to the catalogs estimated for the ALF2 EOS. 

Before continuing, we note that by combining the Bayes factors with simple multiplication we are not making use of the knowledge that all BNSs share the same EOS. The inclusion of this additional constraint would require us to do the analysis on all BNSs in a catalog simultaneously, which would be computationally very demanding. 
While our procedure is sub-optimal, it is a conservative one. Most importantly, we see that 
by using our catalog Bayes factor as a detection statistic for strong phase transitions, 
with 12 sources we can already draw significant conclusions, regardless of the interpretation of 
the Bayes factor.

Returning to Fig.~\ref{fig:PT_NPT_bayes_cat}, we 
see that the phase transition in the KINK EOS is easier to identify than the TWIN EOS, even 
though the TWIN EOS has the more pronounced feature in the $M$-$R$ relation. Moreover, 
in the case of TWIN, the parameters $B_1$ and $B_2$, which indicate on which branch the
stars live, are not redundant, and one would think that this should boost the evidence in that case.
However, the individual $\Lambda_i$ for each component in a BNS are not well measured\footnote{This 
results from the poor measurement of $\delta\tilde\Lambda$ with 
second-generation detectors~\cite{Wade:2014vqa}.}; as a result, the inclusion of the $\{B_i\}$ 
does not necessarily lead to the evidence being elevated. 
In addition, as seen in Fig.~\ref{fig:EOS}, our parameterization can fit the KINK EOS better than the TWIN EOS, which contributes to the higher evidence of KINK compared to TWIN.

\begin{figure*}[t]
\begin{center}
\includegraphics[width=0.49\textwidth]{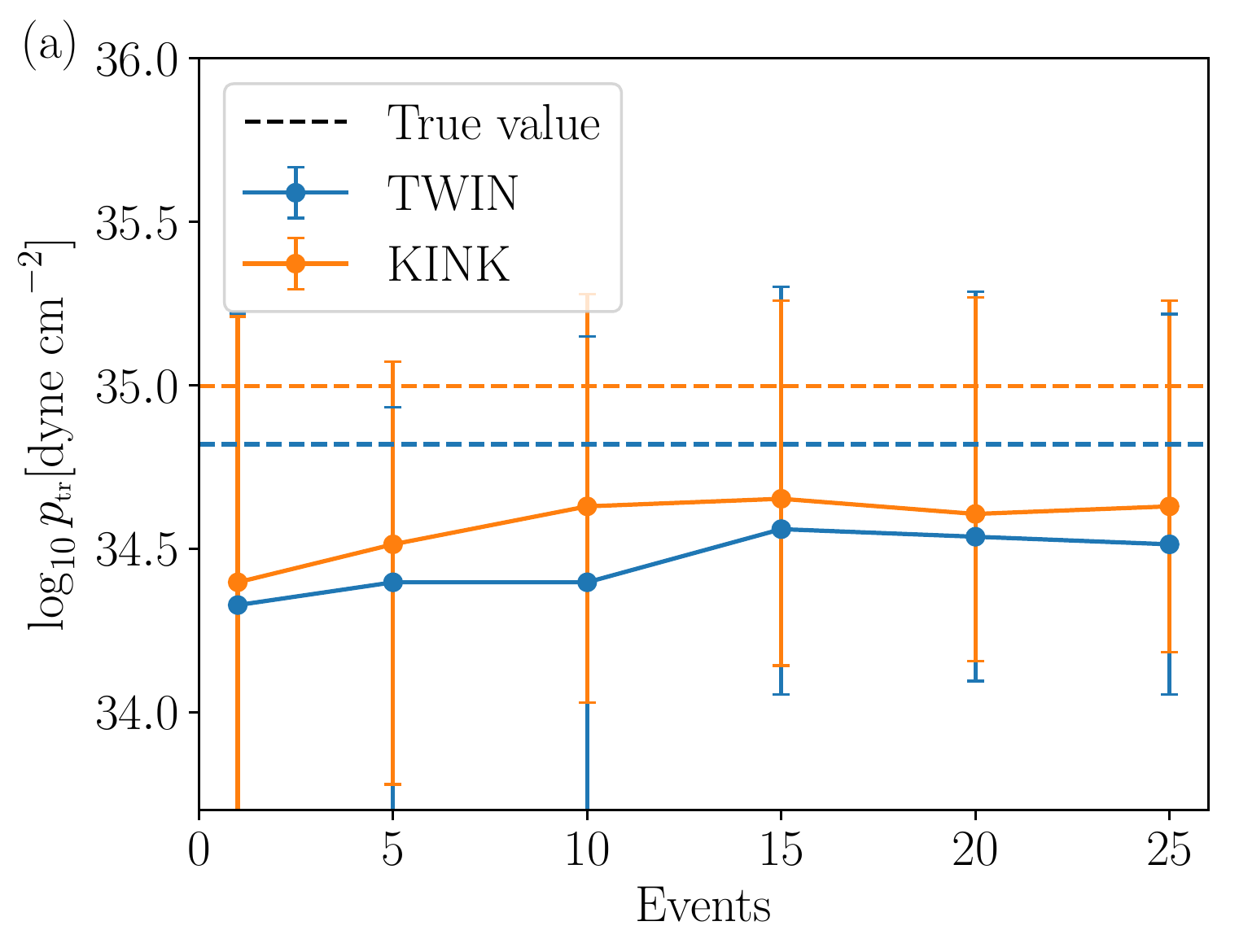}
\includegraphics[width=0.49\textwidth]{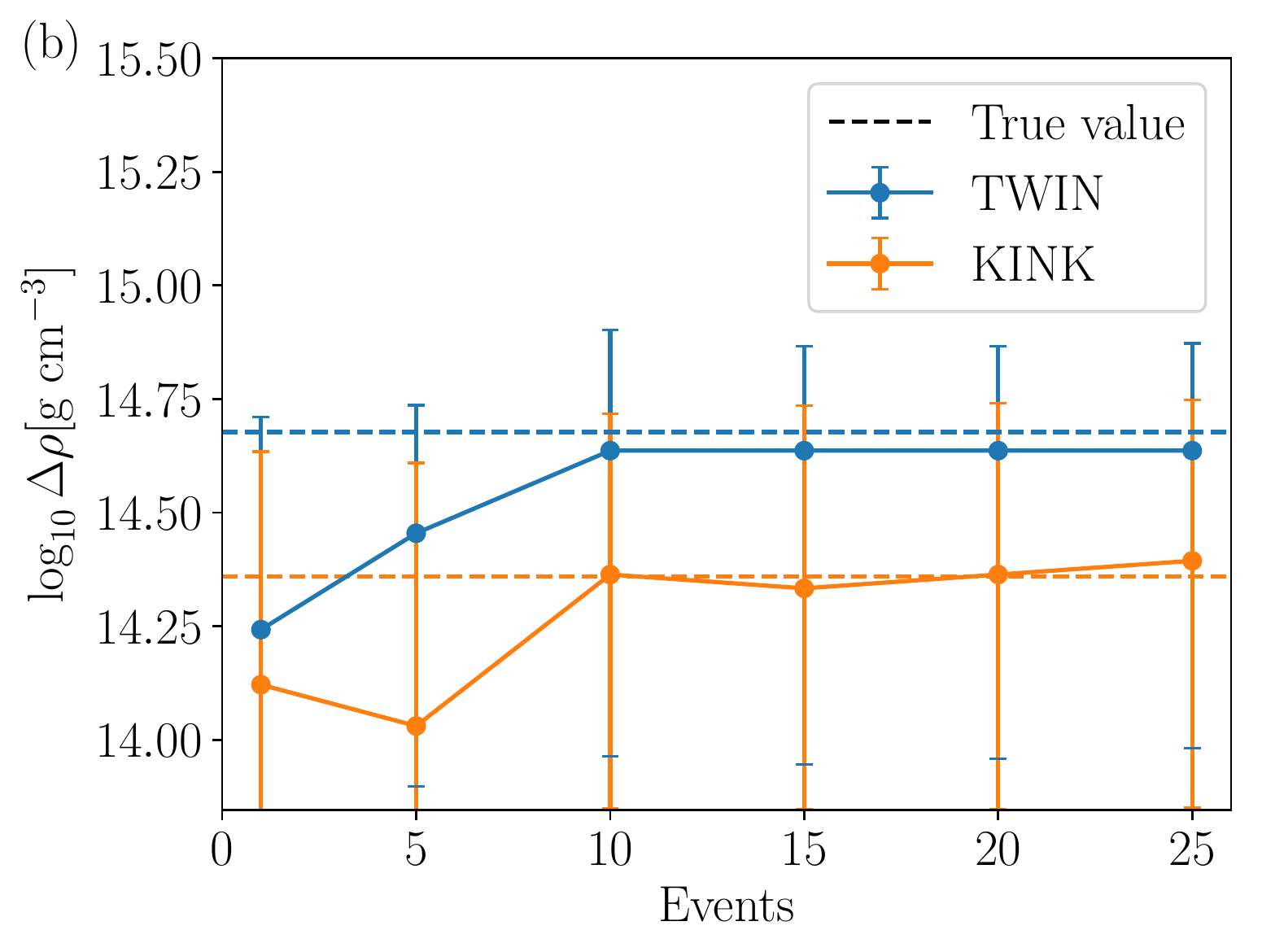}
	\caption{Maximum \textit{a posteriori} (MAP) and $95\%$ credible interval evolution for $\log_{10}\ptr$ (a) and $\log_{10}\Delta\rho$ (b) for the TWIN and KINK EOSs. 
	The dashed lines indicate the true values, which are found to lie within the extracted $95\%$ credible interval for both EOS.}
\label{fig:ptr_trend}
\end{center}
\end{figure*}

\begin{figure*}[t]
\begin{center}
\includegraphics[width=0.49\textwidth]{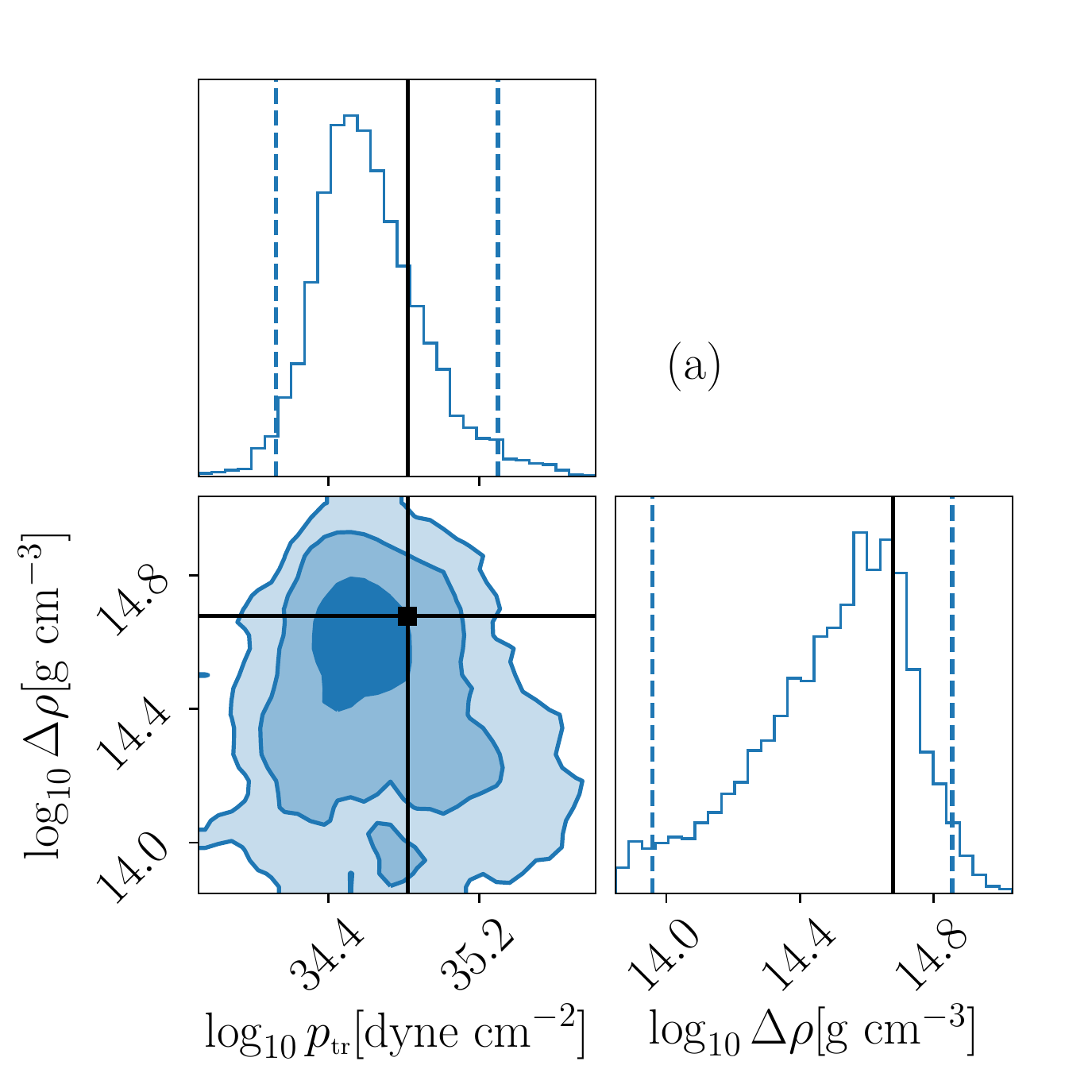}
\includegraphics[width=0.49\textwidth]{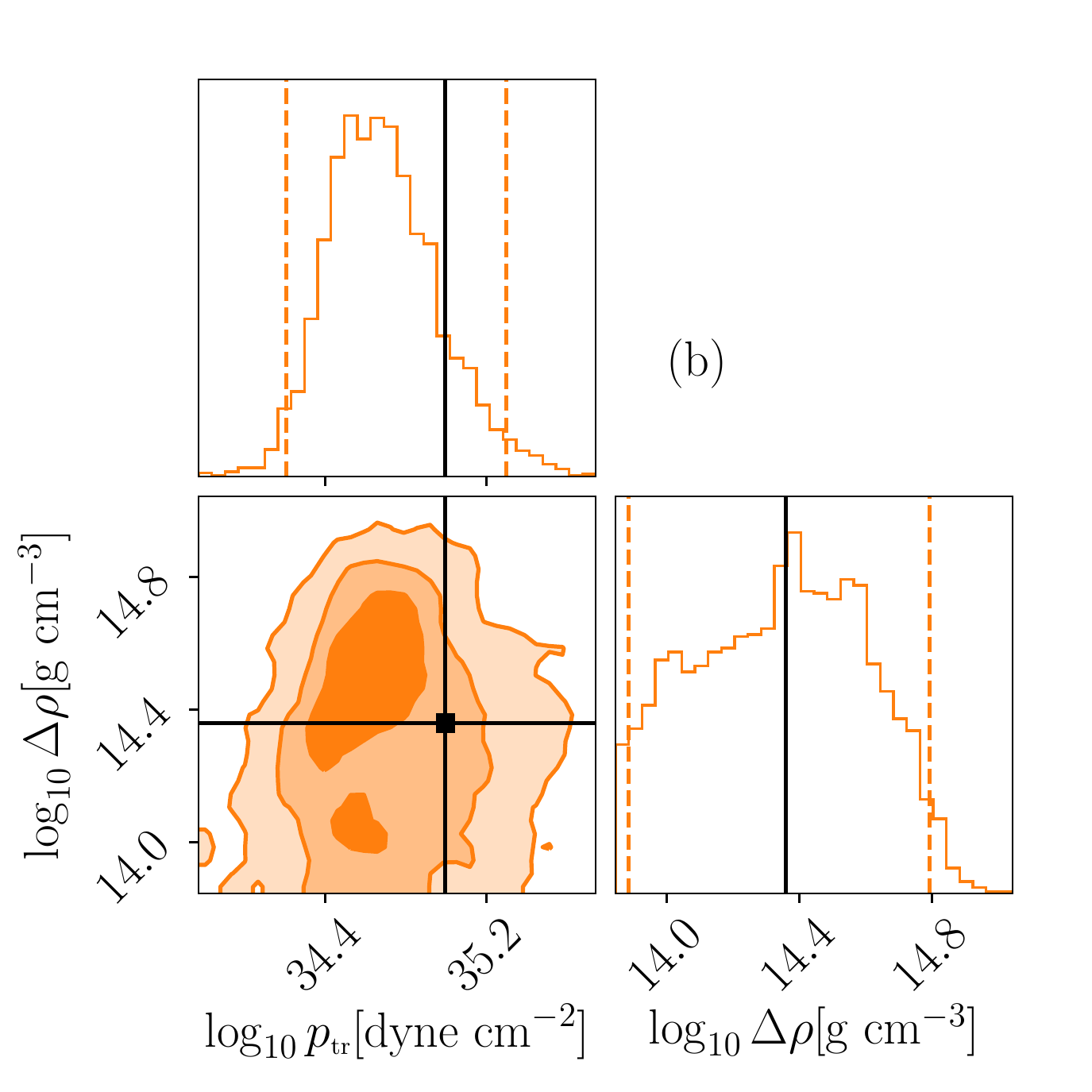}
	\caption{The joint posteriors for $\log_{10}\ptr$ and $\log_{10}\Delta\rho$ with $25$ combined events for the TWIN EOS (a) and the KINK EOS (b). 
	The dashed lines indicate the $95\%$ credible interval and the solid lines the true value. For both EOSs, the true values of the phase transition parameters $(\log_{10}\ptr,\log_{10}\Delta\rho)$ are constrained with $\lesssim10$\% statistical uncertainty.}
\label{fig:ptr_pe}
\end{center}
\end{figure*}

In addition to observing the presence of a phase transition with statistical significance, we can also 
probe its characteristics.  
In particular, the phase transition onset pressure $\ptr$, and the phase transition density jump $\Delta\rho$, can be measured. 
In Fig.~\ref{fig:ptr_trend}, we show the maximum \textit{a posteriori} (MAP) and $95\%$ credible interval evolution for $\log_{10}\ptr$ and $\log_{10}\Delta\rho$ with 
an increasing number of events, where we only include events with 
positive $\ln\mathcal{B}^{\textrm{PT}}_{\textrm{NPT}}$ for TWIN and KINK. Indeed, we expect that
events with positive 
$\ln\mathcal{B}^{\textrm{PT}}_{\textrm{NPT}}$ will tend to have $p_c > \ptr$, and hence
be informative for the estimation of $\log_{10}\ptr$ and $\log_{10}\Delta\rho$; this is something
we will return to momentarily. 
Looking at the results, for both EOSs the true phase transition parameters are recovered within the 
$95\%$ credible interval. 
For $\log_{10}\Delta\rho$, the true value is recovered after $\sim 10$ events are included.
The statistical errors decrease as more events are combined, though the rate of decrease 
appears to slow down after $\sim 10$ events. We note that the credible intervals for the parameters of the phase transition are not dominated by the one or two loudest events. Instead, in order to detect the phase transition, we need to map-out a significant part of the M-$\Lambda$ curve, which requires multiple detections. 

In Fig.~\ref{fig:ptr_pe}, we show the joint posteriors for $\log_{10}\ptr$ and $\log_{10}\Delta\rho$  
with $25$ combined events for the TWIN and KINK EOSs. 
For both EOSs, the phase transition parameters $(\log_{10}\ptr,\log_{10}\Delta\rho)$ 
are measured with $\lesssim 10\%$ statistical uncertainty, 
with the true values lying within the $95\%$ credible interval.

\begin{figure}[t]
\begin{center}
\includegraphics[width=0.99\columnwidth]{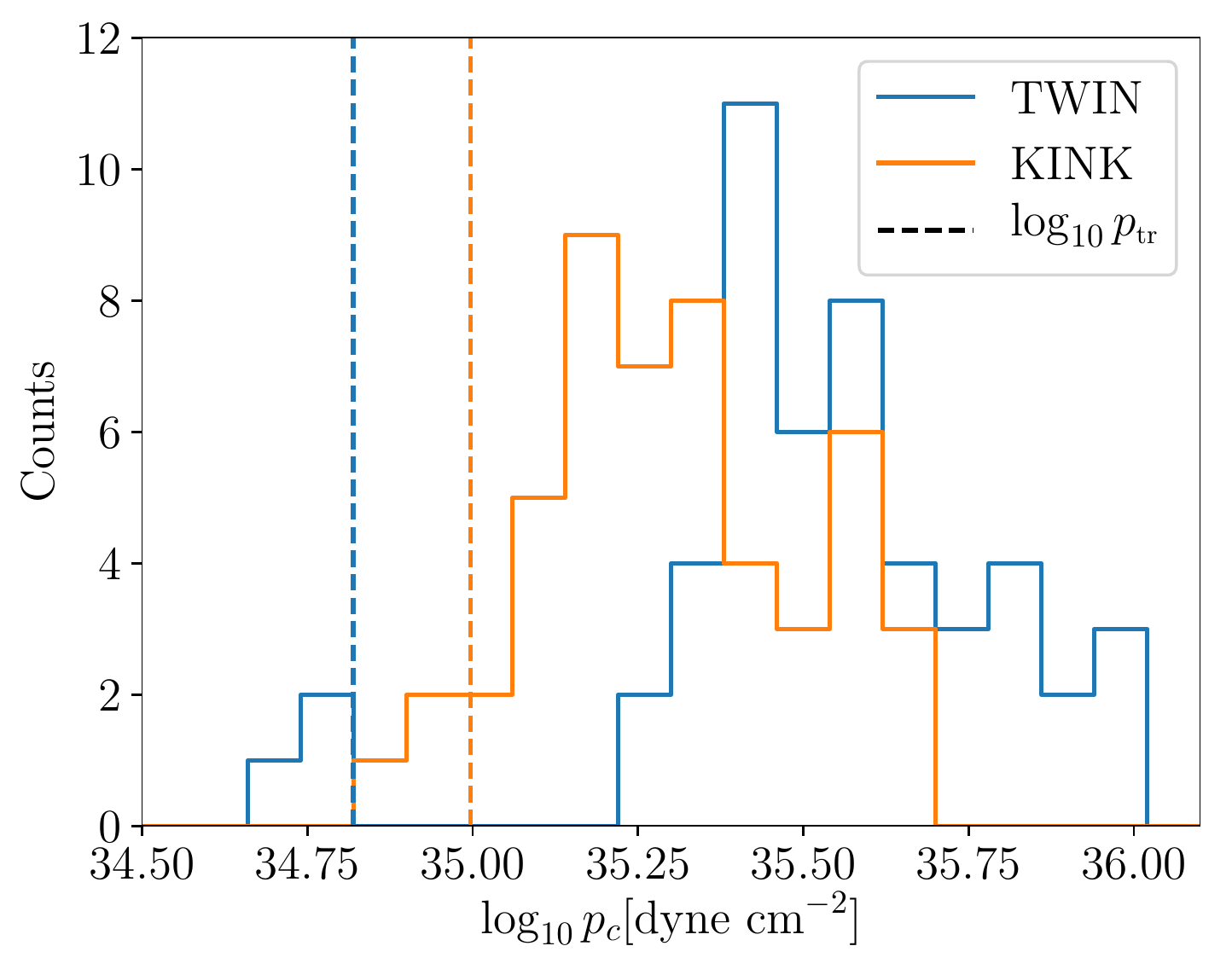}
	\caption{The distribution of the central pressure $p_c$ for the 25 events included in the joint parameter-estimation analysis with the phase transition pressure for the two EOSs indicated by the dashed line. Most of the events have both of the components' central pressure above the phase transition pressure. In particular, none of the events have both components' central pressure below phase transition pressure.}
\label{fig:logpc_hist}
\end{center}
\end{figure}

Finally, for the same 25 events, we show the distribution of the central pressure $p_c$ for both stars 
in Fig.~\ref{fig:logpc_hist}. 
For the majority of the events, both stars have a central pressure above the phase transition 
onset pressure. This matches our expectation from the 
positive $\ln\mathcal{B}^{\textrm{PT}}_{\textrm{NPT}}$ and provides a 
valuable crosscheck for our analysis.

Based on our findings, we conclude that:
\begin{enumerate}[label=(\roman*)]
	\item It is possible to establish the presence of a strong first-order phase transition with twelve BNS observations;
	\item With $\sim 10$ BNSs, the phase transition parameters $(\log_{10}\ptr,\log_{10}\Delta\rho)$ 
	can be measured with $\lesssim 10\%$ statistical uncertainty.
\end{enumerate}

Because we are imposing a strict bound on the $M_{\textrm{TOV}}$ during the stage II analysis, systematics might be induced as suggested in Ref.~\cite{Miller_2019, Landry:2020vaw}.
However, since we are not interested in recovering the full EOS or $M_{\textrm{TOV}}$ but the parameters of the phase transition, the simple hard cut on $M_{\textrm{TOV}}$ is sufficient for the purposes of this analysis. 
Also, no significant systematics are observed with respect to the simulation.

\subsection{Limitations of our analysis}

As we have shown in the previous section, it is possible to confirm the existence of a phase transition and extract its parameters.
However, our approach is limited to such EOSs where the phase transition is Maxwell-like, i.e., it is described by a segment with $c_S=0$.
As the comparison with the ALF2 EOS clearly shows, our approach cannot establish the existence of a phase transition in case a mixed phase appears, that smears out an observable EOS feature.
Due to the masquerade problem~\cite{Alford:2006vz}, macroscopic structure properties ($M$-$R$ or $M$-$\Lambda$ relations) for such EOS cannot be distinguished from purely nucleonic EOS.
As only such properties affect the GW waveforms, the inspiral phase does not provide information on the phase transition in this case.
Should such a case be realized in nature, information might only be obtained from the postmerger GW signal.

Furthermore, we can only observe a Maxwell-like transition that is strong enough to leave an observable feature, e.g., at least a kink, in the $M$-$\Lambda$ curve. 
Should the density jump be too small, the $M$-$\Lambda$ curve would be smooth and, again, the inspiral phase would not provide information on the phase transition.

Finally, our method only works if inspiraling NSs have central pressures above the onset of the phase transition. 
If NS masses in binaries are limited to be around $1.4 \,M_{\odot}$, exploring lower central pressures, but phase transitions appear at much higher pressures in heavier stars, it will only be probed by the postmerger GW signal. 
However, the observation of GW190425 shows that also BNS mergers of heavy NS might be observed by GW interferometers (note that GW190425 could potentially have been a neutron-star--black-hole merger~\cite{Foley:2020kus, Han:2020qmn, Kyutoku:2020xka}).
Should the phase transition appear at low pressures/densities, in NS below $1 \,M_{\odot}$, such that both NS in a binary are hybrid stars, our method might also not be able to identify its presence.
In this case, while the observed macroscopic NS observables represent integrals over the EOS at all densities in the star, and, hence, in principle include information of the phase transition, observations might not be able to distinguish between the ``true" EOS and an EOS that connects smoothly between the low-density nuclear-physics constraints and the observed part of the $M$-$R$ curve.
However, in this case the phase transition onset would likely be between $(1-2) n_{\textrm{sat}}$, at low energy densities, which might be identified using other analysis techniques~\cite{Essick:2020flb}.

These shortcomings highlight that GW observations alone might not be able to answer the 
question of whether phase transitions exists in NSs or not. 
Hence, interdisciplinary studies including both nuclear physics and GW 
astrophysics are crucial.

\section{Analysis of GW170817 and GW190425}
\label{sec:real_GW}

Having shown with simulations that our methodology is in principle capable of uncovering and characterizing 
strong phase transitions, we now apply it to the real signals GW170817~\cite{TheLIGOScientific:2017qsa} 
and GW190425~\cite{Abbott:2020uma}. The former can confidently be assumed to have come from 
a BNS inspiral. We will also assume that the latter was emitted by a BNS. 
For both events, we take the publicly available posterior 
samples~\cite{GW170817_PE_samples,GW190425_PE_samples} as the input for the stage II 
analysis as described in Sec.~\ref{sec:implementation}. 

\begin{table}
\begin{ruledtabular}\begin{tabular}{l|c|c}
	Event & $\ln\mathcal{B}^{\textrm{PT}}_{\textrm{NPT}}$ & KL divergence\\
	\hline
	GW170817 & $0.889\pm0.113$ & $0.809$\\
	GW190425 & $0.441\pm0.085$ & $0.371$\\
	Combined & $1.330\pm0.141$ & $0.865$\\
\end{tabular}\end{ruledtabular}
\caption{The log Bayes factor $\ln\mathcal{B}^{\textrm{PT}}_{\textrm{NPT}}$ and the KL divergence estimated with the two BNS event.}
\label{tab:money_tab}
\end{table}

Table \ref{tab:money_tab} shows the log Bayes factors 
$\ln\mathcal{B}^{\textrm{PT}}_{\textrm{NPT}}$ and the 
Kullback–Leibler (KL) divergence for the two events separately and combined. 
The KL divergence is estimated from the posterior and prior for the EOS parameters, i.e.,
\begin{equation}
	\textrm{KL divergence } = \int \mathcal{P}(\vec{E}_c)\ln\frac{\mathcal{P}(\vec{E}_c)}{\pi(\vec{E}_c)}d\vec{E}_c \,,
\end{equation}
and it quantifies to what extent the posterior distribution is different from the prior distribution. 
A KL divergence of zero indicates that the two distributions are identical.

Based on the values of KL divergence for the two events, it would seem that GW170817 is 
carrying more information regarding the EOS, while GW190425 is not very informative, similar to 
the findings of Ref.~\cite{Landry:2020vaw}. 

In order to make a statistically robust statement, we would need to have a reliable distribution of 
$\ln\mathcal{B}^{\textrm{PT (cat)}}_{\textrm{NPT}}$ in the absence of strong phase transitions, which
we could use as ``background'' to estimate the significance of the ``foreground'' values in Table 
\ref{tab:money_tab}. 
This would require (i) an accurate or at least representative simulated BNS population, and (ii) 
a justifiable representation of EOSs without a strong phase transition. 
One can systematically generate a representative ensemble of EOSs by using parameteric or 
non-parameteric methods and used them to calculate a background distribution for 
$\ln\mathcal{B}^{\textrm{PT (cat)}}_{\textrm{NPT}}$, but the required computational resources 
will be significant. 
Due to the uncertainty in the BNS population and the limitation of available computational resources, 
such an estimation is currently not achievable. 
Instead, we follow the interpretation of the Bayes factor described in Ref.~\cite{Jeffreys61}, 
and find no strong evidence for or against the presence of a strong phase transition.

\begin{figure}[t]
\begin{center}
\includegraphics[width=0.99\columnwidth]{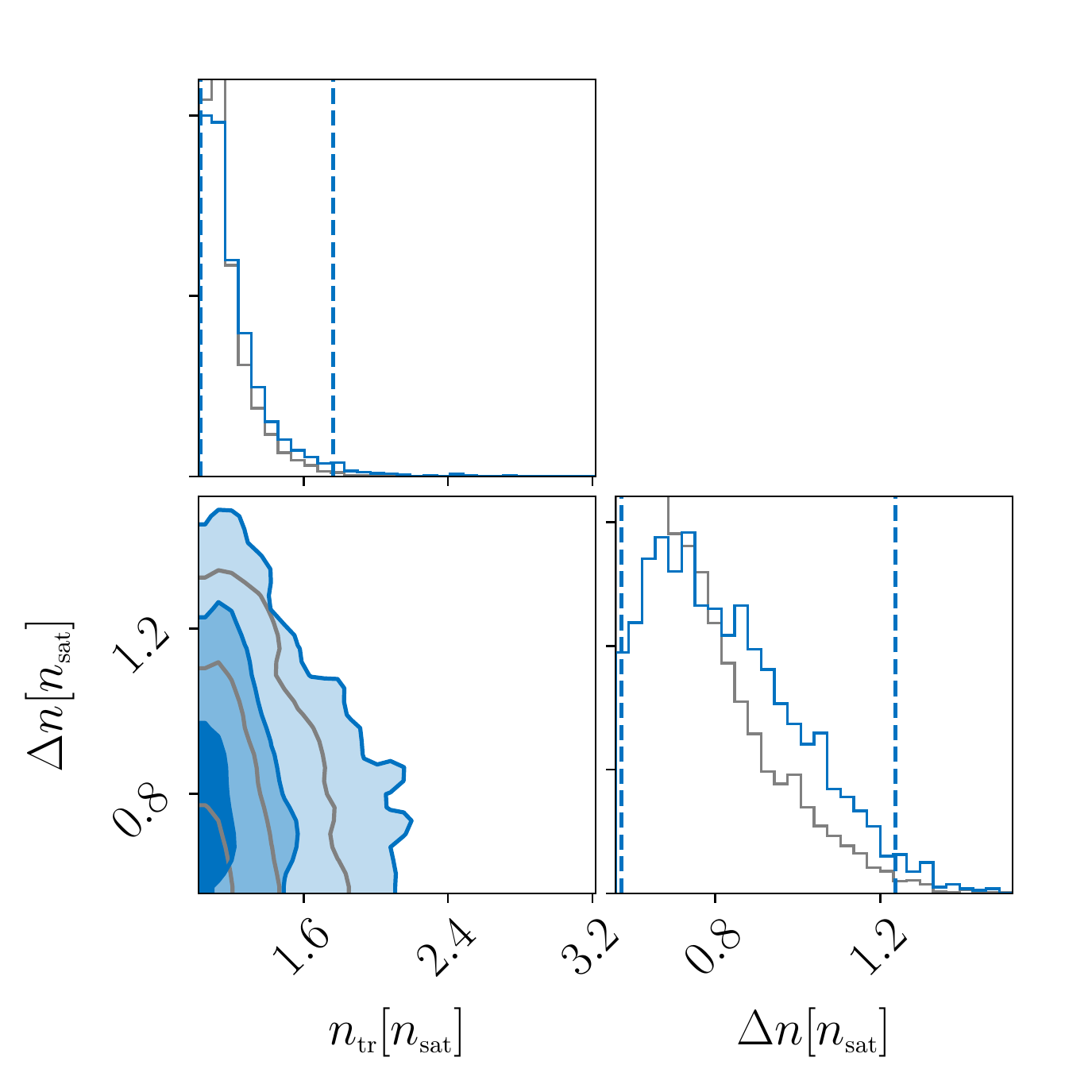}
	\caption{Corner plots showing the posterior distribution of $n_{\small\textrm{tr}}$ and $\Delta n$ in terms of $n_{\small\textrm{sat}}$ with GW170817 and GW190425 combined (\textit{blue}) on top of the prior with heavy pulsars constraint (\textit{grey}). The dashed lines mark the $95\%$ credible intervals. The posterior does not significantly deviate from the prior.}
\label{fig:detection_joint}
\end{center}
\end{figure}

\begin{figure*}[t]
\begin{center}
\includegraphics[width=\textwidth]{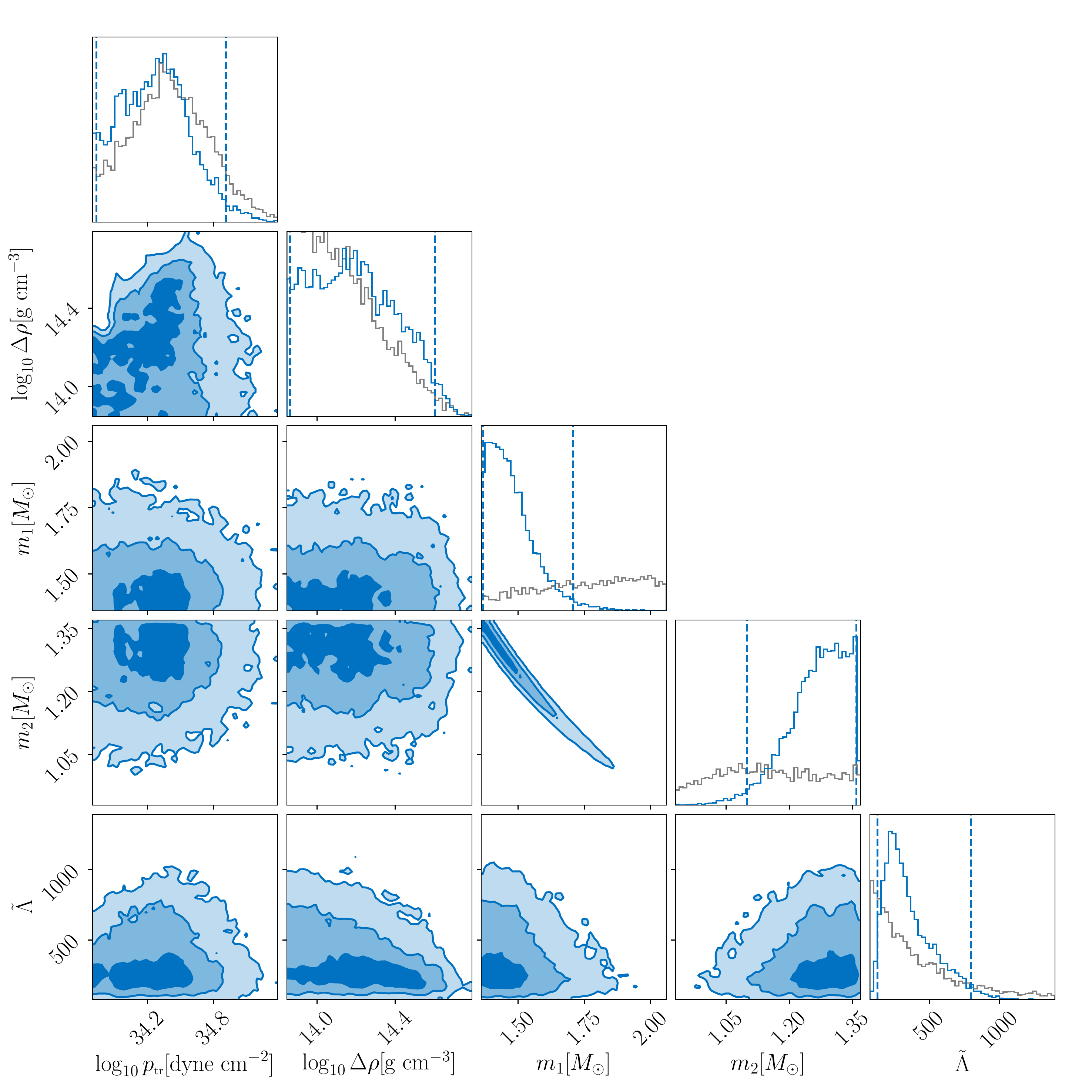}
	\caption{Corner plot showing the posterior distribution of $\log_{10}\ptr$, $\log_{10}\Delta\rho$, $m_1$, $m_2$ and $\tilde{\Lambda}$ for GW170817 (\textit{blue}) on top of the prior distribution arcoss parameters (\textit{grey}). The dashed lines marks the $95\%$ credible intervals for the posterior distribution.}
\label{fig:detection_pe_GW170817}
\end{center}
\end{figure*}

\begin{figure*}[t]
\begin{center}
\includegraphics[width=\textwidth]{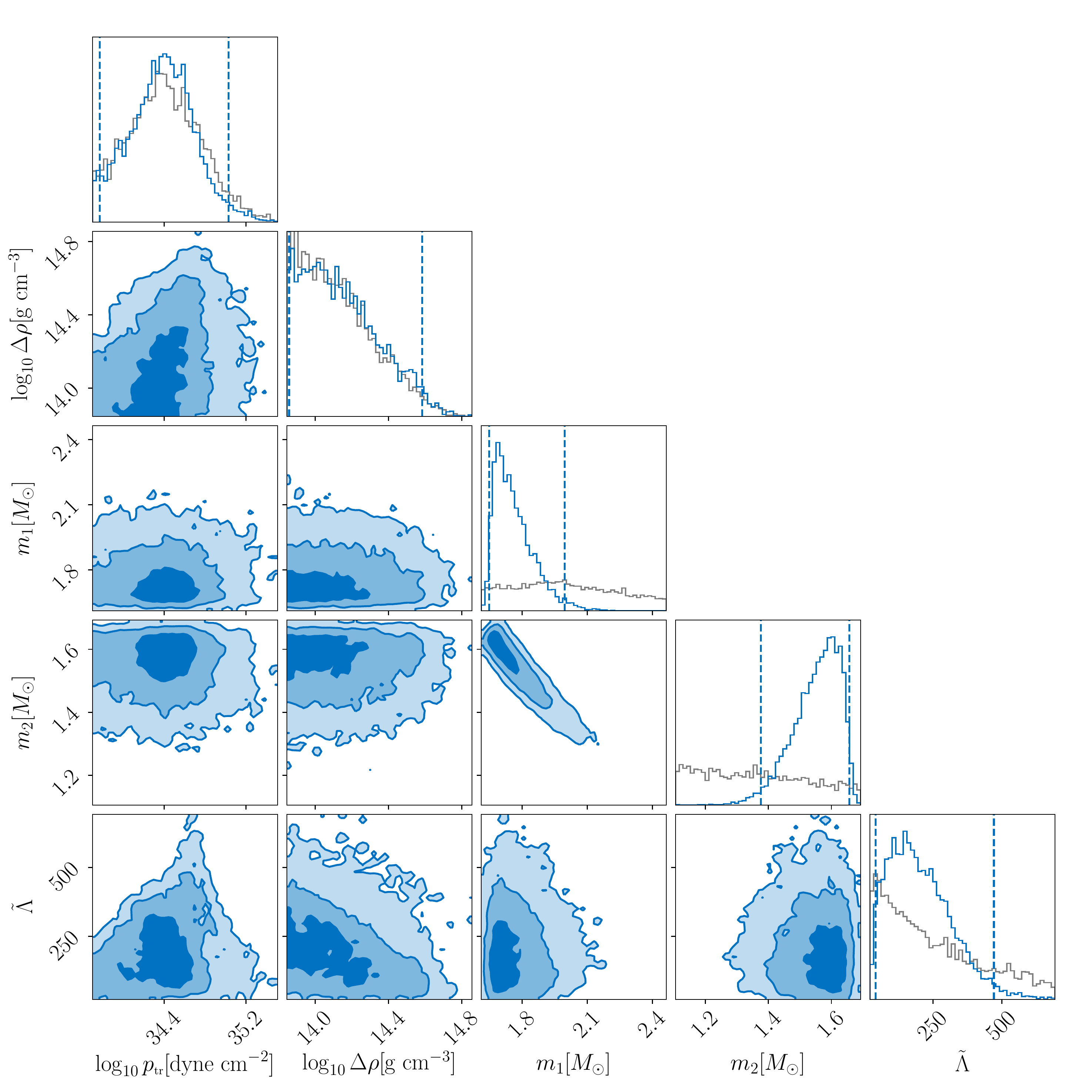}
	\caption{Corner plot showing the posterior distribution of $\log_{10}\ptr$, $\log_{10}\Delta\rho$, $m_1$, $m_2$ and $\tilde{\Lambda}$ for GW190425 (\textit{blue}) on top of the prior distribution arcoss parameters (\textit{grey}). The dashed lines marks the $95\%$ credible intervals for the posterior distribution.}
\label{fig:detection_pe_GW190425}
\end{center}
\end{figure*}

In addition to the Bayes factors, a small value for the KL divergence indicates similarity 
of posterior and prior, and the extracted phase transition parameters are strongly influenced by the priors. 
The combined posteriors for the phase transition onset density, $n_{\small\textrm{tr}}$, and the 
corresponding density jump in terms of $n_{\small\textrm{sat}}$, $\Delta n$, are shown together with the 
prior in Fig.~\ref{fig:detection_joint}.\footnote{Initially we have priors 
$n_{\small\textrm{tr}}\in [1, 4]n_{\small\textrm{sat}}$ and 
$\Delta n \in[0.26,37]n_{\small\textrm{sat}}$, but the presence of heavy pulsars contrained both  
priors to more narrow ranges.}
The posteriors distributions for $\log_{10}\ptr$, $\log_{10}\Delta\rho$, $m_1$, $m_2$, and 
$\tilde{\Lambda}$ for GW170817 and GW190425 are shown in Figs.~\ref{fig:detection_pe_GW170817} 
and~\ref{fig:detection_pe_GW190425}, respectively. 
We conclude that our measurements of $n_{\small\textrm{tr}}$ and $n_{\small\textrm{sat}}$ 
are not very informative, as could be expected based on the KL divergence.

\section{Conclusion}
\label{sec:conclusion}

In this paper, we have presented a reliable way of searching for phase transitions in supranuclear matter using the inspiral waveform of binary NS mergers, which is successful both if the resulting mass-radius relation has only one or multiple stable branches.
In contrast to previous works, which calculated the preference for multiple stable branches to search for strong phase transitions~\cite{PhysRevD.101.063007, Landry:2020vaw, Essick:2020flb}, our approach searches for an extended segment with $c_S=0$, i.e., we do not explicitly assume a multiple-branch feature in the $M$-$R$ curve.

As long as there is some observable imprint of the phase transition in the mass-radius relation, our method can recover injected phase-transition parameters, and, hence, represents an important step forward in the search for a possible phase transition.
We have explicitly demonstrated this by injecting simulated BNS mergers with different equations of state into synthetic stationary Gaussian noise. 
We have shown that our method can detect the presence of phase transitions at $5\sigma$ confidence with 12 signals. 
Moreover, the phase-transition onset pressure and the corresponding density jump $(\log_{10}\ptr,\log_{10}\Delta\rho)$ were recovered with $\lesssim$10\% statistical uncertainty with $\sim$10 events.
Finally, we have applied the method to GW170817 and GW190425, but found no strong evidence for or against the presence of strong phase transitions. 

\begin{acknowledgments}
  We thank M.~Hanauske and L.~Rezzolla for fruitful discussions stimulating this work. 
  We would also like to extend our gratitude to R.~Essick for many helpful comments and discussions, and the very careful reading of our manuscript.
P.T.H.P and C.V.D.B are supported by the research program of the Netherlands Organization for Scientific Research (NWO). The work of I.T. was supported by the U.S. Department of Energy, Office of Science, Office of Nuclear Physics, under Contract No.~DE-AC52-06NA25396, by the NUCLEI SciDAC program, and by the LDRD program at LANL.   
We are thankful for the computing resources provided by the LIGO-Caltech Computing Cluster where our simulations and analysis were carried out, which is supported by National Science Foundation Grants PHY-0757058 and PHY-0823459.
Computational resources have also been provided by the Los Alamos National Laboratory Institutional Computing Program, which is supported by the U.S. Department of Energy National Nuclear Security Administration under Contract No.~89233218CNA000001, and by the National Energy Research Scientific Computing Center (NERSC), which is supported by the U.S. Department of Energy, Office of Science, under Contract No.~DE-AC02-05CH11231.
This research has made use of data, software and/or web tools obtained from the Gravitational 
Wave Open Science Center (https://www.gw-openscience.org), a service of LIGO Laboratory, the 
LIGO Scientific Collaboration and the Virgo Collaboration. LIGO is funded by the U.S. National 
Science Foundation. Virgo is funded by the French Centre National de Recherche Scientifique (CNRS), 
the Italian Istituto Nazionale della Fisica Nucleare (INFN) and the Dutch Nikhef, with contributions 
by Polish and Hungarian institutes.
\end{acknowledgments}

\bibliography{refs}

\begin{thebibliography}{125}%
\makeatletter
\providecommand \@ifxundefined [1]{%
 \@ifx{#1\undefined}
}%
\providecommand \@ifnum [1]{%
 \ifnum #1\expandafter \@firstoftwo
 \else \expandafter \@secondoftwo
 \fi
}%
\providecommand \@ifx [1]{%
 \ifx #1\expandafter \@firstoftwo
 \else \expandafter \@secondoftwo
 \fi
}%
\providecommand \natexlab [1]{#1}%
\providecommand \enquote  [1]{``#1''}%
\providecommand \bibnamefont  [1]{#1}%
\providecommand \bibfnamefont [1]{#1}%
\providecommand \citenamefont [1]{#1}%
\providecommand \href@noop [0]{\@secondoftwo}%
\providecommand \href [0]{\begingroup \@sanitize@url \@href}%
\providecommand \@href[1]{\@@startlink{#1}\@@href}%
\providecommand \@@href[1]{\endgroup#1\@@endlink}%
\providecommand \@sanitize@url [0]{\catcode `\\12\catcode `\$12\catcode
  `\&12\catcode `\#12\catcode `\^12\catcode `\_12\catcode `\%12\relax}%
\providecommand \@@startlink[1]{}%
\providecommand \@@endlink[0]{}%
\providecommand \url  [0]{\begingroup\@sanitize@url \@url }%
\providecommand \@url [1]{\endgroup\@href {#1}{\urlprefix }}%
\providecommand \urlprefix  [0]{URL }%
\providecommand \Eprint [0]{\href }%
\providecommand \doibase [0]{http://dx.doi.org/}%
\providecommand \selectlanguage [0]{\@gobble}%
\providecommand \bibinfo  [0]{\@secondoftwo}%
\providecommand \bibfield  [0]{\@secondoftwo}%
\providecommand \translation [1]{[#1]}%
\providecommand \BibitemOpen [0]{}%
\providecommand \bibitemStop [0]{}%
\providecommand \bibitemNoStop [0]{.\EOS\space}%
\providecommand \EOS [0]{\spacefactor3000\relax}%
\providecommand \BibitemShut  [1]{\csname bibitem#1\endcsname}%
\let\auto@bib@innerbib\@empty
\bibitem [{\citenamefont {Demorest}\ \emph {et~al.}(2010)\citenamefont
  {Demorest}, \citenamefont {Pennucci}, \citenamefont {Ransom}, \citenamefont
  {Roberts},\ and\ \citenamefont {Hessels}}]{Demorest:2010bx}%
  \BibitemOpen
  \bibfield  {author} {\bibinfo {author} {\bibfnamefont {P.}~\bibnamefont
  {Demorest}}, \bibinfo {author} {\bibfnamefont {T.}~\bibnamefont {Pennucci}},
  \bibinfo {author} {\bibfnamefont {S.}~\bibnamefont {Ransom}}, \bibinfo
  {author} {\bibfnamefont {M.}~\bibnamefont {Roberts}}, \ and\ \bibinfo
  {author} {\bibfnamefont {J.}~\bibnamefont {Hessels}},\ }\href {\doibase
  10.1038/nature09466} {\bibfield  {journal} {\bibinfo  {journal} {Nature}\
  }\textbf {\bibinfo {volume} {467}},\ \bibinfo {pages} {1081} (\bibinfo {year}
  {2010})},\ \Eprint {http://arxiv.org/abs/1010.5788} {arXiv:1010.5788
  [astro-ph.HE]} \BibitemShut {NoStop}%
\bibitem [{\citenamefont {Antoniadis}\ \emph {et~al.}(2013)\citenamefont
  {Antoniadis}, \citenamefont {Freire}, \citenamefont {Wex}, \citenamefont
  {Tauris}, \citenamefont {Lynch} \emph {et~al.}}]{Antoniadis:2013pzd}%
  \BibitemOpen
  \bibfield  {author} {\bibinfo {author} {\bibfnamefont {J.}~\bibnamefont
  {Antoniadis}}, \bibinfo {author} {\bibfnamefont {P.~C.}\ \bibnamefont
  {Freire}}, \bibinfo {author} {\bibfnamefont {N.}~\bibnamefont {Wex}},
  \bibinfo {author} {\bibfnamefont {T.~M.}\ \bibnamefont {Tauris}}, \bibinfo
  {author} {\bibfnamefont {R.~S.}\ \bibnamefont {Lynch}},  \emph {et~al.},\
  }\href {\doibase 10.1126/science.1233232} {\bibfield  {journal} {\bibinfo
  {journal} {Science}\ }\textbf {\bibinfo {volume} {340}},\ \bibinfo {pages}
  {6131} (\bibinfo {year} {2013})},\ \Eprint {http://arxiv.org/abs/1304.6875}
  {arXiv:1304.6875 [astro-ph.HE]} \BibitemShut {NoStop}%
\bibitem [{\citenamefont {Cromartie}\ \emph {et~al.}(2019)\citenamefont
  {Cromartie} \emph {et~al.}}]{Cromartie:2019kug}%
  \BibitemOpen
  \bibfield  {author} {\bibinfo {author} {\bibfnamefont {H.~T.}\ \bibnamefont
  {Cromartie}} \emph {et~al.},\ }\href {\doibase 10.1038/s41550-019-0880-2}
  {\bibfield  {journal} {\bibinfo  {journal} {Nat. Astron.}\ }\textbf {\bibinfo
  {volume} {4}},\ \bibinfo {pages} {72} (\bibinfo {year} {2019})},\ \Eprint
  {http://arxiv.org/abs/1904.06759} {arXiv:1904.06759 [astro-ph.HE]}
  \BibitemShut {NoStop}%
\bibitem [{\citenamefont {Steiner}\ \emph
  {et~al.}(2013{\natexlab{a}})\citenamefont {Steiner}, \citenamefont
  {Lattimer},\ and\ \citenamefont {Brown}}]{Steiner:2012xt}%
  \BibitemOpen
  \bibfield  {author} {\bibinfo {author} {\bibfnamefont {A.~W.}\ \bibnamefont
  {Steiner}}, \bibinfo {author} {\bibfnamefont {J.~M.}\ \bibnamefont
  {Lattimer}}, \ and\ \bibinfo {author} {\bibfnamefont {E.~F.}\ \bibnamefont
  {Brown}},\ }\href {\doibase 10.1088/2041-8205/765/1/L5} {\bibfield  {journal}
  {\bibinfo  {journal} {Astrophys. J.}\ }\textbf {\bibinfo {volume} {765}},\
  \bibinfo {pages} {L5} (\bibinfo {year} {2013}{\natexlab{a}})},\ \Eprint
  {http://arxiv.org/abs/1205.6871} {arXiv:1205.6871 [nucl-th]} \BibitemShut
  {NoStop}%
\bibitem [{\citenamefont {Miller}\ \emph
  {et~al.}(2019{\natexlab{a}})\citenamefont {Miller} \emph
  {et~al.}}]{Miller:2019cac}%
  \BibitemOpen
  \bibfield  {author} {\bibinfo {author} {\bibfnamefont {M.}~\bibnamefont
  {Miller}} \emph {et~al.},\ }\href {\doibase 10.3847/2041-8213/ab50c5}
  {\bibfield  {journal} {\bibinfo  {journal} {Astrophys. J. Lett.}\ }\textbf
  {\bibinfo {volume} {887}},\ \bibinfo {pages} {L24} (\bibinfo {year}
  {2019}{\natexlab{a}})},\ \Eprint {http://arxiv.org/abs/1912.05705}
  {arXiv:1912.05705 [astro-ph.HE]} \BibitemShut {NoStop}%
\bibitem [{\citenamefont {Riley}\ \emph {et~al.}(2019)\citenamefont {Riley}
  \emph {et~al.}}]{Riley:2019yda}%
  \BibitemOpen
  \bibfield  {author} {\bibinfo {author} {\bibfnamefont {T.~E.}\ \bibnamefont
  {Riley}} \emph {et~al.},\ }\href {\doibase 10.3847/2041-8213/ab481c}
  {\bibfield  {journal} {\bibinfo  {journal} {Astrophys. J. Lett.}\ }\textbf
  {\bibinfo {volume} {887}},\ \bibinfo {pages} {L21} (\bibinfo {year}
  {2019})},\ \Eprint {http://arxiv.org/abs/1912.05702} {arXiv:1912.05702
  [astro-ph.HE]} \BibitemShut {NoStop}%
\bibitem [{\citenamefont {Ozel}\ and\ \citenamefont
  {Freire}(2016)}]{ozel:2016oaf}%
  \BibitemOpen
  \bibfield  {author} {\bibinfo {author} {\bibfnamefont {F.}~\bibnamefont
  {Ozel}}\ and\ \bibinfo {author} {\bibfnamefont {P.}~\bibnamefont {Freire}},\
  }\href {\doibase 10.1146/annurev-astro-081915-023322} {\bibfield  {journal}
  {\bibinfo  {journal} {Ann. Rev. Astron. Astrophys.}\ }\textbf {\bibinfo
  {volume} {54}},\ \bibinfo {pages} {401} (\bibinfo {year} {2016})},\ \Eprint
  {http://arxiv.org/abs/1603.02698} {arXiv:1603.02698 [astro-ph.HE]}
  \BibitemShut {NoStop}%
\bibitem [{\citenamefont {Aasi}\ \emph {et~al.}(2015)\citenamefont {Aasi} \emph
  {et~al.}}]{TheLIGOScientific:2014jea}%
  \BibitemOpen
  \bibfield  {author} {\bibinfo {author} {\bibfnamefont {J.}~\bibnamefont
  {Aasi}} \emph {et~al.} (\bibinfo {collaboration} {LIGO Scientific}),\ }\href
  {\doibase 10.1088/0264-9381/32/7/074001} {\bibfield  {journal} {\bibinfo
  {journal} {Class. Quant. Grav.}\ }\textbf {\bibinfo {volume} {32}},\ \bibinfo
  {pages} {074001} (\bibinfo {year} {2015})},\ \Eprint
  {http://arxiv.org/abs/1411.4547} {arXiv:1411.4547 [gr-qc]} \BibitemShut
  {NoStop}%
\bibitem [{\citenamefont {Acernese}\ \emph {et~al.}(2015)\citenamefont
  {Acernese} \emph {et~al.}}]{TheVirgo:2014hva}%
  \BibitemOpen
  \bibfield  {author} {\bibinfo {author} {\bibfnamefont {F.}~\bibnamefont
  {Acernese}} \emph {et~al.} (\bibinfo {collaboration} {VIRGO}),\ }\href
  {\doibase 10.1088/0264-9381/32/2/024001} {\bibfield  {journal} {\bibinfo
  {journal} {Class. Quant. Grav.}\ }\textbf {\bibinfo {volume} {32}},\ \bibinfo
  {pages} {024001} (\bibinfo {year} {2015})},\ \Eprint
  {http://arxiv.org/abs/1408.3978} {arXiv:1408.3978 [gr-qc]} \BibitemShut
  {NoStop}%
\bibitem [{\citenamefont {Abbott}\ \emph
  {et~al.}(2017{\natexlab{a}})\citenamefont {Abbott} \emph
  {et~al.}}]{TheLIGOScientific:2017qsa}%
  \BibitemOpen
  \bibfield  {author} {\bibinfo {author} {\bibfnamefont {B.~P.}\ \bibnamefont
  {Abbott}} \emph {et~al.} (\bibinfo {collaboration} {Virgo, LIGO
  Scientific}),\ }\href {\doibase 10.1103/PhysRevLett.119.161101} {\bibfield
  {journal} {\bibinfo  {journal} {Phys. Rev. Lett.}\ }\textbf {\bibinfo
  {volume} {119}},\ \bibinfo {pages} {161101} (\bibinfo {year}
  {2017}{\natexlab{a}})},\ \Eprint {http://arxiv.org/abs/1710.05832}
  {arXiv:1710.05832 [gr-qc]} \BibitemShut {NoStop}%
\bibitem [{GBM(2017)}]{GBM:2017lvd}%
  \BibitemOpen
  \href {\doibase 10.3847/2041-8213/aa91c9} {\bibfield  {journal} {\bibinfo
  {journal} {Astrophys. J.}\ }\textbf {\bibinfo {volume} {848}},\ \bibinfo
  {pages} {L12} (\bibinfo {year} {2017})},\ \Eprint
  {http://arxiv.org/abs/1710.05833} {arXiv:1710.05833 [astro-ph.HE]}
  \BibitemShut {NoStop}%
\bibitem [{\citenamefont {Abbott}\ \emph
  {et~al.}(2017{\natexlab{b}})\citenamefont {Abbott} \emph
  {et~al.}}]{Monitor:2017mdv}%
  \BibitemOpen
  \bibfield  {author} {\bibinfo {author} {\bibfnamefont {B.~P.}\ \bibnamefont
  {Abbott}} \emph {et~al.} (\bibinfo {collaboration} {Virgo, Fermi-GBM,
  INTEGRAL, LIGO Scientific}),\ }\href {\doibase 10.3847/2041-8213/aa920c}
  {\bibfield  {journal} {\bibinfo  {journal} {Astrophys. J.}\ }\textbf
  {\bibinfo {volume} {848}},\ \bibinfo {pages} {L13} (\bibinfo {year}
  {2017}{\natexlab{b}})},\ \Eprint {http://arxiv.org/abs/1710.05834}
  {arXiv:1710.05834 [astro-ph.HE]} \BibitemShut {NoStop}%
\bibitem [{\citenamefont {Read}\ \emph
  {et~al.}(2009{\natexlab{a}})\citenamefont {Read}, \citenamefont {Markakis},
  \citenamefont {Shibata}, \citenamefont {Uryu}, \citenamefont {Creighton}
  \emph {et~al.}}]{Read:2009yp}%
  \BibitemOpen
  \bibfield  {author} {\bibinfo {author} {\bibfnamefont {J.~S.}\ \bibnamefont
  {Read}}, \bibinfo {author} {\bibfnamefont {C.}~\bibnamefont {Markakis}},
  \bibinfo {author} {\bibfnamefont {M.}~\bibnamefont {Shibata}}, \bibinfo
  {author} {\bibfnamefont {K.}~\bibnamefont {Uryu}}, \bibinfo {author}
  {\bibfnamefont {J.~D.}\ \bibnamefont {Creighton}},  \emph {et~al.},\ }\href
  {\doibase 10.1103/PhysRevD.79.124033} {\bibfield  {journal} {\bibinfo
  {journal} {Phys.Rev.}\ }\textbf {\bibinfo {volume} {D79}},\ \bibinfo {pages}
  {124033} (\bibinfo {year} {2009}{\natexlab{a}})},\ \Eprint
  {http://arxiv.org/abs/0901.3258} {arXiv:0901.3258 [gr-qc]} \BibitemShut
  {NoStop}%
\bibitem [{\citenamefont {Markakis}\ \emph {et~al.}(2010)\citenamefont
  {Markakis}, \citenamefont {Read}, \citenamefont {Shibata}, \citenamefont
  {Uryu}, \citenamefont {Creighton},\ and\ \citenamefont
  {Friedman}}]{Markakis:2010mp}%
  \BibitemOpen
  \bibfield  {author} {\bibinfo {author} {\bibfnamefont {C.}~\bibnamefont
  {Markakis}}, \bibinfo {author} {\bibfnamefont {J.~S.}\ \bibnamefont {Read}},
  \bibinfo {author} {\bibfnamefont {M.}~\bibnamefont {Shibata}}, \bibinfo
  {author} {\bibfnamefont {K.}~\bibnamefont {Uryu}}, \bibinfo {author}
  {\bibfnamefont {J.~D.}\ \bibnamefont {Creighton}}, \ and\ \bibinfo {author}
  {\bibfnamefont {J.~L.}\ \bibnamefont {Friedman}},\ }in\ \href {\doibase
  10.1142/9789814374552\_0046} {\emph {\bibinfo {booktitle} {{12th Marcel
  Grossmann Meeting on General Relativity}}}}\ (\bibinfo {year} {2010})\ pp.\
  \bibinfo {pages} {743--745},\ \Eprint {http://arxiv.org/abs/1008.1822}
  {arXiv:1008.1822 [gr-qc]} \BibitemShut {NoStop}%
\bibitem [{\citenamefont {Markakis}\ \emph {et~al.}(2009)\citenamefont
  {Markakis}, \citenamefont {Read}, \citenamefont {Shibata}, \citenamefont
  {Uryu}, \citenamefont {Creighton}, \citenamefont {Friedman},\ and\
  \citenamefont {Lackey}}]{Markakis:2011vd}%
  \BibitemOpen
  \bibfield  {author} {\bibinfo {author} {\bibfnamefont {C.}~\bibnamefont
  {Markakis}}, \bibinfo {author} {\bibfnamefont {J.~S.}\ \bibnamefont {Read}},
  \bibinfo {author} {\bibfnamefont {M.}~\bibnamefont {Shibata}}, \bibinfo
  {author} {\bibfnamefont {K.}~\bibnamefont {Uryu}}, \bibinfo {author}
  {\bibfnamefont {J.~D.}\ \bibnamefont {Creighton}}, \bibinfo {author}
  {\bibfnamefont {J.~L.}\ \bibnamefont {Friedman}}, \ and\ \bibinfo {author}
  {\bibfnamefont {B.~D.}\ \bibnamefont {Lackey}},\ }\href {\doibase
  10.1088/1742-6596/189/1/012024} {\bibfield  {journal} {\bibinfo  {journal}
  {J. Phys. Conf. Ser.}\ }\textbf {\bibinfo {volume} {189}},\ \bibinfo {pages}
  {012024} (\bibinfo {year} {2009})},\ \Eprint {http://arxiv.org/abs/1110.3759}
  {arXiv:1110.3759 [gr-qc]} \BibitemShut {NoStop}%
\bibitem [{\citenamefont {Read}\ \emph {et~al.}(2013)\citenamefont {Read},
  \citenamefont {Baiotti}, \citenamefont {Creighton}, \citenamefont {Friedman},
  \citenamefont {Giacomazzo} \emph {et~al.}}]{Read:2013zra}%
  \BibitemOpen
  \bibfield  {author} {\bibinfo {author} {\bibfnamefont {J.~S.}\ \bibnamefont
  {Read}}, \bibinfo {author} {\bibfnamefont {L.}~\bibnamefont {Baiotti}},
  \bibinfo {author} {\bibfnamefont {J.~D.~E.}\ \bibnamefont {Creighton}},
  \bibinfo {author} {\bibfnamefont {J.~L.}\ \bibnamefont {Friedman}}, \bibinfo
  {author} {\bibfnamefont {B.}~\bibnamefont {Giacomazzo}},  \emph {et~al.},\
  }\href {\doibase 10.1103/PhysRevD.88.044042} {\bibfield  {journal} {\bibinfo
  {journal} {Phys.Rev.}\ }\textbf {\bibinfo {volume} {D88}},\ \bibinfo {pages}
  {044042} (\bibinfo {year} {2013})},\ \Eprint {http://arxiv.org/abs/1306.4065}
  {arXiv:1306.4065 [gr-qc]} \BibitemShut {NoStop}%
\bibitem [{\citenamefont {Del~Pozzo}\ \emph {et~al.}(2013)\citenamefont
  {Del~Pozzo}, \citenamefont {Li}, \citenamefont {Agathos}, \citenamefont {Van
  Den~Broeck},\ and\ \citenamefont {Vitale}}]{DelPozzo:2013ala}%
  \BibitemOpen
  \bibfield  {author} {\bibinfo {author} {\bibfnamefont {W.}~\bibnamefont
  {Del~Pozzo}}, \bibinfo {author} {\bibfnamefont {T.~G.~F.}\ \bibnamefont
  {Li}}, \bibinfo {author} {\bibfnamefont {M.}~\bibnamefont {Agathos}},
  \bibinfo {author} {\bibfnamefont {C.}~\bibnamefont {Van Den~Broeck}}, \ and\
  \bibinfo {author} {\bibfnamefont {S.}~\bibnamefont {Vitale}},\ }\href
  {\doibase 10.1103/PhysRevLett.111.071101} {\bibfield  {journal} {\bibinfo
  {journal} {Phys. Rev. Lett.}\ }\textbf {\bibinfo {volume} {111}},\ \bibinfo
  {pages} {071101} (\bibinfo {year} {2013})},\ \Eprint
  {http://arxiv.org/abs/1307.8338} {arXiv:1307.8338 [gr-qc]} \BibitemShut
  {NoStop}%
\bibitem [{\citenamefont {Lackey}\ and\ \citenamefont
  {Wade}(2015)}]{Lackey:2014fwa}%
  \BibitemOpen
  \bibfield  {author} {\bibinfo {author} {\bibfnamefont {B.~D.}\ \bibnamefont
  {Lackey}}\ and\ \bibinfo {author} {\bibfnamefont {L.}~\bibnamefont {Wade}},\
  }\href {\doibase 10.1103/PhysRevD.91.043002} {\bibfield  {journal} {\bibinfo
  {journal} {Phys.Rev.}\ }\textbf {\bibinfo {volume} {D91}},\ \bibinfo {pages}
  {043002} (\bibinfo {year} {2015})},\ \Eprint {http://arxiv.org/abs/1410.8866}
  {arXiv:1410.8866 [gr-qc]} \BibitemShut {NoStop}%
\bibitem [{\citenamefont {Agathos}\ \emph {et~al.}(2015)\citenamefont
  {Agathos}, \citenamefont {Meidam}, \citenamefont {Del~Pozzo}, \citenamefont
  {Li}, \citenamefont {Tompitak}, \citenamefont {Veitch}, \citenamefont
  {Vitale},\ and\ \citenamefont {Broeck}}]{Agathos:2015uaa}%
  \BibitemOpen
  \bibfield  {author} {\bibinfo {author} {\bibfnamefont {M.}~\bibnamefont
  {Agathos}}, \bibinfo {author} {\bibfnamefont {J.}~\bibnamefont {Meidam}},
  \bibinfo {author} {\bibfnamefont {W.}~\bibnamefont {Del~Pozzo}}, \bibinfo
  {author} {\bibfnamefont {T.~G.~F.}\ \bibnamefont {Li}}, \bibinfo {author}
  {\bibfnamefont {M.}~\bibnamefont {Tompitak}}, \bibinfo {author}
  {\bibfnamefont {J.}~\bibnamefont {Veitch}}, \bibinfo {author} {\bibfnamefont
  {S.}~\bibnamefont {Vitale}}, \ and\ \bibinfo {author} {\bibfnamefont
  {C.~V.~D.}\ \bibnamefont {Broeck}},\ }\href {\doibase
  10.1103/PhysRevD.92.023012} {\bibfield  {journal} {\bibinfo  {journal} {Phys.
  Rev.}\ }\textbf {\bibinfo {volume} {D92}},\ \bibinfo {pages} {023012}
  (\bibinfo {year} {2015})},\ \Eprint {http://arxiv.org/abs/1503.05405}
  {arXiv:1503.05405 [gr-qc]} \BibitemShut {NoStop}%
\bibitem [{\citenamefont {Annala}\ \emph {et~al.}(2018)\citenamefont {Annala},
  \citenamefont {Gorda}, \citenamefont {Kurkela},\ and\ \citenamefont
  {Vuorinen}}]{Annala:2017llu}%
  \BibitemOpen
  \bibfield  {author} {\bibinfo {author} {\bibfnamefont {E.}~\bibnamefont
  {Annala}}, \bibinfo {author} {\bibfnamefont {T.}~\bibnamefont {Gorda}},
  \bibinfo {author} {\bibfnamefont {A.}~\bibnamefont {Kurkela}}, \ and\
  \bibinfo {author} {\bibfnamefont {A.}~\bibnamefont {Vuorinen}},\ }\href
  {\doibase 10.1103/PhysRevLett.120.172703} {\bibfield  {journal} {\bibinfo
  {journal} {Phys. Rev. Lett.}\ }\textbf {\bibinfo {volume} {120}},\ \bibinfo
  {pages} {172703} (\bibinfo {year} {2018})},\ \Eprint
  {http://arxiv.org/abs/1711.02644} {arXiv:1711.02644 [astro-ph.HE]}
  \BibitemShut {NoStop}%
\bibitem [{\citenamefont {Fattoyev}\ \emph {et~al.}(2018)\citenamefont
  {Fattoyev}, \citenamefont {Piekarewicz},\ and\ \citenamefont
  {Horowitz}}]{Fattoyev:2017jql}%
  \BibitemOpen
  \bibfield  {author} {\bibinfo {author} {\bibfnamefont {F.~J.}\ \bibnamefont
  {Fattoyev}}, \bibinfo {author} {\bibfnamefont {J.}~\bibnamefont
  {Piekarewicz}}, \ and\ \bibinfo {author} {\bibfnamefont {C.~J.}\ \bibnamefont
  {Horowitz}},\ }\href {\doibase 10.1103/PhysRevLett.120.172702} {\bibfield
  {journal} {\bibinfo  {journal} {Phys. Rev. Lett.}\ }\textbf {\bibinfo
  {volume} {120}},\ \bibinfo {pages} {172702} (\bibinfo {year} {2018})},\
  \Eprint {http://arxiv.org/abs/1711.06615} {arXiv:1711.06615 [nucl-th]}
  \BibitemShut {NoStop}%
\bibitem [{\citenamefont {Most}\ \emph {et~al.}(2018)\citenamefont {Most},
  \citenamefont {Weih}, \citenamefont {Rezzolla},\ and\ \citenamefont
  {Schaffner-Bielich}}]{Most:2018hfd}%
  \BibitemOpen
  \bibfield  {author} {\bibinfo {author} {\bibfnamefont {E.~R.}\ \bibnamefont
  {Most}}, \bibinfo {author} {\bibfnamefont {L.~R.}\ \bibnamefont {Weih}},
  \bibinfo {author} {\bibfnamefont {L.}~\bibnamefont {Rezzolla}}, \ and\
  \bibinfo {author} {\bibfnamefont {J.}~\bibnamefont {Schaffner-Bielich}},\
  }\href {\doibase 10.1103/PhysRevLett.120.261103} {\bibfield  {journal}
  {\bibinfo  {journal} {Phys. Rev. Lett.}\ }\textbf {\bibinfo {volume} {120}},\
  \bibinfo {pages} {261103} (\bibinfo {year} {2018})},\ \Eprint
  {http://arxiv.org/abs/1803.00549} {arXiv:1803.00549 [gr-qc]} \BibitemShut
  {NoStop}%
\bibitem [{\citenamefont {Lim}\ and\ \citenamefont {Holt}(2018)}]{Lim:2018bkq}%
  \BibitemOpen
  \bibfield  {author} {\bibinfo {author} {\bibfnamefont {Y.}~\bibnamefont
  {Lim}}\ and\ \bibinfo {author} {\bibfnamefont {J.~W.}\ \bibnamefont {Holt}},\
  }\href {\doibase 10.1103/PhysRevLett.121.062701} {\bibfield  {journal}
  {\bibinfo  {journal} {Phys. Rev. Lett.}\ }\textbf {\bibinfo {volume} {121}},\
  \bibinfo {pages} {062701} (\bibinfo {year} {2018})},\ \Eprint
  {http://arxiv.org/abs/1803.02803} {arXiv:1803.02803 [nucl-th]} \BibitemShut
  {NoStop}%
\bibitem [{\citenamefont {Tews}\ \emph
  {et~al.}(2018{\natexlab{a}})\citenamefont {Tews}, \citenamefont {Margueron},\
  and\ \citenamefont {Reddy}}]{Tews:2018chv}%
  \BibitemOpen
  \bibfield  {author} {\bibinfo {author} {\bibfnamefont {I.}~\bibnamefont
  {Tews}}, \bibinfo {author} {\bibfnamefont {J.}~\bibnamefont {Margueron}}, \
  and\ \bibinfo {author} {\bibfnamefont {S.}~\bibnamefont {Reddy}},\ }\href
  {\doibase 10.1103/PhysRevC.98.045804} {\bibfield  {journal} {\bibinfo
  {journal} {Phys. Rev.}\ }\textbf {\bibinfo {volume} {C98}},\ \bibinfo {pages}
  {045804} (\bibinfo {year} {2018}{\natexlab{a}})},\ \Eprint
  {http://arxiv.org/abs/1804.02783} {arXiv:1804.02783 [nucl-th]} \BibitemShut
  {NoStop}%
\bibitem [{\citenamefont {Greif}\ \emph {et~al.}(2019)\citenamefont {Greif},
  \citenamefont {Raaijmakers}, \citenamefont {Hebeler}, \citenamefont
  {Schwenk},\ and\ \citenamefont {Watts}}]{Greif:2018njt}%
  \BibitemOpen
  \bibfield  {author} {\bibinfo {author} {\bibfnamefont {S.}~\bibnamefont
  {Greif}}, \bibinfo {author} {\bibfnamefont {G.}~\bibnamefont {Raaijmakers}},
  \bibinfo {author} {\bibfnamefont {K.}~\bibnamefont {Hebeler}}, \bibinfo
  {author} {\bibfnamefont {A.}~\bibnamefont {Schwenk}}, \ and\ \bibinfo
  {author} {\bibfnamefont {A.}~\bibnamefont {Watts}},\ }\href {\doibase
  10.1093/mnras/stz654} {\bibfield  {journal} {\bibinfo  {journal} {Mon. Not.
  Roy. Astron. Soc.}\ }\textbf {\bibinfo {volume} {485}},\ \bibinfo {pages}
  {5363} (\bibinfo {year} {2019})},\ \Eprint {http://arxiv.org/abs/1812.08188}
  {arXiv:1812.08188 [astro-ph.HE]} \BibitemShut {NoStop}%
\bibitem [{\citenamefont {Raaijmakers}\ \emph {et~al.}(2020)\citenamefont
  {Raaijmakers} \emph {et~al.}}]{Raaijmakers:2019dks}%
  \BibitemOpen
  \bibfield  {author} {\bibinfo {author} {\bibfnamefont {G.}~\bibnamefont
  {Raaijmakers}} \emph {et~al.},\ }\href {\doibase 10.3847/2041-8213/ab822f}
  {\bibfield  {journal} {\bibinfo  {journal} {Astrophys. J. Lett.}\ }\textbf
  {\bibinfo {volume} {893}},\ \bibinfo {pages} {L21} (\bibinfo {year}
  {2020})},\ \Eprint {http://arxiv.org/abs/1912.11031} {arXiv:1912.11031
  [astro-ph.HE]} \BibitemShut {NoStop}%
\bibitem [{\citenamefont {Essick}\ \emph
  {et~al.}(2020{\natexlab{a}})\citenamefont {Essick}, \citenamefont {Landry},\
  and\ \citenamefont {Holz}}]{PhysRevD.101.063007}%
  \BibitemOpen
  \bibfield  {author} {\bibinfo {author} {\bibfnamefont {R.}~\bibnamefont
  {Essick}}, \bibinfo {author} {\bibfnamefont {P.}~\bibnamefont {Landry}}, \
  and\ \bibinfo {author} {\bibfnamefont {D.~E.}\ \bibnamefont {Holz}},\ }\href
  {\doibase 10.1103/PhysRevD.101.063007} {\bibfield  {journal} {\bibinfo
  {journal} {Phys. Rev. D}\ }\textbf {\bibinfo {volume} {101}},\ \bibinfo
  {pages} {063007} (\bibinfo {year} {2020}{\natexlab{a}})}\BibitemShut
  {NoStop}%
\bibitem [{\citenamefont {Landry}\ \emph {et~al.}(2020)\citenamefont {Landry},
  \citenamefont {Essick},\ and\ \citenamefont
  {Chatziioannou}}]{Landry:2020vaw}%
  \BibitemOpen
  \bibfield  {author} {\bibinfo {author} {\bibfnamefont {P.}~\bibnamefont
  {Landry}}, \bibinfo {author} {\bibfnamefont {R.}~\bibnamefont {Essick}}, \
  and\ \bibinfo {author} {\bibfnamefont {K.}~\bibnamefont {Chatziioannou}},\
  }\href {\doibase 10.1103/PhysRevD.101.123007} {\bibfield  {journal} {\bibinfo
   {journal} {Phys. Rev. D}\ }\textbf {\bibinfo {volume} {101}},\ \bibinfo
  {pages} {123007} (\bibinfo {year} {2020})},\ \Eprint
  {http://arxiv.org/abs/2003.04880} {arXiv:2003.04880 [astro-ph.HE]}
  \BibitemShut {NoStop}%
\bibitem [{\citenamefont {Chatziioannou}\ and\ \citenamefont
  {Han}(2020)}]{Chatziioannou:2019yko}%
  \BibitemOpen
  \bibfield  {author} {\bibinfo {author} {\bibfnamefont {K.}~\bibnamefont
  {Chatziioannou}}\ and\ \bibinfo {author} {\bibfnamefont {S.}~\bibnamefont
  {Han}},\ }\href {\doibase 10.1103/PhysRevD.101.044019} {\bibfield  {journal}
  {\bibinfo  {journal} {Phys. Rev. D}\ }\textbf {\bibinfo {volume} {101}},\
  \bibinfo {pages} {044019} (\bibinfo {year} {2020})},\ \Eprint
  {http://arxiv.org/abs/1911.07091} {arXiv:1911.07091 [gr-qc]} \BibitemShut
  {NoStop}%
\bibitem [{\citenamefont {Essick}\ \emph
  {et~al.}(2020{\natexlab{b}})\citenamefont {Essick}, \citenamefont {Tews},
  \citenamefont {Landry}, \citenamefont {Reddy},\ and\ \citenamefont
  {Holz}}]{Essick:2020flb}%
  \BibitemOpen
  \bibfield  {author} {\bibinfo {author} {\bibfnamefont {R.}~\bibnamefont
  {Essick}}, \bibinfo {author} {\bibfnamefont {I.}~\bibnamefont {Tews}},
  \bibinfo {author} {\bibfnamefont {P.}~\bibnamefont {Landry}}, \bibinfo
  {author} {\bibfnamefont {S.}~\bibnamefont {Reddy}}, \ and\ \bibinfo {author}
  {\bibfnamefont {D.~E.}\ \bibnamefont {Holz}},\ }\href@noop {} {\  (\bibinfo
  {year} {2020}{\natexlab{b}})},\ \Eprint {http://arxiv.org/abs/2004.07744}
  {arXiv:2004.07744 [astro-ph.HE]} \BibitemShut {NoStop}%
\bibitem [{\citenamefont {Wysocki}\ \emph {et~al.}(2020)\citenamefont
  {Wysocki}, \citenamefont {O'Shaughnessy}, \citenamefont {Wade},\ and\
  \citenamefont {Lange}}]{wysocki2020inferring}%
  \BibitemOpen
  \bibfield  {author} {\bibinfo {author} {\bibfnamefont {D.}~\bibnamefont
  {Wysocki}}, \bibinfo {author} {\bibfnamefont {R.}~\bibnamefont
  {O'Shaughnessy}}, \bibinfo {author} {\bibfnamefont {L.}~\bibnamefont {Wade}},
  \ and\ \bibinfo {author} {\bibfnamefont {J.}~\bibnamefont {Lange}},\
  }\href@noop {} {\enquote {\bibinfo {title} {Inferring the neutron star
  equation of state simultaneously with the population of merging neutron
  stars},}\ } (\bibinfo {year} {2020}),\ \Eprint
  {http://arxiv.org/abs/2001.01747} {arXiv:2001.01747 [gr-qc]} \BibitemShut
  {NoStop}%
\bibitem [{\citenamefont {Dietrich}\ \emph {et~al.}(2020)\citenamefont
  {Dietrich}, \citenamefont {Coughlin}, \citenamefont {Pang}, \citenamefont
  {Bulla}, \citenamefont {Heinzel}, \citenamefont {Issa}, \citenamefont
  {Tews},\ and\ \citenamefont {Antier}}]{Dietrich:2020lps}%
  \BibitemOpen
  \bibfield  {author} {\bibinfo {author} {\bibfnamefont {T.}~\bibnamefont
  {Dietrich}}, \bibinfo {author} {\bibfnamefont {M.~W.}\ \bibnamefont
  {Coughlin}}, \bibinfo {author} {\bibfnamefont {P.~T.}\ \bibnamefont {Pang}},
  \bibinfo {author} {\bibfnamefont {M.}~\bibnamefont {Bulla}}, \bibinfo
  {author} {\bibfnamefont {J.}~\bibnamefont {Heinzel}}, \bibinfo {author}
  {\bibfnamefont {L.}~\bibnamefont {Issa}}, \bibinfo {author} {\bibfnamefont
  {I.}~\bibnamefont {Tews}}, \ and\ \bibinfo {author} {\bibfnamefont
  {S.}~\bibnamefont {Antier}},\ }\href@noop {} {\  (\bibinfo {year} {2020})},\
  \Eprint {http://arxiv.org/abs/2002.11355} {arXiv:2002.11355 [astro-ph.HE]}
  \BibitemShut {NoStop}%
\bibitem [{\citenamefont {Chatziioannou}(2020)}]{Chatziioannou:2020pqz}%
  \BibitemOpen
  \bibfield  {author} {\bibinfo {author} {\bibfnamefont {K.}~\bibnamefont
  {Chatziioannou}},\ }\href@noop {} {\  (\bibinfo {year} {2020})},\ \Eprint
  {http://arxiv.org/abs/2006.03168} {arXiv:2006.03168 [gr-qc]} \BibitemShut
  {NoStop}%
\bibitem [{\citenamefont {Dai}\ \emph {et~al.}(2018)\citenamefont {Dai},
  \citenamefont {Venumadhav},\ and\ \citenamefont {Zackay}}]{Dai:2018dca}%
  \BibitemOpen
  \bibfield  {author} {\bibinfo {author} {\bibfnamefont {L.}~\bibnamefont
  {Dai}}, \bibinfo {author} {\bibfnamefont {T.}~\bibnamefont {Venumadhav}}, \
  and\ \bibinfo {author} {\bibfnamefont {B.}~\bibnamefont {Zackay}},\
  }\href@noop {} {\bibfield  {journal} {\bibinfo  {journal} {arXiv:
  1806.08793}\ } (\bibinfo {year} {2018})}\BibitemShut {NoStop}%
\bibitem [{\citenamefont {De}\ \emph {et~al.}(2018)\citenamefont {De},
  \citenamefont {Finstad}, \citenamefont {Lattimer}, \citenamefont {Brown},
  \citenamefont {Berger},\ and\ \citenamefont {Biwer}}]{De:2018uhw}%
  \BibitemOpen
  \bibfield  {author} {\bibinfo {author} {\bibfnamefont {S.}~\bibnamefont
  {De}}, \bibinfo {author} {\bibfnamefont {D.}~\bibnamefont {Finstad}},
  \bibinfo {author} {\bibfnamefont {J.~M.}\ \bibnamefont {Lattimer}}, \bibinfo
  {author} {\bibfnamefont {D.~A.}\ \bibnamefont {Brown}}, \bibinfo {author}
  {\bibfnamefont {E.}~\bibnamefont {Berger}}, \ and\ \bibinfo {author}
  {\bibfnamefont {C.~M.}\ \bibnamefont {Biwer}},\ }\href {\doibase
  10.1103/PhysRevLett.121.091102} {\bibfield  {journal} {\bibinfo  {journal}
  {Phys. Rev. Lett.}\ }\textbf {\bibinfo {volume} {121}},\ \bibinfo {pages}
  {091102} (\bibinfo {year} {2018})},\ \Eprint
  {http://arxiv.org/abs/1804.08583} {arXiv:1804.08583 [astro-ph.HE]}
  \BibitemShut {NoStop}%
\bibitem [{\citenamefont {Abbott}\ \emph
  {et~al.}(2019{\natexlab{a}})\citenamefont {Abbott} \emph
  {et~al.}}]{Abbott:2018wiz}%
  \BibitemOpen
  \bibfield  {author} {\bibinfo {author} {\bibfnamefont {B.}~\bibnamefont
  {Abbott}} \emph {et~al.} (\bibinfo {collaboration} {LIGO Scientific,
  Virgo}),\ }\href {\doibase 10.1103/PhysRevX.9.011001} {\bibfield  {journal}
  {\bibinfo  {journal} {Phys. Rev. X}\ }\textbf {\bibinfo {volume} {9}},\
  \bibinfo {pages} {011001} (\bibinfo {year} {2019}{\natexlab{a}})},\ \Eprint
  {http://arxiv.org/abs/1805.11579} {arXiv:1805.11579 [gr-qc]} \BibitemShut
  {NoStop}%
\bibitem [{\citenamefont {Abbott}\ \emph {et~al.}(2018)\citenamefont {Abbott}
  \emph {et~al.}}]{Abbott:2018exr}%
  \BibitemOpen
  \bibfield  {author} {\bibinfo {author} {\bibfnamefont {B.~P.}\ \bibnamefont
  {Abbott}} \emph {et~al.} (\bibinfo {collaboration} {Virgo, LIGO
  Scientific}),\ }\href {\doibase 10.1103/PhysRevLett.121.161101} {\bibfield
  {journal} {\bibinfo  {journal} {Phys. Rev. Lett.}\ }\textbf {\bibinfo
  {volume} {121}},\ \bibinfo {pages} {161101} (\bibinfo {year} {2018})},\
  \Eprint {http://arxiv.org/abs/1805.11581} {arXiv:1805.11581 [gr-qc]}
  \BibitemShut {NoStop}%
\bibitem [{\citenamefont {Abbott}\ \emph
  {et~al.}(2019{\natexlab{b}})\citenamefont {Abbott} \emph
  {et~al.}}]{LIGOScientific:2018mvr}%
  \BibitemOpen
  \bibfield  {author} {\bibinfo {author} {\bibfnamefont {B.}~\bibnamefont
  {Abbott}} \emph {et~al.} (\bibinfo {collaboration} {LIGO Scientific,
  Virgo}),\ }\href {\doibase 10.1103/PhysRevX.9.031040} {\bibfield  {journal}
  {\bibinfo  {journal} {Phys. Rev. X}\ }\textbf {\bibinfo {volume} {9}},\
  \bibinfo {pages} {031040} (\bibinfo {year} {2019}{\natexlab{b}})},\ \Eprint
  {http://arxiv.org/abs/1811.12907} {arXiv:1811.12907 [astro-ph.HE]}
  \BibitemShut {NoStop}%
\bibitem [{\citenamefont {Radice}\ \emph {et~al.}(2018)\citenamefont {Radice},
  \citenamefont {Perego}, \citenamefont {Zappa},\ and\ \citenamefont
  {Bernuzzi}}]{Radice:2017lry}%
  \BibitemOpen
  \bibfield  {author} {\bibinfo {author} {\bibfnamefont {D.}~\bibnamefont
  {Radice}}, \bibinfo {author} {\bibfnamefont {A.}~\bibnamefont {Perego}},
  \bibinfo {author} {\bibfnamefont {F.}~\bibnamefont {Zappa}}, \ and\ \bibinfo
  {author} {\bibfnamefont {S.}~\bibnamefont {Bernuzzi}},\ }\href {\doibase
  10.3847/2041-8213/aaa402} {\bibfield  {journal} {\bibinfo  {journal}
  {Astrophys. J.}\ }\textbf {\bibinfo {volume} {852}},\ \bibinfo {pages} {L29}
  (\bibinfo {year} {2018})},\ \Eprint {http://arxiv.org/abs/1711.03647}
  {arXiv:1711.03647 [astro-ph.HE]} \BibitemShut {NoStop}%
\bibitem [{\citenamefont {Bauswein}\ \emph {et~al.}(2017)\citenamefont
  {Bauswein}, \citenamefont {Just}, \citenamefont {Janka},\ and\ \citenamefont
  {Stergioulas}}]{Bauswein:2017vtn}%
  \BibitemOpen
  \bibfield  {author} {\bibinfo {author} {\bibfnamefont {A.}~\bibnamefont
  {Bauswein}}, \bibinfo {author} {\bibfnamefont {O.}~\bibnamefont {Just}},
  \bibinfo {author} {\bibfnamefont {H.-T.}\ \bibnamefont {Janka}}, \ and\
  \bibinfo {author} {\bibfnamefont {N.}~\bibnamefont {Stergioulas}},\ }\href
  {\doibase 10.3847/2041-8213/aa9994} {\bibfield  {journal} {\bibinfo
  {journal} {Astrophys. J.}\ }\textbf {\bibinfo {volume} {850}},\ \bibinfo
  {pages} {L34} (\bibinfo {year} {2017})},\ \Eprint
  {http://arxiv.org/abs/1710.06843} {arXiv:1710.06843 [astro-ph.HE]}
  \BibitemShut {NoStop}%
\bibitem [{\citenamefont {Margalit}\ and\ \citenamefont
  {Metzger}(2017)}]{Margalit:2017dij}%
  \BibitemOpen
  \bibfield  {author} {\bibinfo {author} {\bibfnamefont {B.}~\bibnamefont
  {Margalit}}\ and\ \bibinfo {author} {\bibfnamefont {B.}~\bibnamefont
  {Metzger}},\ }\href@noop {} {\  (\bibinfo {year} {2017})},\ \Eprint
  {http://arxiv.org/abs/1710.05938} {arXiv:1710.05938 [astro-ph.HE]}
  \BibitemShut {NoStop}%
\bibitem [{\citenamefont {Coughlin}\ \emph
  {et~al.}(2018{\natexlab{a}})\citenamefont {Coughlin}, \citenamefont
  {Dietrich}, \citenamefont {Doctor}, \citenamefont {Kasen}, \citenamefont
  {Coughlin}, \citenamefont {Jerkstrand}, \citenamefont {Leloudas},
  \citenamefont {McBrien}, \citenamefont {Metzger}, \citenamefont
  {O’Shaughnessy},\ and\ \citenamefont {Smartt}}]{Coughlin:2018miv}%
  \BibitemOpen
  \bibfield  {author} {\bibinfo {author} {\bibfnamefont {M.~W.}\ \bibnamefont
  {Coughlin}}, \bibinfo {author} {\bibfnamefont {T.}~\bibnamefont {Dietrich}},
  \bibinfo {author} {\bibfnamefont {Z.}~\bibnamefont {Doctor}}, \bibinfo
  {author} {\bibfnamefont {D.}~\bibnamefont {Kasen}}, \bibinfo {author}
  {\bibfnamefont {S.}~\bibnamefont {Coughlin}}, \bibinfo {author}
  {\bibfnamefont {A.}~\bibnamefont {Jerkstrand}}, \bibinfo {author}
  {\bibfnamefont {G.}~\bibnamefont {Leloudas}}, \bibinfo {author}
  {\bibfnamefont {O.}~\bibnamefont {McBrien}}, \bibinfo {author} {\bibfnamefont
  {B.~D.}\ \bibnamefont {Metzger}}, \bibinfo {author} {\bibfnamefont
  {R.}~\bibnamefont {O’Shaughnessy}}, \ and\ \bibinfo {author} {\bibfnamefont
  {S.~J.}\ \bibnamefont {Smartt}},\ }\href {\doibase 10.1093/mnras/sty2174}
  {\bibfield  {journal} {\bibinfo  {journal} {Monthly Notices of the Royal
  Astronomical Society}\ }\textbf {\bibinfo {volume} {480}},\ \bibinfo {pages}
  {3871} (\bibinfo {year} {2018}{\natexlab{a}})},\ \Eprint
  {http://arxiv.org/abs/1805.09371} {arXiv:1805.09371} \BibitemShut {NoStop}%
\bibitem [{\citenamefont {Radice}\ and\ \citenamefont
  {Dai}(2018)}]{Radice:2018ozg}%
  \BibitemOpen
  \bibfield  {author} {\bibinfo {author} {\bibfnamefont {D.}~\bibnamefont
  {Radice}}\ and\ \bibinfo {author} {\bibfnamefont {L.}~\bibnamefont {Dai}},\
  }\href@noop {} {\  (\bibinfo {year} {2018})},\ \Eprint
  {http://arxiv.org/abs/1810.12917} {arXiv:1810.12917 [astro-ph.HE]}
  \BibitemShut {NoStop}%
\bibitem [{\citenamefont {Coughlin}\ \emph
  {et~al.}(2018{\natexlab{b}})\citenamefont {Coughlin}, \citenamefont
  {Dietrich}, \citenamefont {Margalit},\ and\ \citenamefont
  {Metzger}}]{Coughlin:2018fis}%
  \BibitemOpen
  \bibfield  {author} {\bibinfo {author} {\bibfnamefont {M.~W.}\ \bibnamefont
  {Coughlin}}, \bibinfo {author} {\bibfnamefont {T.}~\bibnamefont {Dietrich}},
  \bibinfo {author} {\bibfnamefont {B.}~\bibnamefont {Margalit}}, \ and\
  \bibinfo {author} {\bibfnamefont {B.~D.}\ \bibnamefont {Metzger}},\
  }\href@noop {} {\  (\bibinfo {year} {2018}{\natexlab{b}})},\ \Eprint
  {http://arxiv.org/abs/1812.04803} {arXiv:1812.04803 [astro-ph.HE]}
  \BibitemShut {NoStop}%
\bibitem [{\citenamefont {Capano}\ \emph {et~al.}(2019)\citenamefont {Capano},
  \citenamefont {Tews}, \citenamefont {Brown}, \citenamefont {Margalit},
  \citenamefont {De}, \citenamefont {Kumar}, \citenamefont {Brown},
  \citenamefont {Krishnan},\ and\ \citenamefont {Reddy}}]{Capano:2019eae}%
  \BibitemOpen
  \bibfield  {author} {\bibinfo {author} {\bibfnamefont {C.~D.}\ \bibnamefont
  {Capano}}, \bibinfo {author} {\bibfnamefont {I.}~\bibnamefont {Tews}},
  \bibinfo {author} {\bibfnamefont {S.~M.}\ \bibnamefont {Brown}}, \bibinfo
  {author} {\bibfnamefont {B.}~\bibnamefont {Margalit}}, \bibinfo {author}
  {\bibfnamefont {S.}~\bibnamefont {De}}, \bibinfo {author} {\bibfnamefont
  {S.}~\bibnamefont {Kumar}}, \bibinfo {author} {\bibfnamefont {D.~A.}\
  \bibnamefont {Brown}}, \bibinfo {author} {\bibfnamefont {B.}~\bibnamefont
  {Krishnan}}, \ and\ \bibinfo {author} {\bibfnamefont {S.}~\bibnamefont
  {Reddy}},\ }\href {\doibase 10.1038/s41550-020-1014-6} {\  (\bibinfo {year}
  {2019}),\ 10.1038/s41550-020-1014-6},\ \Eprint
  {http://arxiv.org/abs/1908.10352} {arXiv:1908.10352 [astro-ph.HE]}
  \BibitemShut {NoStop}%
\bibitem [{\citenamefont {Rezzolla}\ \emph {et~al.}(2018)\citenamefont
  {Rezzolla}, \citenamefont {Most},\ and\ \citenamefont
  {Weih}}]{Rezzolla:2017aly}%
  \BibitemOpen
  \bibfield  {author} {\bibinfo {author} {\bibfnamefont {L.}~\bibnamefont
  {Rezzolla}}, \bibinfo {author} {\bibfnamefont {E.~R.}\ \bibnamefont {Most}},
  \ and\ \bibinfo {author} {\bibfnamefont {L.~R.}\ \bibnamefont {Weih}},\
  }\href {\doibase 10.3847/2041-8213/aaa401} {\bibfield  {journal} {\bibinfo
  {journal} {Astrophys. J.}\ }\textbf {\bibinfo {volume} {852}},\ \bibinfo
  {pages} {L25} (\bibinfo {year} {2018})},\ \Eprint
  {http://arxiv.org/abs/1711.00314} {arXiv:1711.00314 [astro-ph.HE]}
  \BibitemShut {NoStop}%
\bibitem [{\citenamefont {Shibata}\ \emph {et~al.}(2017)\citenamefont
  {Shibata}, \citenamefont {Fujibayashi}, \citenamefont {Hotokezaka},
  \citenamefont {Kiuchi}, \citenamefont {Kyutoku}, \citenamefont {Sekiguchi},\
  and\ \citenamefont {Tanaka}}]{Shibata:2017xdx}%
  \BibitemOpen
  \bibfield  {author} {\bibinfo {author} {\bibfnamefont {M.}~\bibnamefont
  {Shibata}}, \bibinfo {author} {\bibfnamefont {S.}~\bibnamefont
  {Fujibayashi}}, \bibinfo {author} {\bibfnamefont {K.}~\bibnamefont
  {Hotokezaka}}, \bibinfo {author} {\bibfnamefont {K.}~\bibnamefont {Kiuchi}},
  \bibinfo {author} {\bibfnamefont {K.}~\bibnamefont {Kyutoku}}, \bibinfo
  {author} {\bibfnamefont {Y.}~\bibnamefont {Sekiguchi}}, \ and\ \bibinfo
  {author} {\bibfnamefont {M.}~\bibnamefont {Tanaka}},\ }\href {\doibase
  10.1103/PhysRevD.96.123012} {\bibfield  {journal} {\bibinfo  {journal} {Phys.
  Rev.}\ }\textbf {\bibinfo {volume} {D96}},\ \bibinfo {pages} {123012}
  (\bibinfo {year} {2017})},\ \Eprint {http://arxiv.org/abs/1710.07579}
  {arXiv:1710.07579 [astro-ph.HE]} \BibitemShut {NoStop}%
\bibitem [{\citenamefont {Glendenning}(1992)}]{Glendenning:1992vb}%
  \BibitemOpen
  \bibfield  {author} {\bibinfo {author} {\bibfnamefont {N.~K.}\ \bibnamefont
  {Glendenning}},\ }\href {\doibase 10.1103/PhysRevD.46.1274} {\bibfield
  {journal} {\bibinfo  {journal} {Phys. Rev. D}\ }\textbf {\bibinfo {volume}
  {46}},\ \bibinfo {pages} {1274} (\bibinfo {year} {1992})}\BibitemShut
  {NoStop}%
\bibitem [{\citenamefont {Alford}(2001)}]{Alford:2001dt}%
  \BibitemOpen
  \bibfield  {author} {\bibinfo {author} {\bibfnamefont {M.~G.}\ \bibnamefont
  {Alford}},\ }\href {\doibase 10.1146/annurev.nucl.51.101701.132449}
  {\bibfield  {journal} {\bibinfo  {journal} {Ann. Rev. Nucl. Part. Sci.}\
  }\textbf {\bibinfo {volume} {51}},\ \bibinfo {pages} {131} (\bibinfo {year}
  {2001})},\ \Eprint {http://arxiv.org/abs/hep-ph/0102047}
  {arXiv:hep-ph/0102047} \BibitemShut {NoStop}%
\bibitem [{\citenamefont {Alford}\ \emph {et~al.}(2007)\citenamefont {Alford},
  \citenamefont {Blaschke}, \citenamefont {Drago}, \citenamefont {Klahn},
  \citenamefont {Pagliara},\ and\ \citenamefont
  {Schaffner-Bielich}}]{Alford:2006vz}%
  \BibitemOpen
  \bibfield  {author} {\bibinfo {author} {\bibfnamefont {M.}~\bibnamefont
  {Alford}}, \bibinfo {author} {\bibfnamefont {D.}~\bibnamefont {Blaschke}},
  \bibinfo {author} {\bibfnamefont {A.}~\bibnamefont {Drago}}, \bibinfo
  {author} {\bibfnamefont {T.}~\bibnamefont {Klahn}}, \bibinfo {author}
  {\bibfnamefont {G.}~\bibnamefont {Pagliara}}, \ and\ \bibinfo {author}
  {\bibfnamefont {J.}~\bibnamefont {Schaffner-Bielich}},\ }\href {\doibase
  10.1038/nature05582} {\bibfield  {journal} {\bibinfo  {journal} {Nature}\
  }\textbf {\bibinfo {volume} {445}},\ \bibinfo {pages} {E7} (\bibinfo {year}
  {2007})},\ \Eprint {http://arxiv.org/abs/astro-ph/0606524}
  {arXiv:astro-ph/0606524} \BibitemShut {NoStop}%
\bibitem [{\citenamefont {Alford}\ \emph {et~al.}(2005)\citenamefont {Alford},
  \citenamefont {Braby}, \citenamefont {Paris},\ and\ \citenamefont
  {Reddy}}]{Alford:2004pf}%
  \BibitemOpen
  \bibfield  {author} {\bibinfo {author} {\bibfnamefont {M.}~\bibnamefont
  {Alford}}, \bibinfo {author} {\bibfnamefont {M.}~\bibnamefont {Braby}},
  \bibinfo {author} {\bibfnamefont {M.}~\bibnamefont {Paris}}, \ and\ \bibinfo
  {author} {\bibfnamefont {S.}~\bibnamefont {Reddy}},\ }\href {\doibase
  10.1086/430902} {\bibfield  {journal} {\bibinfo  {journal} {Astrophys. J.}\
  }\textbf {\bibinfo {volume} {629}},\ \bibinfo {pages} {969} (\bibinfo {year}
  {2005})},\ \Eprint {http://arxiv.org/abs/nucl-th/0411016}
  {arXiv:nucl-th/0411016} \BibitemShut {NoStop}%
\bibitem [{\citenamefont {Alford}\ \emph {et~al.}(2013)\citenamefont {Alford},
  \citenamefont {Han},\ and\ \citenamefont {Prakash}}]{Alford:2013aca}%
  \BibitemOpen
  \bibfield  {author} {\bibinfo {author} {\bibfnamefont {M.~G.}\ \bibnamefont
  {Alford}}, \bibinfo {author} {\bibfnamefont {S.}~\bibnamefont {Han}}, \ and\
  \bibinfo {author} {\bibfnamefont {M.}~\bibnamefont {Prakash}},\ }\href
  {\doibase 10.1103/PhysRevD.88.083013} {\bibfield  {journal} {\bibinfo
  {journal} {Phys. Rev. D}\ }\textbf {\bibinfo {volume} {88}},\ \bibinfo
  {pages} {083013} (\bibinfo {year} {2013})},\ \Eprint
  {http://arxiv.org/abs/1302.4732} {arXiv:1302.4732 [astro-ph.SR]} \BibitemShut
  {NoStop}%
\bibitem [{\citenamefont {Benic}\ \emph {et~al.}(2015)\citenamefont {Benic},
  \citenamefont {Blaschke}, \citenamefont {Alvarez-Castillo}, \citenamefont
  {Fischer},\ and\ \citenamefont {Typel}}]{Benic:2014jia}%
  \BibitemOpen
  \bibfield  {author} {\bibinfo {author} {\bibfnamefont {S.}~\bibnamefont
  {Benic}}, \bibinfo {author} {\bibfnamefont {D.}~\bibnamefont {Blaschke}},
  \bibinfo {author} {\bibfnamefont {D.~E.}\ \bibnamefont {Alvarez-Castillo}},
  \bibinfo {author} {\bibfnamefont {T.}~\bibnamefont {Fischer}}, \ and\
  \bibinfo {author} {\bibfnamefont {S.}~\bibnamefont {Typel}},\ }\href
  {\doibase 10.1051/0004-6361/201425318} {\bibfield  {journal} {\bibinfo
  {journal} {Astron. Astrophys.}\ }\textbf {\bibinfo {volume} {577}},\ \bibinfo
  {pages} {A40} (\bibinfo {year} {2015})},\ \Eprint
  {http://arxiv.org/abs/1411.2856} {arXiv:1411.2856 [astro-ph.HE]} \BibitemShut
  {NoStop}%
\bibitem [{\citenamefont {Alford}\ and\ \citenamefont
  {Sedrakian}(2017)}]{Alford:2017qgh}%
  \BibitemOpen
  \bibfield  {author} {\bibinfo {author} {\bibfnamefont {M.~G.}\ \bibnamefont
  {Alford}}\ and\ \bibinfo {author} {\bibfnamefont {A.}~\bibnamefont
  {Sedrakian}},\ }\href {\doibase 10.1103/PhysRevLett.119.161104} {\bibfield
  {journal} {\bibinfo  {journal} {Phys. Rev. Lett.}\ }\textbf {\bibinfo
  {volume} {119}},\ \bibinfo {pages} {161104} (\bibinfo {year} {2017})},\
  \Eprint {http://arxiv.org/abs/1706.01592} {arXiv:1706.01592 [astro-ph.HE]}
  \BibitemShut {NoStop}%
\bibitem [{\citenamefont {Bauswein}\ \emph
  {et~al.}(2019{\natexlab{a}})\citenamefont {Bauswein}, \citenamefont
  {Bastian}, \citenamefont {Blaschke}, \citenamefont {Chatziioannou},
  \citenamefont {Clark}, \citenamefont {Fischer},\ and\ \citenamefont
  {Oertel}}]{Bauswein:2018bma}%
  \BibitemOpen
  \bibfield  {author} {\bibinfo {author} {\bibfnamefont {A.}~\bibnamefont
  {Bauswein}}, \bibinfo {author} {\bibfnamefont {N.-U.~F.}\ \bibnamefont
  {Bastian}}, \bibinfo {author} {\bibfnamefont {D.~B.}\ \bibnamefont
  {Blaschke}}, \bibinfo {author} {\bibfnamefont {K.}~\bibnamefont
  {Chatziioannou}}, \bibinfo {author} {\bibfnamefont {J.~A.}\ \bibnamefont
  {Clark}}, \bibinfo {author} {\bibfnamefont {T.}~\bibnamefont {Fischer}}, \
  and\ \bibinfo {author} {\bibfnamefont {M.}~\bibnamefont {Oertel}},\ }\href
  {\doibase 10.1103/PhysRevLett.122.061102} {\bibfield  {journal} {\bibinfo
  {journal} {Phys. Rev. Lett.}\ }\textbf {\bibinfo {volume} {122}},\ \bibinfo
  {pages} {061102} (\bibinfo {year} {2019}{\natexlab{a}})},\ \Eprint
  {http://arxiv.org/abs/1809.01116} {arXiv:1809.01116 [astro-ph.HE]}
  \BibitemShut {NoStop}%
\bibitem [{\citenamefont {Most}\ \emph {et~al.}(2019)\citenamefont {Most},
  \citenamefont {Papenfort}, \citenamefont {Dexheimer}, \citenamefont
  {Hanauske}, \citenamefont {Schramm}, \citenamefont {Stöcker},\ and\
  \citenamefont {Rezzolla}}]{Most:2018eaw}%
  \BibitemOpen
  \bibfield  {author} {\bibinfo {author} {\bibfnamefont {E.~R.}\ \bibnamefont
  {Most}}, \bibinfo {author} {\bibfnamefont {L.~J.}\ \bibnamefont {Papenfort}},
  \bibinfo {author} {\bibfnamefont {V.}~\bibnamefont {Dexheimer}}, \bibinfo
  {author} {\bibfnamefont {M.}~\bibnamefont {Hanauske}}, \bibinfo {author}
  {\bibfnamefont {S.}~\bibnamefont {Schramm}}, \bibinfo {author} {\bibfnamefont
  {H.}~\bibnamefont {Stöcker}}, \ and\ \bibinfo {author} {\bibfnamefont
  {L.}~\bibnamefont {Rezzolla}},\ }\href {\doibase
  10.1103/PhysRevLett.122.061101} {\bibfield  {journal} {\bibinfo  {journal}
  {Phys. Rev. Lett.}\ }\textbf {\bibinfo {volume} {122}},\ \bibinfo {pages}
  {061101} (\bibinfo {year} {2019})},\ \Eprint
  {http://arxiv.org/abs/1807.03684} {arXiv:1807.03684 [astro-ph.HE]}
  \BibitemShut {NoStop}%
\bibitem [{\citenamefont {Bauswein}\ \emph
  {et~al.}(2019{\natexlab{b}})\citenamefont {Bauswein}, \citenamefont
  {Friedrich~Bastian}, \citenamefont {Blaschke}, \citenamefont {Chatziioannou},
  \citenamefont {Clark}, \citenamefont {Fischer}, \citenamefont {Janka},
  \citenamefont {Just}, \citenamefont {Oertel},\ and\ \citenamefont
  {Stergioulas}}]{Bauswein:2019skm}%
  \BibitemOpen
  \bibfield  {author} {\bibinfo {author} {\bibfnamefont {A.}~\bibnamefont
  {Bauswein}}, \bibinfo {author} {\bibfnamefont {N.-U.}\ \bibnamefont
  {Friedrich~Bastian}}, \bibinfo {author} {\bibfnamefont {D.}~\bibnamefont
  {Blaschke}}, \bibinfo {author} {\bibfnamefont {K.}~\bibnamefont
  {Chatziioannou}}, \bibinfo {author} {\bibfnamefont {J.~A.}\ \bibnamefont
  {Clark}}, \bibinfo {author} {\bibfnamefont {T.}~\bibnamefont {Fischer}},
  \bibinfo {author} {\bibfnamefont {H.-T.}\ \bibnamefont {Janka}}, \bibinfo
  {author} {\bibfnamefont {O.}~\bibnamefont {Just}}, \bibinfo {author}
  {\bibfnamefont {M.}~\bibnamefont {Oertel}}, \ and\ \bibinfo {author}
  {\bibfnamefont {N.}~\bibnamefont {Stergioulas}},\ }\href {\doibase
  10.1063/1.5117803} {\bibfield  {journal} {\bibinfo  {journal} {AIP Conf.
  Proc.}\ }\textbf {\bibinfo {volume} {2127}},\ \bibinfo {pages} {020013}
  (\bibinfo {year} {2019}{\natexlab{b}})},\ \Eprint
  {http://arxiv.org/abs/1904.01306} {arXiv:1904.01306 [astro-ph.HE]}
  \BibitemShut {NoStop}%
\bibitem [{\citenamefont {Most}\ \emph {et~al.}(2020)\citenamefont {Most},
  \citenamefont {Jens~Papenfort}, \citenamefont {Dexheimer}, \citenamefont
  {Hanauske}, \citenamefont {Stoecker},\ and\ \citenamefont
  {Rezzolla}}]{Most:2019onn}%
  \BibitemOpen
  \bibfield  {author} {\bibinfo {author} {\bibfnamefont {E.~R.}\ \bibnamefont
  {Most}}, \bibinfo {author} {\bibfnamefont {L.}~\bibnamefont
  {Jens~Papenfort}}, \bibinfo {author} {\bibfnamefont {V.}~\bibnamefont
  {Dexheimer}}, \bibinfo {author} {\bibfnamefont {M.}~\bibnamefont {Hanauske}},
  \bibinfo {author} {\bibfnamefont {H.}~\bibnamefont {Stoecker}}, \ and\
  \bibinfo {author} {\bibfnamefont {L.}~\bibnamefont {Rezzolla}},\ }\href
  {\doibase 10.1140/epja/s10050-020-00073-4} {\bibfield  {journal} {\bibinfo
  {journal} {Eur. Phys. J. A}\ }\textbf {\bibinfo {volume} {56}},\ \bibinfo
  {pages} {59} (\bibinfo {year} {2020})},\ \Eprint
  {http://arxiv.org/abs/1910.13893} {arXiv:1910.13893 [astro-ph.HE]}
  \BibitemShut {NoStop}%
\bibitem [{\citenamefont {Weih}\ \emph {et~al.}(2020)\citenamefont {Weih},
  \citenamefont {Hanauske},\ and\ \citenamefont {Rezzolla}}]{Weih:2019xvw}%
  \BibitemOpen
  \bibfield  {author} {\bibinfo {author} {\bibfnamefont {L.~R.}\ \bibnamefont
  {Weih}}, \bibinfo {author} {\bibfnamefont {M.}~\bibnamefont {Hanauske}}, \
  and\ \bibinfo {author} {\bibfnamefont {L.}~\bibnamefont {Rezzolla}},\ }\href
  {\doibase 10.1103/PhysRevLett.124.171103} {\bibfield  {journal} {\bibinfo
  {journal} {Phys. Rev. Lett.}\ }\textbf {\bibinfo {volume} {124}},\ \bibinfo
  {pages} {171103} (\bibinfo {year} {2020})},\ \Eprint
  {http://arxiv.org/abs/1912.09340} {arXiv:1912.09340 [gr-qc]} \BibitemShut
  {NoStop}%
\bibitem [{\citenamefont {Blacker}\ \emph {et~al.}(2020)\citenamefont
  {Blacker}, \citenamefont {Bastian}, \citenamefont {Bauswein}, \citenamefont
  {Blaschke}, \citenamefont {Fischer}, \citenamefont {Oertel}, \citenamefont
  {Soultanis},\ and\ \citenamefont {Typel}}]{Blacker:2020nlq}%
  \BibitemOpen
  \bibfield  {author} {\bibinfo {author} {\bibfnamefont {S.}~\bibnamefont
  {Blacker}}, \bibinfo {author} {\bibfnamefont {N.-U.~F.}\ \bibnamefont
  {Bastian}}, \bibinfo {author} {\bibfnamefont {A.}~\bibnamefont {Bauswein}},
  \bibinfo {author} {\bibfnamefont {D.~B.}\ \bibnamefont {Blaschke}}, \bibinfo
  {author} {\bibfnamefont {T.}~\bibnamefont {Fischer}}, \bibinfo {author}
  {\bibfnamefont {M.}~\bibnamefont {Oertel}}, \bibinfo {author} {\bibfnamefont
  {T.}~\bibnamefont {Soultanis}}, \ and\ \bibinfo {author} {\bibfnamefont
  {S.}~\bibnamefont {Typel}},\ }\href@noop {} {\  (\bibinfo {year} {2020})},\
  \Eprint {http://arxiv.org/abs/2006.03789} {arXiv:2006.03789 [astro-ph.HE]}
  \BibitemShut {NoStop}%
\bibitem [{\citenamefont {Montana}\ \emph {et~al.}(2019)\citenamefont
  {Montana}, \citenamefont {Tolos}, \citenamefont {Hanauske},\ and\
  \citenamefont {Rezzolla}}]{Montana:2018bkb}%
  \BibitemOpen
  \bibfield  {author} {\bibinfo {author} {\bibfnamefont {G.}~\bibnamefont
  {Montana}}, \bibinfo {author} {\bibfnamefont {L.}~\bibnamefont {Tolos}},
  \bibinfo {author} {\bibfnamefont {M.}~\bibnamefont {Hanauske}}, \ and\
  \bibinfo {author} {\bibfnamefont {L.}~\bibnamefont {Rezzolla}},\ }\href
  {\doibase 10.1103/PhysRevD.99.103009} {\bibfield  {journal} {\bibinfo
  {journal} {Phys. Rev. D}\ }\textbf {\bibinfo {volume} {99}},\ \bibinfo
  {pages} {103009} (\bibinfo {year} {2019})},\ \Eprint
  {http://arxiv.org/abs/1811.10929} {arXiv:1811.10929 [astro-ph.HE]}
  \BibitemShut {NoStop}%
\bibitem [{\citenamefont {Gieg}\ \emph {et~al.}(2019)\citenamefont {Gieg},
  \citenamefont {Dietrich},\ and\ \citenamefont {Ujevic}}]{Gieg:2019yzq}%
  \BibitemOpen
  \bibfield  {author} {\bibinfo {author} {\bibfnamefont {H.}~\bibnamefont
  {Gieg}}, \bibinfo {author} {\bibfnamefont {T.}~\bibnamefont {Dietrich}}, \
  and\ \bibinfo {author} {\bibfnamefont {M.}~\bibnamefont {Ujevic}},\ }\href
  {\doibase 10.3390/particles2030023} {\bibfield  {journal} {\bibinfo
  {journal} {Particles}\ }\textbf {\bibinfo {volume} {2}},\ \bibinfo {pages}
  {365} (\bibinfo {year} {2019})},\ \Eprint {http://arxiv.org/abs/1908.03135}
  {arXiv:1908.03135 [gr-qc]} \BibitemShut {NoStop}%
\bibitem [{\citenamefont {Flanagan}\ and\ \citenamefont
  {Hinderer}(2008)}]{Flanagan:2007ix}%
  \BibitemOpen
  \bibfield  {author} {\bibinfo {author} {\bibfnamefont {E.~E.}\ \bibnamefont
  {Flanagan}}\ and\ \bibinfo {author} {\bibfnamefont {T.}~\bibnamefont
  {Hinderer}},\ }\href {\doibase 10.1103/PhysRevD.77.021502} {\bibfield
  {journal} {\bibinfo  {journal} {Phys.Rev.}\ }\textbf {\bibinfo {volume}
  {D77}},\ \bibinfo {pages} {021502} (\bibinfo {year} {2008})},\ \Eprint
  {http://arxiv.org/abs/0709.1915} {arXiv:0709.1915 [astro-ph]} \BibitemShut
  {NoStop}%
\bibitem [{\citenamefont {Hinderer}(2008)}]{Hinderer:2007mb}%
  \BibitemOpen
  \bibfield  {author} {\bibinfo {author} {\bibfnamefont {T.}~\bibnamefont
  {Hinderer}},\ }\href {\doibase 10.1086/533487} {\bibfield  {journal}
  {\bibinfo  {journal} {Astrophys.J.}\ }\textbf {\bibinfo {volume} {677}},\
  \bibinfo {pages} {1216} (\bibinfo {year} {2008})},\ \Eprint
  {http://arxiv.org/abs/0711.2420} {arXiv:0711.2420 [astro-ph]} \BibitemShut
  {NoStop}%
\bibitem [{\citenamefont {Hinderer}\ \emph {et~al.}(2010)\citenamefont
  {Hinderer}, \citenamefont {Lackey}, \citenamefont {Lang},\ and\ \citenamefont
  {Read}}]{Hinderer:2009ca}%
  \BibitemOpen
  \bibfield  {author} {\bibinfo {author} {\bibfnamefont {T.}~\bibnamefont
  {Hinderer}}, \bibinfo {author} {\bibfnamefont {B.~D.}\ \bibnamefont
  {Lackey}}, \bibinfo {author} {\bibfnamefont {R.~N.}\ \bibnamefont {Lang}}, \
  and\ \bibinfo {author} {\bibfnamefont {J.~S.}\ \bibnamefont {Read}},\ }\href
  {\doibase 10.1103/PhysRevD.81.123016} {\bibfield  {journal} {\bibinfo
  {journal} {Phys. Rev.}\ }\textbf {\bibinfo {volume} {D81}},\ \bibinfo {pages}
  {123016} (\bibinfo {year} {2010})},\ \Eprint {http://arxiv.org/abs/0911.3535}
  {arXiv:0911.3535 [astro-ph.HE]} \BibitemShut {NoStop}%
\bibitem [{\citenamefont {Yagi}\ and\ \citenamefont
  {Yunes}(2017)}]{Yagi:2016bkt}%
  \BibitemOpen
  \bibfield  {author} {\bibinfo {author} {\bibfnamefont {K.}~\bibnamefont
  {Yagi}}\ and\ \bibinfo {author} {\bibfnamefont {N.}~\bibnamefont {Yunes}},\
  }\href {\doibase 10.1016/j.physrep.2017.03.002} {\bibfield  {journal}
  {\bibinfo  {journal} {Phys. Rept.}\ }\textbf {\bibinfo {volume} {681}},\
  \bibinfo {pages} {1} (\bibinfo {year} {2017})},\ \Eprint
  {http://arxiv.org/abs/1608.02582} {arXiv:1608.02582 [gr-qc]} \BibitemShut
  {NoStop}%
\bibitem [{\citenamefont {Chatziioannou}\ \emph {et~al.}(2018)\citenamefont
  {Chatziioannou}, \citenamefont {Haster},\ and\ \citenamefont
  {Zimmerman}}]{Chatziioannou:2018vzf}%
  \BibitemOpen
  \bibfield  {author} {\bibinfo {author} {\bibfnamefont {K.}~\bibnamefont
  {Chatziioannou}}, \bibinfo {author} {\bibfnamefont {C.-J.}\ \bibnamefont
  {Haster}}, \ and\ \bibinfo {author} {\bibfnamefont {A.}~\bibnamefont
  {Zimmerman}},\ }\href {\doibase 10.1103/PhysRevD.97.104036} {\bibfield
  {journal} {\bibinfo  {journal} {Phys. Rev.}\ }\textbf {\bibinfo {volume}
  {D97}},\ \bibinfo {pages} {104036} (\bibinfo {year} {2018})},\ \Eprint
  {http://arxiv.org/abs/1804.03221} {arXiv:1804.03221 [gr-qc]} \BibitemShut
  {NoStop}%
\bibitem [{\citenamefont {Annala}\ \emph {et~al.}(2020)\citenamefont {Annala},
  \citenamefont {Gorda}, \citenamefont {Kurkela}, \citenamefont {Nättilä},\
  and\ \citenamefont {Vuorinen}}]{Annala:2019puf}%
  \BibitemOpen
  \bibfield  {author} {\bibinfo {author} {\bibfnamefont {E.}~\bibnamefont
  {Annala}}, \bibinfo {author} {\bibfnamefont {T.}~\bibnamefont {Gorda}},
  \bibinfo {author} {\bibfnamefont {A.}~\bibnamefont {Kurkela}}, \bibinfo
  {author} {\bibfnamefont {J.}~\bibnamefont {Nättilä}}, \ and\ \bibinfo
  {author} {\bibfnamefont {A.}~\bibnamefont {Vuorinen}},\ }\href {\doibase
  10.1038/s41567-020-0914-9} {\bibfield  {journal} {\bibinfo  {journal} {Nature
  Phys.}\ } (\bibinfo {year} {2020}),\ 10.1038/s41567-020-0914-9},\ \Eprint
  {http://arxiv.org/abs/1903.09121} {arXiv:1903.09121 [astro-ph.HE]}
  \BibitemShut {NoStop}%
\bibitem [{\citenamefont {Douchin}\ and\ \citenamefont
  {Haensel}(2001)}]{Douchin:2001sv}%
  \BibitemOpen
  \bibfield  {author} {\bibinfo {author} {\bibfnamefont {F.}~\bibnamefont
  {Douchin}}\ and\ \bibinfo {author} {\bibfnamefont {P.}~\bibnamefont
  {Haensel}},\ }\href@noop {} {\bibfield  {journal} {\bibinfo  {journal}
  {Astron. Astrophys.}\ }\textbf {\bibinfo {volume} {380}},\ \bibinfo {pages}
  {151} (\bibinfo {year} {2001})},\ \Eprint
  {http://arxiv.org/abs/astro-ph/0111092} {astro-ph/0111092} \BibitemShut
  {NoStop}%
\bibitem [{\citenamefont {Typel}\ \emph {et~al.}(2010)\citenamefont {Typel},
  \citenamefont {Ropke}, \citenamefont {Klahn}, \citenamefont {Blaschke},\ and\
  \citenamefont {Wolter}}]{Typel:2009sy}%
  \BibitemOpen
  \bibfield  {author} {\bibinfo {author} {\bibfnamefont {S.}~\bibnamefont
  {Typel}}, \bibinfo {author} {\bibfnamefont {G.}~\bibnamefont {Ropke}},
  \bibinfo {author} {\bibfnamefont {T.}~\bibnamefont {Klahn}}, \bibinfo
  {author} {\bibfnamefont {D.}~\bibnamefont {Blaschke}}, \ and\ \bibinfo
  {author} {\bibfnamefont {H.~H.}\ \bibnamefont {Wolter}},\ }\href {\doibase
  10.1103/PhysRevC.81.015803} {\bibfield  {journal} {\bibinfo  {journal} {Phys.
  Rev.}\ }\textbf {\bibinfo {volume} {C81}},\ \bibinfo {pages} {015803}
  (\bibinfo {year} {2010})},\ \Eprint {http://arxiv.org/abs/0908.2344}
  {arXiv:0908.2344 [nucl-th]} \BibitemShut {NoStop}%
\bibitem [{\citenamefont {Steiner}\ \emph
  {et~al.}(2013{\natexlab{b}})\citenamefont {Steiner}, \citenamefont {Hempel},\
  and\ \citenamefont {Fischer}}]{steiner:2012rk}%
  \BibitemOpen
  \bibfield  {author} {\bibinfo {author} {\bibfnamefont {A.~W.}\ \bibnamefont
  {Steiner}}, \bibinfo {author} {\bibfnamefont {M.}~\bibnamefont {Hempel}}, \
  and\ \bibinfo {author} {\bibfnamefont {T.}~\bibnamefont {Fischer}},\ }\href
  {\doibase 10.1088/0004-637X/774/1/17} {\bibfield  {journal} {\bibinfo
  {journal} {Astrophys. J.}\ }\textbf {\bibinfo {volume} {774}},\ \bibinfo
  {pages} {17} (\bibinfo {year} {2013}{\natexlab{b}})},\ \Eprint
  {http://arxiv.org/abs/1207.2184} {arXiv:1207.2184 [astro-ph.SR]} \BibitemShut
  {NoStop}%
\bibitem [{\citenamefont {Gandolfi}\ \emph {et~al.}(2012)\citenamefont
  {Gandolfi}, \citenamefont {Carlson},\ and\ \citenamefont
  {Reddy}}]{Gandolfi:2011xu}%
  \BibitemOpen
  \bibfield  {author} {\bibinfo {author} {\bibfnamefont {S.}~\bibnamefont
  {Gandolfi}}, \bibinfo {author} {\bibfnamefont {J.}~\bibnamefont {Carlson}}, \
  and\ \bibinfo {author} {\bibfnamefont {S.}~\bibnamefont {Reddy}},\ }\href
  {\doibase 10.1103/PhysRevC.85.032801} {\bibfield  {journal} {\bibinfo
  {journal} {Phys. Rev. C}\ }\textbf {\bibinfo {volume} {85}},\ \bibinfo
  {pages} {032801} (\bibinfo {year} {2012})},\ \Eprint
  {http://arxiv.org/abs/1101.1921} {arXiv:1101.1921 [nucl-th]} \BibitemShut
  {NoStop}%
\bibitem [{\citenamefont {Hebeler}\ \emph {et~al.}(2013)\citenamefont
  {Hebeler}, \citenamefont {Lattimer}, \citenamefont {Pethick},\ and\
  \citenamefont {Schwenk}}]{Hebeler:2013nza}%
  \BibitemOpen
  \bibfield  {author} {\bibinfo {author} {\bibfnamefont {K.}~\bibnamefont
  {Hebeler}}, \bibinfo {author} {\bibfnamefont {J.}~\bibnamefont {Lattimer}},
  \bibinfo {author} {\bibfnamefont {C.}~\bibnamefont {Pethick}}, \ and\
  \bibinfo {author} {\bibfnamefont {A.}~\bibnamefont {Schwenk}},\ }\href
  {\doibase 10.1088/0004-637X/773/1/11} {\bibfield  {journal} {\bibinfo
  {journal} {Astrophys. J.}\ }\textbf {\bibinfo {volume} {773}},\ \bibinfo
  {pages} {11} (\bibinfo {year} {2013})},\ \Eprint
  {http://arxiv.org/abs/1303.4662} {arXiv:1303.4662 [astro-ph.SR]} \BibitemShut
  {NoStop}%
\bibitem [{\citenamefont {Carbone}\ \emph {et~al.}(2013)\citenamefont
  {Carbone}, \citenamefont {Cipollone}, \citenamefont {Barbieri}, \citenamefont
  {Rios},\ and\ \citenamefont {Polls}}]{Carbone:2013eqa}%
  \BibitemOpen
  \bibfield  {author} {\bibinfo {author} {\bibfnamefont {A.}~\bibnamefont
  {Carbone}}, \bibinfo {author} {\bibfnamefont {A.}~\bibnamefont {Cipollone}},
  \bibinfo {author} {\bibfnamefont {C.}~\bibnamefont {Barbieri}}, \bibinfo
  {author} {\bibfnamefont {A.}~\bibnamefont {Rios}}, \ and\ \bibinfo {author}
  {\bibfnamefont {A.}~\bibnamefont {Polls}},\ }\href {\doibase
  10.1103/PhysRevC.88.054326} {\bibfield  {journal} {\bibinfo  {journal} {Phys.
  Rev. C}\ }\textbf {\bibinfo {volume} {88}},\ \bibinfo {pages} {054326}
  (\bibinfo {year} {2013})},\ \Eprint {http://arxiv.org/abs/1310.3688}
  {arXiv:1310.3688 [nucl-th]} \BibitemShut {NoStop}%
\bibitem [{\citenamefont {Hagen}\ \emph {et~al.}(2014)\citenamefont {Hagen},
  \citenamefont {Papenbrock}, \citenamefont {Ekström}, \citenamefont {Wendt},
  \citenamefont {Baardsen}, \citenamefont {Gandolfi}, \citenamefont
  {Hjorth-Jensen},\ and\ \citenamefont {Horowitz}}]{Hagen:2013yba}%
  \BibitemOpen
  \bibfield  {author} {\bibinfo {author} {\bibfnamefont {G.}~\bibnamefont
  {Hagen}}, \bibinfo {author} {\bibfnamefont {T.}~\bibnamefont {Papenbrock}},
  \bibinfo {author} {\bibfnamefont {A.}~\bibnamefont {Ekström}}, \bibinfo
  {author} {\bibfnamefont {K.}~\bibnamefont {Wendt}}, \bibinfo {author}
  {\bibfnamefont {G.}~\bibnamefont {Baardsen}}, \bibinfo {author}
  {\bibfnamefont {S.}~\bibnamefont {Gandolfi}}, \bibinfo {author}
  {\bibfnamefont {M.}~\bibnamefont {Hjorth-Jensen}}, \ and\ \bibinfo {author}
  {\bibfnamefont {C.}~\bibnamefont {Horowitz}},\ }\href {\doibase
  10.1103/PhysRevC.89.014319} {\bibfield  {journal} {\bibinfo  {journal} {Phys.
  Rev. C}\ }\textbf {\bibinfo {volume} {89}},\ \bibinfo {pages} {014319}
  (\bibinfo {year} {2014})},\ \Eprint {http://arxiv.org/abs/1311.2925}
  {arXiv:1311.2925 [nucl-th]} \BibitemShut {NoStop}%
\bibitem [{\citenamefont {Holt}\ and\ \citenamefont
  {Kaiser}(2017)}]{Holt:2016pjb}%
  \BibitemOpen
  \bibfield  {author} {\bibinfo {author} {\bibfnamefont {J.}~\bibnamefont
  {Holt}}\ and\ \bibinfo {author} {\bibfnamefont {N.}~\bibnamefont {Kaiser}},\
  }\href {\doibase 10.1103/PhysRevC.95.034326} {\bibfield  {journal} {\bibinfo
  {journal} {Phys. Rev. C}\ }\textbf {\bibinfo {volume} {95}},\ \bibinfo
  {pages} {034326} (\bibinfo {year} {2017})},\ \Eprint
  {http://arxiv.org/abs/1612.04309} {arXiv:1612.04309 [nucl-th]} \BibitemShut
  {NoStop}%
\bibitem [{\citenamefont {Drischler}\ \emph {et~al.}(2019)\citenamefont
  {Drischler}, \citenamefont {Hebeler},\ and\ \citenamefont
  {Schwenk}}]{Drischler:2017wtt}%
  \BibitemOpen
  \bibfield  {author} {\bibinfo {author} {\bibfnamefont {C.}~\bibnamefont
  {Drischler}}, \bibinfo {author} {\bibfnamefont {K.}~\bibnamefont {Hebeler}},
  \ and\ \bibinfo {author} {\bibfnamefont {A.}~\bibnamefont {Schwenk}},\ }\href
  {\doibase 10.1103/PhysRevLett.122.042501} {\bibfield  {journal} {\bibinfo
  {journal} {Phys. Rev. Lett.}\ }\textbf {\bibinfo {volume} {122}},\ \bibinfo
  {pages} {042501} (\bibinfo {year} {2019})},\ \Eprint
  {http://arxiv.org/abs/1710.08220} {arXiv:1710.08220 [nucl-th]} \BibitemShut
  {NoStop}%
\bibitem [{\citenamefont {Tews}\ \emph
  {et~al.}(2018{\natexlab{b}})\citenamefont {Tews}, \citenamefont {Carlson},
  \citenamefont {Gandolfi},\ and\ \citenamefont {Reddy}}]{Tews:2018kmu}%
  \BibitemOpen
  \bibfield  {author} {\bibinfo {author} {\bibfnamefont {I.}~\bibnamefont
  {Tews}}, \bibinfo {author} {\bibfnamefont {J.}~\bibnamefont {Carlson}},
  \bibinfo {author} {\bibfnamefont {S.}~\bibnamefont {Gandolfi}}, \ and\
  \bibinfo {author} {\bibfnamefont {S.}~\bibnamefont {Reddy}},\ }\href
  {\doibase 10.3847/1538-4357/aac267} {\bibfield  {journal} {\bibinfo
  {journal} {Astrophys. J.}\ }\textbf {\bibinfo {volume} {860}},\ \bibinfo
  {pages} {149} (\bibinfo {year} {2018}{\natexlab{b}})},\ \Eprint
  {http://arxiv.org/abs/1801.01923} {arXiv:1801.01923 [nucl-th]} \BibitemShut
  {NoStop}%
\bibitem [{\citenamefont {Epelbaum}\ \emph {et~al.}(2009)\citenamefont
  {Epelbaum}, \citenamefont {Hammer},\ and\ \citenamefont
  {Meissner}}]{Epelbaum:2008ga}%
  \BibitemOpen
  \bibfield  {author} {\bibinfo {author} {\bibfnamefont {E.}~\bibnamefont
  {Epelbaum}}, \bibinfo {author} {\bibfnamefont {H.-W.}\ \bibnamefont
  {Hammer}}, \ and\ \bibinfo {author} {\bibfnamefont {U.-G.}\ \bibnamefont
  {Meissner}},\ }\href {\doibase 10.1103/RevModPhys.81.1773} {\bibfield
  {journal} {\bibinfo  {journal} {Rev. Mod. Phys.}\ }\textbf {\bibinfo {volume}
  {81}},\ \bibinfo {pages} {1773} (\bibinfo {year} {2009})},\ \Eprint
  {http://arxiv.org/abs/0811.1338} {arXiv:0811.1338 [nucl-th]} \BibitemShut
  {NoStop}%
\bibitem [{\citenamefont {Machleidt}\ and\ \citenamefont
  {Entem}(2011)}]{Machleidt:2011zz}%
  \BibitemOpen
  \bibfield  {author} {\bibinfo {author} {\bibfnamefont {R.}~\bibnamefont
  {Machleidt}}\ and\ \bibinfo {author} {\bibfnamefont {D.}~\bibnamefont
  {Entem}},\ }\href {\doibase 10.1016/j.physrep.2011.02.001} {\bibfield
  {journal} {\bibinfo  {journal} {Phys. Rept.}\ }\textbf {\bibinfo {volume}
  {503}},\ \bibinfo {pages} {1} (\bibinfo {year} {2011})},\ \Eprint
  {http://arxiv.org/abs/1105.2919} {arXiv:1105.2919 [nucl-th]} \BibitemShut
  {NoStop}%
\bibitem [{\citenamefont {Epelbaum}\ \emph {et~al.}(2015)\citenamefont
  {Epelbaum}, \citenamefont {Krebs},\ and\ \citenamefont
  {Meißner}}]{Epelbaum:2014efa}%
  \BibitemOpen
  \bibfield  {author} {\bibinfo {author} {\bibfnamefont {E.}~\bibnamefont
  {Epelbaum}}, \bibinfo {author} {\bibfnamefont {H.}~\bibnamefont {Krebs}}, \
  and\ \bibinfo {author} {\bibfnamefont {U.}~\bibnamefont {Meißner}},\ }\href
  {\doibase 10.1140/epja/i2015-15053-8} {\bibfield  {journal} {\bibinfo
  {journal} {Eur. Phys. J. A}\ }\textbf {\bibinfo {volume} {51}},\ \bibinfo
  {pages} {53} (\bibinfo {year} {2015})},\ \Eprint
  {http://arxiv.org/abs/1412.0142} {arXiv:1412.0142 [nucl-th]} \BibitemShut
  {NoStop}%
\bibitem [{\citenamefont {Drischler}\ \emph {et~al.}(2020)\citenamefont
  {Drischler}, \citenamefont {Melendez}, \citenamefont {Furnstahl},\ and\
  \citenamefont {Phillips}}]{Drischler:2020yad}%
  \BibitemOpen
  \bibfield  {author} {\bibinfo {author} {\bibfnamefont {C.}~\bibnamefont
  {Drischler}}, \bibinfo {author} {\bibfnamefont {J.}~\bibnamefont {Melendez}},
  \bibinfo {author} {\bibfnamefont {R.}~\bibnamefont {Furnstahl}}, \ and\
  \bibinfo {author} {\bibfnamefont {D.}~\bibnamefont {Phillips}},\ }\href@noop
  {} {\  (\bibinfo {year} {2020})},\ \Eprint {http://arxiv.org/abs/2004.07805}
  {arXiv:2004.07805 [nucl-th]} \BibitemShut {NoStop}%
\bibitem [{\citenamefont {Read}\ \emph
  {et~al.}(2009{\natexlab{b}})\citenamefont {Read}, \citenamefont {Lackey},
  \citenamefont {Owen},\ and\ \citenamefont {Friedman}}]{Read:2008iy}%
  \BibitemOpen
  \bibfield  {author} {\bibinfo {author} {\bibfnamefont {J.~S.}\ \bibnamefont
  {Read}}, \bibinfo {author} {\bibfnamefont {B.~D.}\ \bibnamefont {Lackey}},
  \bibinfo {author} {\bibfnamefont {B.~J.}\ \bibnamefont {Owen}}, \ and\
  \bibinfo {author} {\bibfnamefont {J.~L.}\ \bibnamefont {Friedman}},\ }\href
  {\doibase 10.1103/PhysRevD.79.124032} {\bibfield  {journal} {\bibinfo
  {journal} {Phys. Rev.}\ }\textbf {\bibinfo {volume} {D79}},\ \bibinfo {pages}
  {124032} (\bibinfo {year} {2009}{\natexlab{b}})},\ \Eprint
  {http://arxiv.org/abs/0812.2163} {arXiv:0812.2163 [astro-ph]} \BibitemShut
  {NoStop}%
\bibitem [{\citenamefont {Landry}\ and\ \citenamefont
  {Essick}(2019)}]{Landry:2018prl}%
  \BibitemOpen
  \bibfield  {author} {\bibinfo {author} {\bibfnamefont {P.}~\bibnamefont
  {Landry}}\ and\ \bibinfo {author} {\bibfnamefont {R.}~\bibnamefont
  {Essick}},\ }\href {\doibase 10.1103/PhysRevD.99.084049} {\bibfield
  {journal} {\bibinfo  {journal} {Phys. Rev. D}\ }\textbf {\bibinfo {volume}
  {99}},\ \bibinfo {pages} {084049} (\bibinfo {year} {2019})},\ \Eprint
  {http://arxiv.org/abs/1811.12529} {arXiv:1811.12529 [gr-qc]} \BibitemShut
  {NoStop}%
\bibitem [{\citenamefont {Tews}(2017)}]{Tews:2016ofv}%
  \BibitemOpen
  \bibfield  {author} {\bibinfo {author} {\bibfnamefont {I.}~\bibnamefont
  {Tews}},\ }\href {\doibase 10.1103/PhysRevC.95.015803} {\bibfield  {journal}
  {\bibinfo  {journal} {Phys. Rev. C}\ }\textbf {\bibinfo {volume} {95}},\
  \bibinfo {pages} {015803} (\bibinfo {year} {2017})},\ \Eprint
  {http://arxiv.org/abs/1607.06998} {arXiv:1607.06998 [nucl-th]} \BibitemShut
  {NoStop}%
\bibitem [{\citenamefont {Damour}\ \emph {et~al.}(2012)\citenamefont {Damour},
  \citenamefont {Nagar},\ and\ \citenamefont {Villain}}]{Damour:2012yf}%
  \BibitemOpen
  \bibfield  {author} {\bibinfo {author} {\bibfnamefont {T.}~\bibnamefont
  {Damour}}, \bibinfo {author} {\bibfnamefont {A.}~\bibnamefont {Nagar}}, \
  and\ \bibinfo {author} {\bibfnamefont {L.}~\bibnamefont {Villain}},\ }\href
  {\doibase 10.1103/PhysRevD.85.123007} {\bibfield  {journal} {\bibinfo
  {journal} {Phys.Rev.}\ }\textbf {\bibinfo {volume} {D85}},\ \bibinfo {pages}
  {123007} (\bibinfo {year} {2012})},\ \Eprint {http://arxiv.org/abs/1203.4352}
  {arXiv:1203.4352 [gr-qc]} \BibitemShut {NoStop}%
\bibitem [{\citenamefont {Yagi}\ and\ \citenamefont {Yunes}(2016)}]{Yagi_2016}%
  \BibitemOpen
  \bibfield  {author} {\bibinfo {author} {\bibfnamefont {K.}~\bibnamefont
  {Yagi}}\ and\ \bibinfo {author} {\bibfnamefont {N.}~\bibnamefont {Yunes}},\
  }\href {\doibase 10.1088/0264-9381/33/13/13lt01} {\bibfield  {journal}
  {\bibinfo  {journal} {Classical and Quantum Gravity}\ }\textbf {\bibinfo
  {volume} {33}},\ \bibinfo {pages} {13LT01} (\bibinfo {year}
  {2016})}\BibitemShut {NoStop}%
\bibitem [{\citenamefont {{Shoemaker}}\ \emph {et~al.}(2010)\citenamefont
  {{Shoemaker}} \emph {et~al.}}]{advLIGOcurves}%
  \BibitemOpen
  \bibfield  {author} {\bibinfo {author} {\bibfnamefont {D.}~\bibnamefont
  {{Shoemaker}}} \emph {et~al.} (\bibinfo {collaboration} {LIGO Scientific
  Collaboration}),\ }\href@noop {} {\bibfield  {journal} {\bibinfo  {journal}
  {LIGO-T0900288, https://dcc.ligo.org/cgi-bin/DocDB/ShowDocument?docid=2974}\
  } (\bibinfo {year} {2010})}\BibitemShut {NoStop}%
\bibitem [{\citenamefont {Chatziioannou}\ \emph {et~al.}(2017)\citenamefont
  {Chatziioannou}, \citenamefont {Clark}, \citenamefont {Bauswein},
  \citenamefont {Millhouse}, \citenamefont {Littenberg},\ and\ \citenamefont
  {Cornish}}]{Chatziioannou:2017ixj}%
  \BibitemOpen
  \bibfield  {author} {\bibinfo {author} {\bibfnamefont {K.}~\bibnamefont
  {Chatziioannou}}, \bibinfo {author} {\bibfnamefont {J.~A.}\ \bibnamefont
  {Clark}}, \bibinfo {author} {\bibfnamefont {A.}~\bibnamefont {Bauswein}},
  \bibinfo {author} {\bibfnamefont {M.}~\bibnamefont {Millhouse}}, \bibinfo
  {author} {\bibfnamefont {T.~B.}\ \bibnamefont {Littenberg}}, \ and\ \bibinfo
  {author} {\bibfnamefont {N.}~\bibnamefont {Cornish}},\ }\href {\doibase
  10.1103/PhysRevD.96.124035} {\bibfield  {journal} {\bibinfo  {journal} {Phys.
  Rev.}\ }\textbf {\bibinfo {volume} {D96}},\ \bibinfo {pages} {124035}
  (\bibinfo {year} {2017})},\ \Eprint {http://arxiv.org/abs/1711.00040}
  {arXiv:1711.00040 [gr-qc]} \BibitemShut {NoStop}%
\bibitem [{\citenamefont {Bose}\ \emph {et~al.}(2018)\citenamefont {Bose},
  \citenamefont {Chakravarti}, \citenamefont {Rezzolla}, \citenamefont
  {Sathyaprakash},\ and\ \citenamefont {Takami}}]{Bose:2017jvk}%
  \BibitemOpen
  \bibfield  {author} {\bibinfo {author} {\bibfnamefont {S.}~\bibnamefont
  {Bose}}, \bibinfo {author} {\bibfnamefont {K.}~\bibnamefont {Chakravarti}},
  \bibinfo {author} {\bibfnamefont {L.}~\bibnamefont {Rezzolla}}, \bibinfo
  {author} {\bibfnamefont {B.~S.}\ \bibnamefont {Sathyaprakash}}, \ and\
  \bibinfo {author} {\bibfnamefont {K.}~\bibnamefont {Takami}},\ }\href
  {\doibase 10.1103/PhysRevLett.120.031102} {\bibfield  {journal} {\bibinfo
  {journal} {Phys. Rev. Lett.}\ }\textbf {\bibinfo {volume} {120}},\ \bibinfo
  {pages} {031102} (\bibinfo {year} {2018})},\ \Eprint
  {http://arxiv.org/abs/1705.10850} {arXiv:1705.10850 [gr-qc]} \BibitemShut
  {NoStop}%
\bibitem [{\citenamefont {Torres-Rivas}\ \emph {et~al.}(2019)\citenamefont
  {Torres-Rivas}, \citenamefont {Chatziioannou}, \citenamefont {Bauswein},\
  and\ \citenamefont {Clark}}]{Torres-Rivas:2018svp}%
  \BibitemOpen
  \bibfield  {author} {\bibinfo {author} {\bibfnamefont {A.}~\bibnamefont
  {Torres-Rivas}}, \bibinfo {author} {\bibfnamefont {K.}~\bibnamefont
  {Chatziioannou}}, \bibinfo {author} {\bibfnamefont {A.}~\bibnamefont
  {Bauswein}}, \ and\ \bibinfo {author} {\bibfnamefont {J.~A.}\ \bibnamefont
  {Clark}},\ }\href {\doibase 10.1103/PhysRevD.99.044014} {\bibfield  {journal}
  {\bibinfo  {journal} {Phys. Rev. D}\ }\textbf {\bibinfo {volume} {99}},\
  \bibinfo {pages} {044014} (\bibinfo {year} {2019})},\ \Eprint
  {http://arxiv.org/abs/1811.08931} {arXiv:1811.08931 [gr-qc]} \BibitemShut
  {NoStop}%
\bibitem [{\citenamefont {Easter}\ \emph {et~al.}(2019)\citenamefont {Easter},
  \citenamefont {Lasky}, \citenamefont {Casey}, \citenamefont {Rezzolla},\ and\
  \citenamefont {Takami}}]{Easter:2018pqy}%
  \BibitemOpen
  \bibfield  {author} {\bibinfo {author} {\bibfnamefont {P.~J.}\ \bibnamefont
  {Easter}}, \bibinfo {author} {\bibfnamefont {P.~D.}\ \bibnamefont {Lasky}},
  \bibinfo {author} {\bibfnamefont {A.~R.}\ \bibnamefont {Casey}}, \bibinfo
  {author} {\bibfnamefont {L.}~\bibnamefont {Rezzolla}}, \ and\ \bibinfo
  {author} {\bibfnamefont {K.}~\bibnamefont {Takami}},\ }\href {\doibase
  10.1103/PhysRevD.100.043005} {\bibfield  {journal} {\bibinfo  {journal}
  {Phys. Rev. D}\ }\textbf {\bibinfo {volume} {100}},\ \bibinfo {pages}
  {043005} (\bibinfo {year} {2019})},\ \Eprint
  {http://arxiv.org/abs/1811.11183} {arXiv:1811.11183 [gr-qc]} \BibitemShut
  {NoStop}%
\bibitem [{\citenamefont {Tsang}\ \emph {et~al.}(2019)\citenamefont {Tsang},
  \citenamefont {Dietrich},\ and\ \citenamefont {Van
  Den~Broeck}}]{Tsang:2019esi}%
  \BibitemOpen
  \bibfield  {author} {\bibinfo {author} {\bibfnamefont {K.~W.}\ \bibnamefont
  {Tsang}}, \bibinfo {author} {\bibfnamefont {T.}~\bibnamefont {Dietrich}}, \
  and\ \bibinfo {author} {\bibfnamefont {C.}~\bibnamefont {Van Den~Broeck}},\
  }\href {\doibase 10.1103/PhysRevD.100.044047} {\bibfield  {journal} {\bibinfo
   {journal} {Phys. Rev. D}\ }\textbf {\bibinfo {volume} {100}},\ \bibinfo
  {pages} {044047} (\bibinfo {year} {2019})},\ \Eprint
  {http://arxiv.org/abs/1907.02424} {arXiv:1907.02424 [gr-qc]} \BibitemShut
  {NoStop}%
\bibitem [{\citenamefont {Breschi}\ \emph {et~al.}(2019)\citenamefont
  {Breschi}, \citenamefont {Bernuzzi}, \citenamefont {Zappa}, \citenamefont
  {Agathos}, \citenamefont {Perego}, \citenamefont {Radice},\ and\
  \citenamefont {Nagar}}]{Breschi:2019srl}%
  \BibitemOpen
  \bibfield  {author} {\bibinfo {author} {\bibfnamefont {M.}~\bibnamefont
  {Breschi}}, \bibinfo {author} {\bibfnamefont {S.}~\bibnamefont {Bernuzzi}},
  \bibinfo {author} {\bibfnamefont {F.}~\bibnamefont {Zappa}}, \bibinfo
  {author} {\bibfnamefont {M.}~\bibnamefont {Agathos}}, \bibinfo {author}
  {\bibfnamefont {A.}~\bibnamefont {Perego}}, \bibinfo {author} {\bibfnamefont
  {D.}~\bibnamefont {Radice}}, \ and\ \bibinfo {author} {\bibfnamefont
  {A.}~\bibnamefont {Nagar}},\ }\href {\doibase 10.1103/PhysRevD.100.104029}
  {\bibfield  {journal} {\bibinfo  {journal} {Phys. Rev. D}\ }\textbf {\bibinfo
  {volume} {100}},\ \bibinfo {pages} {104029} (\bibinfo {year} {2019})},\
  \Eprint {http://arxiv.org/abs/1908.11418} {arXiv:1908.11418 [gr-qc]}
  \BibitemShut {NoStop}%
\bibitem [{\citenamefont {Alvarez-Castillo}\ and\ \citenamefont
  {Blaschke}(2017)}]{Alvarez_Castillo_2017}%
  \BibitemOpen
  \bibfield  {author} {\bibinfo {author} {\bibfnamefont {D.~E.}\ \bibnamefont
  {Alvarez-Castillo}}\ and\ \bibinfo {author} {\bibfnamefont {D.~B.}\
  \bibnamefont {Blaschke}},\ }\href {\doibase 10.1103/physrevc.96.045809}
  {\bibfield  {journal} {\bibinfo  {journal} {Physical Review C}\ }\textbf
  {\bibinfo {volume} {96}} (\bibinfo {year} {2017}),\
  10.1103/physrevc.96.045809}\BibitemShut {NoStop}%
\bibitem [{\citenamefont {Carney}\ \emph {et~al.}(2018)\citenamefont {Carney},
  \citenamefont {Wade},\ and\ \citenamefont {Irwin}}]{PhysRevD.98.063004}%
  \BibitemOpen
  \bibfield  {author} {\bibinfo {author} {\bibfnamefont {M.~F.}\ \bibnamefont
  {Carney}}, \bibinfo {author} {\bibfnamefont {L.~E.}\ \bibnamefont {Wade}}, \
  and\ \bibinfo {author} {\bibfnamefont {B.~S.}\ \bibnamefont {Irwin}},\ }\href
  {\doibase 10.1103/PhysRevD.98.063004} {\bibfield  {journal} {\bibinfo
  {journal} {Phys. Rev. D}\ }\textbf {\bibinfo {volume} {98}},\ \bibinfo
  {pages} {063004} (\bibinfo {year} {2018})}\BibitemShut {NoStop}%
\bibitem [{\citenamefont {Gamba}\ \emph {et~al.}(2019)\citenamefont {Gamba},
  \citenamefont {Read},\ and\ \citenamefont {Wade}}]{Gamba_2019}%
  \BibitemOpen
  \bibfield  {author} {\bibinfo {author} {\bibfnamefont {R.}~\bibnamefont
  {Gamba}}, \bibinfo {author} {\bibfnamefont {J.~S.}\ \bibnamefont {Read}}, \
  and\ \bibinfo {author} {\bibfnamefont {L.~E.}\ \bibnamefont {Wade}},\ }\href
  {\doibase 10.1088/1361-6382/ab5ba4} {\bibfield  {journal} {\bibinfo
  {journal} {Classical and Quantum Gravity}\ }\textbf {\bibinfo {volume}
  {37}},\ \bibinfo {pages} {025008} (\bibinfo {year} {2019})}\BibitemShut
  {NoStop}%
\bibitem [{\citenamefont {Wade}\ \emph {et~al.}(2014)\citenamefont {Wade},
  \citenamefont {Creighton}, \citenamefont {Ochsner}, \citenamefont {Lackey},
  \citenamefont {Farr}, \citenamefont {Littenberg},\ and\ \citenamefont
  {Raymond}}]{Wade:2014vqa}%
  \BibitemOpen
  \bibfield  {author} {\bibinfo {author} {\bibfnamefont {L.}~\bibnamefont
  {Wade}}, \bibinfo {author} {\bibfnamefont {J.~D.~E.}\ \bibnamefont
  {Creighton}}, \bibinfo {author} {\bibfnamefont {E.}~\bibnamefont {Ochsner}},
  \bibinfo {author} {\bibfnamefont {B.~D.}\ \bibnamefont {Lackey}}, \bibinfo
  {author} {\bibfnamefont {B.~F.}\ \bibnamefont {Farr}}, \bibinfo {author}
  {\bibfnamefont {T.~B.}\ \bibnamefont {Littenberg}}, \ and\ \bibinfo {author}
  {\bibfnamefont {V.}~\bibnamefont {Raymond}},\ }\href {\doibase
  10.1103/PhysRevD.89.103012} {\bibfield  {journal} {\bibinfo  {journal} {Phys.
  Rev.}\ }\textbf {\bibinfo {volume} {D89}},\ \bibinfo {pages} {103012}
  (\bibinfo {year} {2014})},\ \Eprint {http://arxiv.org/abs/1402.5156}
  {arXiv:1402.5156 [gr-qc]} \BibitemShut {NoStop}%
\bibitem [{\citenamefont {Veitch}\ \emph {et~al.}(2015)\citenamefont {Veitch}
  \emph {et~al.}}]{Veitch:2014wba}%
  \BibitemOpen
  \bibfield  {author} {\bibinfo {author} {\bibfnamefont {J.}~\bibnamefont
  {Veitch}} \emph {et~al.},\ }\href {\doibase 10.1103/PhysRevD.91.042003}
  {\bibfield  {journal} {\bibinfo  {journal} {Phys. Rev.}\ }\textbf {\bibinfo
  {volume} {D91}},\ \bibinfo {pages} {042003} (\bibinfo {year} {2015})},\
  \Eprint {http://arxiv.org/abs/1409.7215} {arXiv:1409.7215 [gr-qc]}
  \BibitemShut {NoStop}%
\bibitem [{\citenamefont {Abbott}\ \emph {et~al.}(2020)\citenamefont {Abbott}
  \emph {et~al.}}]{Abbott:2020uma}%
  \BibitemOpen
  \bibfield  {author} {\bibinfo {author} {\bibfnamefont {B.}~\bibnamefont
  {Abbott}} \emph {et~al.} (\bibinfo {collaboration} {LIGO Scientific,
  Virgo}),\ }\href {\doibase 10.3847/2041-8213/ab75f5} {\bibfield  {journal}
  {\bibinfo  {journal} {Astrophys. J. Lett.}\ }\textbf {\bibinfo {volume}
  {892}},\ \bibinfo {pages} {L3} (\bibinfo {year} {2020})},\ \Eprint
  {http://arxiv.org/abs/2001.01761} {arXiv:2001.01761 [astro-ph.HE]}
  \BibitemShut {NoStop}%
\bibitem [{\citenamefont {Sathyaprakash}\ and\ \citenamefont
  {Dhurandhar}(1991)}]{Sathyaprakash:1991mt}%
  \BibitemOpen
  \bibfield  {author} {\bibinfo {author} {\bibfnamefont {B.~S.}\ \bibnamefont
  {Sathyaprakash}}\ and\ \bibinfo {author} {\bibfnamefont {S.~V.}\ \bibnamefont
  {Dhurandhar}},\ }\href {\doibase 10.1103/PhysRevD.44.3819} {\bibfield
  {journal} {\bibinfo  {journal} {Phys. Rev.}\ }\textbf {\bibinfo {volume}
  {D44}},\ \bibinfo {pages} {3819} (\bibinfo {year} {1991})}\BibitemShut
  {NoStop}%
\bibitem [{\citenamefont {Blanchet}\ \emph {et~al.}(1995)\citenamefont
  {Blanchet}, \citenamefont {Damour}, \citenamefont {Iyer}, \citenamefont
  {Will},\ and\ \citenamefont {Wiseman}}]{Blanchet:1995ez}%
  \BibitemOpen
  \bibfield  {author} {\bibinfo {author} {\bibfnamefont {L.}~\bibnamefont
  {Blanchet}}, \bibinfo {author} {\bibfnamefont {T.}~\bibnamefont {Damour}},
  \bibinfo {author} {\bibfnamefont {B.~R.}\ \bibnamefont {Iyer}}, \bibinfo
  {author} {\bibfnamefont {C.~M.}\ \bibnamefont {Will}}, \ and\ \bibinfo
  {author} {\bibfnamefont {A.}~\bibnamefont {Wiseman}},\ }\href {\doibase
  10.1103/PhysRevLett.74.3515} {\bibfield  {journal} {\bibinfo  {journal}
  {Phys.Rev.Lett.}\ }\textbf {\bibinfo {volume} {74}},\ \bibinfo {pages} {3515}
  (\bibinfo {year} {1995})}\BibitemShut {NoStop}%
\bibitem [{\citenamefont {Damour}\ \emph {et~al.}(2001)\citenamefont {Damour},
  \citenamefont {Jaranowski},\ and\ \citenamefont {Schaefer}}]{Damour:2001bu}%
  \BibitemOpen
  \bibfield  {author} {\bibinfo {author} {\bibfnamefont {T.}~\bibnamefont
  {Damour}}, \bibinfo {author} {\bibfnamefont {P.}~\bibnamefont {Jaranowski}},
  \ and\ \bibinfo {author} {\bibfnamefont {G.}~\bibnamefont {Schaefer}},\
  }\href {\doibase 10.1016/S0370-2693(01)00642-6} {\bibfield  {journal}
  {\bibinfo  {journal} {Phys. Lett. B}\ }\textbf {\bibinfo {volume} {513}},\
  \bibinfo {pages} {147} (\bibinfo {year} {2001})},\ \Eprint
  {http://arxiv.org/abs/gr-qc/0105038} {arXiv:gr-qc/0105038} \BibitemShut
  {NoStop}%
\bibitem [{\citenamefont {Goldberger}\ and\ \citenamefont
  {Rothstein}(2006)}]{Goldberger:2004jt}%
  \BibitemOpen
  \bibfield  {author} {\bibinfo {author} {\bibfnamefont {W.~D.}\ \bibnamefont
  {Goldberger}}\ and\ \bibinfo {author} {\bibfnamefont {I.~Z.}\ \bibnamefont
  {Rothstein}},\ }\href {\doibase 10.1103/PhysRevD.73.104029} {\bibfield
  {journal} {\bibinfo  {journal} {Phys. Rev. D}\ }\textbf {\bibinfo {volume}
  {73}},\ \bibinfo {pages} {104029} (\bibinfo {year} {2006})},\ \Eprint
  {http://arxiv.org/abs/hep-th/0409156} {arXiv:hep-th/0409156} \BibitemShut
  {NoStop}%
\bibitem [{\citenamefont {Blanchet}\ \emph {et~al.}(2005)\citenamefont
  {Blanchet}, \citenamefont {Damour}, \citenamefont {Esposito-Farese},\ and\
  \citenamefont {Iyer}}]{Blanchet:2005tk}%
  \BibitemOpen
  \bibfield  {author} {\bibinfo {author} {\bibfnamefont {L.}~\bibnamefont
  {Blanchet}}, \bibinfo {author} {\bibfnamefont {T.}~\bibnamefont {Damour}},
  \bibinfo {author} {\bibfnamefont {G.}~\bibnamefont {Esposito-Farese}}, \ and\
  \bibinfo {author} {\bibfnamefont {B.~R.}\ \bibnamefont {Iyer}},\ }\href
  {\doibase 10.1103/PhysRevD.71.124004} {\bibfield  {journal} {\bibinfo
  {journal} {Phys. Rev. D}\ }\textbf {\bibinfo {volume} {71}},\ \bibinfo
  {pages} {124004} (\bibinfo {year} {2005})},\ \Eprint
  {http://arxiv.org/abs/gr-qc/0503044} {arXiv:gr-qc/0503044} \BibitemShut
  {NoStop}%
\bibitem [{\citenamefont {Boh{\'e}}\ \emph {et~al.}(2013)\citenamefont
  {Boh{\'e}}, \citenamefont {Marsat},\ and\ \citenamefont
  {Blanchet}}]{Bohe:2013cla}%
  \BibitemOpen
  \bibfield  {author} {\bibinfo {author} {\bibfnamefont {A.}~\bibnamefont
  {Boh{\'e}}}, \bibinfo {author} {\bibfnamefont {S.}~\bibnamefont {Marsat}}, \
  and\ \bibinfo {author} {\bibfnamefont {L.}~\bibnamefont {Blanchet}},\ }\href
  {\doibase 10.1088/0264-9381/30/13/135009} {\bibfield  {journal} {\bibinfo
  {journal} {Class. Quant. Grav.}\ }\textbf {\bibinfo {volume} {30}},\ \bibinfo
  {pages} {135009} (\bibinfo {year} {2013})},\ \Eprint
  {http://arxiv.org/abs/1303.7412} {arXiv:1303.7412 [gr-qc]} \BibitemShut
  {NoStop}%
\bibitem [{\citenamefont {Arun}\ \emph {et~al.}(2009)\citenamefont {Arun},
  \citenamefont {Buonanno}, \citenamefont {Faye},\ and\ \citenamefont
  {Ochsner}}]{Arun:2008kb}%
  \BibitemOpen
  \bibfield  {author} {\bibinfo {author} {\bibfnamefont {K.~G.}\ \bibnamefont
  {Arun}}, \bibinfo {author} {\bibfnamefont {A.}~\bibnamefont {Buonanno}},
  \bibinfo {author} {\bibfnamefont {G.}~\bibnamefont {Faye}}, \ and\ \bibinfo
  {author} {\bibfnamefont {E.}~\bibnamefont {Ochsner}},\ }\href {\doibase
  10.1103/PhysRevD.79.104023, 10.1103/PhysRevD.84.049901} {\bibfield  {journal}
  {\bibinfo  {journal} {Phys. Rev.}\ }\textbf {\bibinfo {volume} {D79}},\
  \bibinfo {pages} {104023} (\bibinfo {year} {2009})},\ \bibinfo {note}
  {[Erratum: Phys. Rev.D84,049901(2011)]},\ \Eprint
  {http://arxiv.org/abs/0810.5336} {arXiv:0810.5336 [gr-qc]} \BibitemShut
  {NoStop}%
\bibitem [{\citenamefont {Mikoczi}\ \emph {et~al.}(2005)\citenamefont
  {Mikoczi}, \citenamefont {Vasuth},\ and\ \citenamefont
  {Gergely}}]{Mikoczi:2005dn}%
  \BibitemOpen
  \bibfield  {author} {\bibinfo {author} {\bibfnamefont {B.}~\bibnamefont
  {Mikoczi}}, \bibinfo {author} {\bibfnamefont {M.}~\bibnamefont {Vasuth}}, \
  and\ \bibinfo {author} {\bibfnamefont {L.~A.}\ \bibnamefont {Gergely}},\
  }\href {\doibase 10.1103/PhysRevD.71.124043} {\bibfield  {journal} {\bibinfo
  {journal} {Phys. Rev. D}\ }\textbf {\bibinfo {volume} {71}},\ \bibinfo
  {pages} {124043} (\bibinfo {year} {2005})},\ \Eprint
  {http://arxiv.org/abs/astro-ph/0504538} {arXiv:astro-ph/0504538} \BibitemShut
  {NoStop}%
\bibitem [{\citenamefont {Boh{\'e}}\ \emph {et~al.}(2015)\citenamefont
  {Boh{\'e}}, \citenamefont {Faye}, \citenamefont {Marsat},\ and\ \citenamefont
  {Porter}}]{Bohe:2015ana}%
  \BibitemOpen
  \bibfield  {author} {\bibinfo {author} {\bibfnamefont {A.}~\bibnamefont
  {Boh{\'e}}}, \bibinfo {author} {\bibfnamefont {G.}~\bibnamefont {Faye}},
  \bibinfo {author} {\bibfnamefont {S.}~\bibnamefont {Marsat}}, \ and\ \bibinfo
  {author} {\bibfnamefont {E.~K.}\ \bibnamefont {Porter}},\ }\href {\doibase
  10.1088/0264-9381/32/19/195010} {\bibfield  {journal} {\bibinfo  {journal}
  {Class. Quant. Grav.}\ }\textbf {\bibinfo {volume} {32}},\ \bibinfo {pages}
  {195010} (\bibinfo {year} {2015})},\ \Eprint
  {http://arxiv.org/abs/1501.01529} {arXiv:1501.01529 [gr-qc]} \BibitemShut
  {NoStop}%
\bibitem [{\citenamefont {Mishra}\ \emph {et~al.}(2016)\citenamefont {Mishra},
  \citenamefont {Kela}, \citenamefont {Arun},\ and\ \citenamefont
  {Faye}}]{Mishra:2016whh}%
  \BibitemOpen
  \bibfield  {author} {\bibinfo {author} {\bibfnamefont {C.~K.}\ \bibnamefont
  {Mishra}}, \bibinfo {author} {\bibfnamefont {A.}~\bibnamefont {Kela}},
  \bibinfo {author} {\bibfnamefont {K.}~\bibnamefont {Arun}}, \ and\ \bibinfo
  {author} {\bibfnamefont {G.}~\bibnamefont {Faye}},\ }\href {\doibase
  10.1103/PhysRevD.93.084054} {\bibfield  {journal} {\bibinfo  {journal} {Phys.
  Rev. D}\ }\textbf {\bibinfo {volume} {93}},\ \bibinfo {pages} {084054}
  (\bibinfo {year} {2016})},\ \Eprint {http://arxiv.org/abs/1601.05588}
  {arXiv:1601.05588 [gr-qc]} \BibitemShut {NoStop}%
\bibitem [{\citenamefont {Vines}\ \emph {et~al.}(2011)\citenamefont {Vines},
  \citenamefont {Flanagan},\ and\ \citenamefont {Hinderer}}]{Vines:2011ud}%
  \BibitemOpen
  \bibfield  {author} {\bibinfo {author} {\bibfnamefont {J.}~\bibnamefont
  {Vines}}, \bibinfo {author} {\bibfnamefont {E.~E.}\ \bibnamefont {Flanagan}},
  \ and\ \bibinfo {author} {\bibfnamefont {T.}~\bibnamefont {Hinderer}},\
  }\href {\doibase 10.1103/PhysRevD.83.084051} {\bibfield  {journal} {\bibinfo
  {journal} {Phys. Rev.}\ }\textbf {\bibinfo {volume} {D83}},\ \bibinfo {pages}
  {084051} (\bibinfo {year} {2011})},\ \Eprint {http://arxiv.org/abs/1101.1673}
  {arXiv:1101.1673 [gr-qc]} \BibitemShut {NoStop}%
\bibitem [{\citenamefont {Samajdar}\ and\ \citenamefont
  {Dietrich}(2020)}]{Samajdar:2020xrd}%
  \BibitemOpen
  \bibfield  {author} {\bibinfo {author} {\bibfnamefont {A.}~\bibnamefont
  {Samajdar}}\ and\ \bibinfo {author} {\bibfnamefont {T.}~\bibnamefont
  {Dietrich}},\ }\href@noop {} {\  (\bibinfo {year} {2020})},\ \Eprint
  {http://arxiv.org/abs/2002.07918} {arXiv:2002.07918 [gr-qc]} \BibitemShut
  {NoStop}%
\bibitem [{\citenamefont {Samajdar}\ and\ \citenamefont
  {Dietrich}(2019)}]{Samajdar:2019ulq}%
  \BibitemOpen
  \bibfield  {author} {\bibinfo {author} {\bibfnamefont {A.}~\bibnamefont
  {Samajdar}}\ and\ \bibinfo {author} {\bibfnamefont {T.}~\bibnamefont
  {Dietrich}},\ }\href {\doibase 10.1103/PhysRevD.100.024046} {\bibfield
  {journal} {\bibinfo  {journal} {Phys. Rev. D}\ }\textbf {\bibinfo {volume}
  {100}},\ \bibinfo {pages} {024046} (\bibinfo {year} {2019})},\ \Eprint
  {http://arxiv.org/abs/1905.03118} {arXiv:1905.03118 [gr-qc]} \BibitemShut
  {NoStop}%
\bibitem [{\citenamefont {O'Shaughnessy}\ \emph {et~al.}(2008)\citenamefont
  {O'Shaughnessy}, \citenamefont {Kim}, \citenamefont {Kalogera},\ and\
  \citenamefont {Belczynski}}]{O_Shaughnessy_2008}%
  \BibitemOpen
  \bibfield  {author} {\bibinfo {author} {\bibfnamefont {R.}~\bibnamefont
  {O'Shaughnessy}}, \bibinfo {author} {\bibfnamefont {C.}~\bibnamefont {Kim}},
  \bibinfo {author} {\bibfnamefont {V.}~\bibnamefont {Kalogera}}, \ and\
  \bibinfo {author} {\bibfnamefont {K.}~\bibnamefont {Belczynski}},\ }\href
  {\doibase 10.1086/523620} {\bibfield  {journal} {\bibinfo  {journal} {The
  Astrophysical Journal}\ }\textbf {\bibinfo {volume} {672}},\ \bibinfo {pages}
  {479} (\bibinfo {year} {2008})}\BibitemShut {NoStop}%
\bibitem [{\citenamefont {Skilling}(2006)}]{skilling2006}%
  \BibitemOpen
  \bibfield  {author} {\bibinfo {author} {\bibfnamefont {J.}~\bibnamefont
  {Skilling}},\ }\href {\doibase 10.1214/06-BA127} {\bibfield  {journal}
  {\bibinfo  {journal} {Bayesian Anal.}\ }\textbf {\bibinfo {volume} {1}},\
  \bibinfo {pages} {833} (\bibinfo {year} {2006})}\BibitemShut {NoStop}%
\bibitem [{\citenamefont {{LIGO Scientific Collaboration}}(2018)}]{lalsuite}%
  \BibitemOpen
  \bibfield  {author} {\bibinfo {author} {\bibnamefont {{LIGO Scientific
  Collaboration}}},\ }\href {\doibase 10.7935/GT1W-FZ16} {\enquote {\bibinfo
  {title} {{LIGO} {A}lgorithm {L}ibrary - {LALS}uite},}\ }\bibinfo
  {howpublished} {free software (GPL)} (\bibinfo {year} {2018})\BibitemShut
  {NoStop}%
\bibitem [{\citenamefont {Feroz}\ \emph {et~al.}(2009)\citenamefont {Feroz},
  \citenamefont {Hobson},\ and\ \citenamefont {Bridges}}]{Feroz_2009}%
  \BibitemOpen
  \bibfield  {author} {\bibinfo {author} {\bibfnamefont {F.}~\bibnamefont
  {Feroz}}, \bibinfo {author} {\bibfnamefont {M.~P.}\ \bibnamefont {Hobson}}, \
  and\ \bibinfo {author} {\bibfnamefont {M.}~\bibnamefont {Bridges}},\ }\href
  {\doibase 10.1111/j.1365-2966.2009.14548.x} {\bibfield  {journal} {\bibinfo
  {journal} {Monthly Notices of the Royal Astronomical Society}\ }\textbf
  {\bibinfo {volume} {398}},\ \bibinfo {pages} {1601–1614} (\bibinfo {year}
  {2009})}\BibitemShut {NoStop}%
\bibitem [{\citenamefont {{Buchner, J.}}\ \emph {et~al.}(2014)\citenamefont
  {{Buchner, J.}}, \citenamefont {{Georgakakis, A.}}, \citenamefont {{Nandra,
  K.}}, \citenamefont {{Hsu, L.}}, \citenamefont {{Rangel, C.}}, \citenamefont
  {{Brightman, M.}}, \citenamefont {{Merloni, A.}}, \citenamefont {{Salvato,
  M.}}, \citenamefont {{Donley, J.}},\ and\ \citenamefont {{Kocevski,
  D.}}}]{pymultinest1}%
  \BibitemOpen
  \bibfield  {author} {\bibinfo {author} {\bibnamefont {{Buchner, J.}}},
  \bibinfo {author} {\bibnamefont {{Georgakakis, A.}}}, \bibinfo {author}
  {\bibnamefont {{Nandra, K.}}}, \bibinfo {author} {\bibnamefont {{Hsu, L.}}},
  \bibinfo {author} {\bibnamefont {{Rangel, C.}}}, \bibinfo {author}
  {\bibnamefont {{Brightman, M.}}}, \bibinfo {author} {\bibnamefont {{Merloni,
  A.}}}, \bibinfo {author} {\bibnamefont {{Salvato, M.}}}, \bibinfo {author}
  {\bibnamefont {{Donley, J.}}}, \ and\ \bibinfo {author} {\bibnamefont
  {{Kocevski, D.}}},\ }\href {\doibase 10.1051/0004-6361/201322971} {\bibfield
  {journal} {\bibinfo  {journal} {A\&A}\ }\textbf {\bibinfo {volume} {564}},\
  \bibinfo {pages} {A125} (\bibinfo {year} {2014})}\BibitemShut {NoStop}%
\bibitem [{\citenamefont {Miller}\ \emph
  {et~al.}(2019{\natexlab{b}})\citenamefont {Miller}, \citenamefont
  {Chirenti},\ and\ \citenamefont {Lamb}}]{Miller_2019}%
  \BibitemOpen
  \bibfield  {author} {\bibinfo {author} {\bibfnamefont {M.~C.}\ \bibnamefont
  {Miller}}, \bibinfo {author} {\bibfnamefont {C.}~\bibnamefont {Chirenti}}, \
  and\ \bibinfo {author} {\bibfnamefont {F.~K.}\ \bibnamefont {Lamb}},\ }\href
  {\doibase 10.3847/1538-4357/ab4ef9} {\bibfield  {journal} {\bibinfo
  {journal} {The Astrophysical Journal}\ }\textbf {\bibinfo {volume} {888}},\
  \bibinfo {pages} {12} (\bibinfo {year} {2019}{\natexlab{b}})}\BibitemShut
  {NoStop}%
\bibitem [{\citenamefont {Foley}\ \emph {et~al.}(2020)\citenamefont {Foley},
  \citenamefont {Coulter}, \citenamefont {Kilpatrick}, \citenamefont {Piro},
  \citenamefont {Ramirez-Ruiz},\ and\ \citenamefont {Schwab}}]{Foley:2020kus}%
  \BibitemOpen
  \bibfield  {author} {\bibinfo {author} {\bibfnamefont {R.~J.}\ \bibnamefont
  {Foley}}, \bibinfo {author} {\bibfnamefont {D.~A.}\ \bibnamefont {Coulter}},
  \bibinfo {author} {\bibfnamefont {C.~D.}\ \bibnamefont {Kilpatrick}},
  \bibinfo {author} {\bibfnamefont {A.~L.}\ \bibnamefont {Piro}}, \bibinfo
  {author} {\bibfnamefont {E.}~\bibnamefont {Ramirez-Ruiz}}, \ and\ \bibinfo
  {author} {\bibfnamefont {J.}~\bibnamefont {Schwab}},\ }\href {\doibase
  10.1093/mnras/staa725} {\bibfield  {journal} {\bibinfo  {journal} {Mon. Not.
  Roy. Astron. Soc.}\ }\textbf {\bibinfo {volume} {494}},\ \bibinfo {pages}
  {190} (\bibinfo {year} {2020})},\ \Eprint {http://arxiv.org/abs/2002.00956}
  {arXiv:2002.00956 [astro-ph.HE]} \BibitemShut {NoStop}%
\bibitem [{\citenamefont {Han}\ \emph {et~al.}(2020)\citenamefont {Han},
  \citenamefont {Tang}, \citenamefont {Hu}, \citenamefont {Li}, \citenamefont
  {Jiang}, \citenamefont {Jin}, \citenamefont {Fan},\ and\ \citenamefont
  {Wei}}]{Han:2020qmn}%
  \BibitemOpen
  \bibfield  {author} {\bibinfo {author} {\bibfnamefont {M.-Z.}\ \bibnamefont
  {Han}}, \bibinfo {author} {\bibfnamefont {S.-P.}\ \bibnamefont {Tang}},
  \bibinfo {author} {\bibfnamefont {Y.-M.}\ \bibnamefont {Hu}}, \bibinfo
  {author} {\bibfnamefont {Y.-J.}\ \bibnamefont {Li}}, \bibinfo {author}
  {\bibfnamefont {J.-L.}\ \bibnamefont {Jiang}}, \bibinfo {author}
  {\bibfnamefont {Z.-P.}\ \bibnamefont {Jin}}, \bibinfo {author} {\bibfnamefont
  {Y.-Z.}\ \bibnamefont {Fan}}, \ and\ \bibinfo {author} {\bibfnamefont
  {D.-M.}\ \bibnamefont {Wei}},\ }\href {\doibase 10.3847/2041-8213/ab745a}
  {\bibfield  {journal} {\bibinfo  {journal} {Astrophys. J. Lett.}\ }\textbf
  {\bibinfo {volume} {891}},\ \bibinfo {pages} {L5} (\bibinfo {year} {2020})},\
  \Eprint {http://arxiv.org/abs/2001.07882} {arXiv:2001.07882 [astro-ph.HE]}
  \BibitemShut {NoStop}%
\bibitem [{\citenamefont {Kyutoku}\ \emph {et~al.}(2020)\citenamefont
  {Kyutoku}, \citenamefont {Fujibayashi}, \citenamefont {Hayashi},
  \citenamefont {Kawaguchi}, \citenamefont {Kiuchi}, \citenamefont {Shibata},\
  and\ \citenamefont {Tanaka}}]{Kyutoku:2020xka}%
  \BibitemOpen
  \bibfield  {author} {\bibinfo {author} {\bibfnamefont {K.}~\bibnamefont
  {Kyutoku}}, \bibinfo {author} {\bibfnamefont {S.}~\bibnamefont
  {Fujibayashi}}, \bibinfo {author} {\bibfnamefont {K.}~\bibnamefont
  {Hayashi}}, \bibinfo {author} {\bibfnamefont {K.}~\bibnamefont {Kawaguchi}},
  \bibinfo {author} {\bibfnamefont {K.}~\bibnamefont {Kiuchi}}, \bibinfo
  {author} {\bibfnamefont {M.}~\bibnamefont {Shibata}}, \ and\ \bibinfo
  {author} {\bibfnamefont {M.}~\bibnamefont {Tanaka}},\ }\href {\doibase
  10.3847/2041-8213/ab6e70} {\bibfield  {journal} {\bibinfo  {journal}
  {Astrophys. J. Lett.}\ }\textbf {\bibinfo {volume} {890}},\ \bibinfo {pages}
  {L4} (\bibinfo {year} {2020})},\ \Eprint {http://arxiv.org/abs/2001.04474}
  {arXiv:2001.04474 [astro-ph.HE]} \BibitemShut {NoStop}%
\bibitem [{\citenamefont {Collaboration}\ and\ \citenamefont {the
  Virgo~Collaboration}(2019)}]{GW170817_PE_samples}%
  \BibitemOpen
  \bibfield  {author} {\bibinfo {author} {\bibfnamefont {T.~L.~S.}\
  \bibnamefont {Collaboration}}\ and\ \bibinfo {author} {\bibnamefont {the
  Virgo~Collaboration}},\ }\href@noop {} {\enquote {\bibinfo {title} {Parameter
  estimation sample release for gwtc-1},}\ }\bibinfo {howpublished}
  {\url{https://dcc.ligo.org/LIGO-P1800370/public}} (\bibinfo {year}
  {2019})\BibitemShut {NoStop}%
\bibitem [{\citenamefont {Collaboration}\ and\ \citenamefont {the
  Virgo~Collaboration}(2020)}]{GW190425_PE_samples}%
  \BibitemOpen
  \bibfield  {author} {\bibinfo {author} {\bibfnamefont {T.~L.~S.}\
  \bibnamefont {Collaboration}}\ and\ \bibinfo {author} {\bibnamefont {the
  Virgo~Collaboration}},\ }\href@noop {} {\enquote {\bibinfo {title} {Parameter
  estimation sample release for gw190425},}\ }\bibinfo {howpublished}
  {\url{https://dcc.ligo.org/LIGO-P2000026/public}} (\bibinfo {year}
  {2020})\BibitemShut {NoStop}%
\bibitem [{\citenamefont {Jeffreys}(1961)}]{Jeffreys61}%
  \BibitemOpen
  \bibfield  {author} {\bibinfo {author} {\bibfnamefont {H.}~\bibnamefont
  {Jeffreys}},\ }\href@noop {} {\emph {\bibinfo {title} {Theory of
  Probability}}},\ \bibinfo {edition} {3rd}\ ed.\ (\bibinfo  {publisher}
  {Oxford},\ \bibinfo {address} {Oxford, England},\ \bibinfo {year}
  {1961})\BibitemShut {NoStop}%
\end{thebibliography}%

\end{document}